\documentclass[12pt]{article}
\usepackage{epsf,epsfig,cite,axodraw}

\newcount\timecount
\newcount\hours \newcount\minutes  \newcount\temp \newcount\pmhours

\hours = \time
\divide\hours by 60
\temp = \hours
\multiply\temp by 60
\minutes = \time
\advance\minutes by -\temp
\def\hour{\the\hours}
\def\minute{\ifnum\minutes<10 0\the\minutes
            \else\the\minutes\fi}
\def\clock{
\ifnum\hours=0 12:\minute\ AM
\else\ifnum\hours<12 \hour:\minute\ AM
      \else\ifnum\hours=12 12:\minute\ PM
            \else\ifnum\hours>12
                 \pmhours=\hours
                 \advance\pmhours by -12
                 \the\pmhours:\minute\ PM
                 \fi
            \fi
      \fi
\fi
}

\def\monthname{\relax\ifcase\month 0/\or January\or February\or
   March\or April\or May\or June\or July\or August\or September\or
   October\or November\or December\else\number\month/\fi}

\def\bold#1{\setbox0=\hbox{$#1$}%
     \kern-.025em\copy0\kern-\wd0
     \kern.05em\copy0\kern-\wd0
     \kern-.025em\raise.0433em\box0 }

\def\gappeq{\mathrel{\rlap {\raise.5ex\hbox{$>$}}
{\lower.5ex\hbox{$\sim$}}}}

\def\lappeq{\mathrel{\rlap{\raise.5ex\hbox{$<$}}
{\lower.5ex\hbox{$\sim$}}}}

\input paperdef

\begin{document}
\begin{titlepage}
\pagestyle{empty}
\baselineskip=21pt
\begin{flushright}
hep-ph/0602220 \hfill
CERN--PH--TH/2006-028\\
DCPT/06/24, IPPP/06/12 \hfill
UMN--TH--2432/06, FTPI--MINN--06/05
\end{flushright}
\vskip 0.3in
\begin{center}
{\large{\bf Phenomenological Indications of the Scale of Supersymmetry}}
\end{center}
\begin{center}
\vskip 0.05in
{{\bf John Ellis}$^1$, 
{\bf Sven Heinemeyer}$^2$,
{\bf Keith A.~Olive}$^{3}$
and {\bf Georg Weiglein}$^{4}$}\\
\vskip 0.05in
{\it
$^1${TH Division, Physics Department, CERN, Geneva, Switzerland}\\
$^2$Depto.\ de F\'isica Te\'orica, Universidad de Zaragoza, 50009 Zaragoza,
Spain \\
$^3${William I.\ Fine Theoretical Physics Institute,\\
University of Minnesota, Minneapolis, MN~55455, USA}\\
$^4${IPPP, University of Durham, Durham DH1~3LE, UK}\\
}
\vskip 0.1in
{\bf Abstract}
\end{center}
\baselineskip=18pt \noindent

{\small
Electroweak precision measurements can provide indirect information about
the possible scale of supersymmetry already at the present level of
accuracy.  We update the present-day sensitivities of precision data using
$\mt = 172.7 \pm 2.9 \gev$ for the experimental value of the top-quark
mass, within 
the constrained minimal supersymmetric extension of the Standard Model
(CMSSM), in which there are three independent soft supersymmetry-breaking
parameters $m_{1/2}, m_0$ and $A_0$. In addition to $\MW$ and $\sweff$,
the analysis is based on $(g-2)_\mu$, $\br(b \to s \ga)$ and the lightest
MSSM Higgs boson mass, $\Mh$. Assuming initially that the lightest
supersymmetric particle (LSP) is a neutralino, we display the CMSSM
results as functions of $m_{1/2}$, fixing $m_0$ so as to obtain the cold
dark matter density allowed by WMAP and other cosmological data for
specific values of $A_0$, $\tb$ and $\mu > 0$. 
For a sample value of $\tb$ we analyze how the global $\chi^2$
function would change following a possible future evolution of
the experimental central value of $\mt$ and its error.
In a second step, we extend
the analysis to other constrained versions of the MSSM: the NUHM in which
the soft supersymmetry-breaking contributions to the Higgs masses are
independent and the Higgs mixing parameter $\mu$ and the pseudoscalar
Higgs mass 
$\MA$ become additional free parameters compared to the CMSSM, a VCMSSM in
which the bilinear soft supersymmetry breaking parameter $B_0 = A_0 -m_0$,
and the GDM in which the LSP is the gravitino. In all scenarios we find
indications for relatively light soft supersymmetry-breaking masses,
offering good prospects for the LHC and the ILC, and in some cases
also for the Tevatron. 

}


\vskip 0.15in
\leftline{CERN--PH--TH/2006-028}
\leftline{February 2006}
\end{titlepage}
\baselineskip=18pt


\section{Introduction}

We have recently analyzed the indications provided by current experimental
data concerning the possible scale of
supersymmetry~\cite{ehow3,LCWS05ehow3} within the framework of the minimal
supersymmetric extension of the Standard Model (MSSM)~\cite{susy,susy2},
assuming that the soft supersymmetry-breaking scalar masses $m_0$, gaugino
masses $m_{1/2}$ and tri-linear parameters $A_0$ were each constrained to
be universal at the input GUT scale, with the gravitino heavy and the
lightest supersymmetric particle (LSP) being the lightest neutralino
$\neu{1}$, a framework often referred to as the constrained
MSSM (CMSSM). 
However, this is not the only possible scenario for supersymmetric
phenomenology. For example, the soft supersymmetry-breaking scalar masses
$m_0$ might not be universal, in particular those of the MSSM Higgs
bosons, a framework we term the NUHM~\cite{nonu,NUHM}. Alternatively, one may
postulate supplementary relations for the soft tri- and bilinear
supersymmetry-breaking parameters $A_0, B_0$ such as those inspired by
specific supergravity scenarios, a framework we term the VCMSSM \cite{VCMSSM}.
Additionally, if one assumes universality between $m_0$ and the gravitino
mass, as in minimal supergravity (mSUGRA), the gravitino might be the LSP
and constitute the dark matter \cite{ekn}, a framework known as the
GDM~\cite{GDM,eov,feng}. 

It is well known that predicting the masses of supersymmetric particles
using precision low-energy data is more difficult than it was for the top
quark or even the Higgs boson. This is because the Standard Model (SM) is
renormalizable, so decoupling theorems imply that many low-energy
observables are insensitive to heavy sparticles~\cite{decoupling}. On
the other hand, 
supersymmetry may provide an important contribution to loop effects that
are rare or forbidden within the Standard Model. In fact, we found
previously~\cite{ehow3} that present data on the electroweak precision
observables $\MW$ and $\sweff$, as well as the loop induced quantities
$(g-2)_\mu$ and $\br(b \to s \ga)$ (see \citere{PomssmRep} for a review),
may already be providing interesting indirect information on the scale of
supersymmetry breaking, at least within the context of the CMSSM with a
neutralino LSP. In that framework, the range of $m_0$ is very restricted
by the cold dark matter density $\Omega_\chi h^2$ determined by WMAP and
other observations, for any set of assumed values of $\tb, m_{1/2}$ and
the trilinear soft supersymmetry-breaking parameter
$A_0$~\cite{WMAPstrips,them}.  We fixed $m_0$ so as to satisfy this
density constraint, $0.094 < \Omega_{\rm CDM} h^2 < 0.129$~\cite{WMAP},
and then analyzed the indirect information as a function of $m_{1/2}$ for
$\tb = 10, 50$. This was done for various discrete values of $A_0$ and as
a scan in the ($m_{1/2}, A_0$) plane.

Within the CMSSM and using the (then) preferred range $\mt = 178.0 \pm 4.3
\gev$~\cite{oldmt}, we found previously~\cite{ehow3,LCWS05ehow3} a
preference for low values of $m_{1/2}$, particularly for $\tb = 10$, that
exhibited also a moderate sensitivity to $A_0$. Our first step in this
paper is to update our previous analysis, taking into account the
newer preferred range $\mt = 172.7 \pm 2.9 \gev$~\cite{newmt}, and
providing a {\it vade mecum} for understanding the implications of any
further evolution in the preferred range and experimental error
of $\mt$. 
The new experimental value
of $\mt$ reduces substantially the mass expected for the lightest MSSM
Higgs boson, $\Mh$, for any given values of $m_{1/2}, m_0, \tb$ and $A_0$,
strengthening the constraints on $m_{1/2}$. 
We therefore improve our analysis by incorporating
the full likelihood information provided by the final
results of the LEP search for a Standard Model-like Higgs
boson~\cite{LEPHiggsSM,LEPHiggsMSSM}.

Other recent analyses~\cite{other} in the framework of the CMSSM differ
from our analysis by the omission of certain observables such as $\MW$,
$\sweff$ or $\Mh$, or in their treatment of the 95\% C.L.\ exclusion
bound for $\Mh$.  The other analyses find a preference for 
somewhat larger $\tb$,
mostly due to the fact that $\MW$ and $\sweff$ are either ignored or
treated differently.

The main purpose of the present paper is to analyze the
sensitivity of the preference for a low value of $m_{1/2}$ to some of 
the restrictive assumptions we
introduced into the analysis, exploring the ranges of parameters that
would be preferred in alternative NUHM, VCMSSM and GDM scenarios.

The NUHM has two additional parameters as compared to the
CMSSM, namely the degrees of non-universality of the soft
supersymmetry-breaking scalar masses for the two Higgs doublets \cite{NUHM}. 
They can be traded for two quantities measurable at low energies,
such as the Higgs mixing parameter $\mu$
and the $\cp$-odd Higgs boson mass, $\MA$. We explore here the
possible sensitivities to these parameters within the NUHM. It would
take prohibitive effort to analyze systematically all this
multi-dimensional parameter space. Therefore, we focus here on
analyzing a limited number of NUHM scenarios, corresponding to
two-dimensional subspaces of parameters that generalize specific 
favoured CMSSM
scenarios, with the idea of exploring whether the dependences on the 
additional NUHM variables are capable of modifying significantly the
CMSSM preference for relatively small values of $m_{1/2}$ and exploring 
possible preferences for the values of other model parameters.

On the other hand, in very constrained variants of the MSSM (VCMSSM) in
which one postulates a relation between the tri- and bilinear soft
supersymmetry-breaking parameters: $A_0 = B_0 + m_0$,%
\footnote{
Our notation for the $A_0$ and $B_0$ parameters follows that
which is standard in supergravity models (see e.g.\ \citere{susy}),
namely the coupling in the scalar potential is given by
$A_0 \, g^{(3)} + B_0 \, g^{(2)}$ for the tri- and bi-linear
superpotential terms 
$g^{(3)}$ and  $g^{(2)}$, respectively.  This differs from the sign
convention used in many publicly available codes, see e.g.\ \citere{spa}. 
}%
~motivated by simple
supergravity, the dimensionality of the model parameter space is reduced
compared with that in the CMSSM. The supersymmetric vacuum conditions then
fix the ratio of MSSM Higgs vacuum expectation values $\tb$ as a
function of $m_{1/2}, m_0$ and $A_0$ \cite{VCMSSM}. We study 
the cases $A_0/m_0 = 0, 0.75, 3 - \sqrt{3}$ and 2, which are
compatible with neutralino dark matter for extended ranges of $m_{1/2}$,
and we discuss the preferred ranges of $m_{1/2}$ and $\tb$ in each case.

In general, yet another relevant parameter, namely the gravitino mass,
must be taken into account, leading to the possibility that the LSP is the
gravitino, in which case it would provide
dark matter, the GDM scenario. In order to simplify the analysis of GDM 
in a motivated manner, we restrict our attention to scenarios 
inspired by minimal
supergravity (mSUGRA), in which the gravitino mass is
constrained to equal $m_0$ at the input GUT scale, and the trilinear
and bilinear soft supersymmetry-breaking parameters are again related by
$A_0 = B_0 + m_0$.  In the cases we analyze in this paper, namely 
$A_0/m_0 = 0, 3/4, 3 - \sqrt{3}, 2$, the regions~%
\footnote{
The case $A_0 = 3 - \sqrt{3}$ is motivated by the
simplest Polonyi model of Planck-scale supersymmetry breaking
\cite{pol}.
}%
~of the $(m_{1/2}, m_0)$ plane allowed by cosmological constraints then take
the form of wedges located at smaller values of $m_0$ than those
allowed in CMSSM scenarios \cite{GDM,eov}. We scan here some of the
GDM wedges allowed by  
cosmology, exploring whether the new ranges of $m_0$ may lead to
preferences for different values of $m_{1/2}$.

We have performed $\chi^2$ fits in all scenarios, and
our main results are as follows. Within the CMSSM, we find that the new,
lower value of $\mt$ and new treatment of the constraints from the LEP
Higgs search do not change
greatly the values of $m_{1/2}$ that were preferred previously
\cite{ehow3,LCWS05ehow3}.  For example, 
the 90\% C.L.\ upper bound on $m_{1/2}$ that we obtain for $\tb = 10$ 
is shifted slightly upwards by about $50 \gev$. 
The minimum value of $\chi^2$ for the global fit is 
increased, however, primarily because of the increased impact of the LEP
$\Mh$ constraint on the CMSSM parameter space. The tension between
$\Mh$ and the precision electroweak observables 
would become severe for $\mt < 170 \gev$. The minimum values of $\chi^2$
for $\tb = 10$ and 50 are now very similar.
We find that the minimum $\chi^2$ values remain approximately the same also
for the intermediate values $\tb = 20$ and $\tb = 35$.
On the other hand, the upper limit on
$m_{1/2}$ could be increased by as much as about 20\%
by possible future changes in
the preferred central value of $\mt$ and likely reductions in its error
(assuming that the experimental results and theoretical predictions for 
the precision observables are otherwise unchanged),
but remains relatively small, in general.

Within the NUHM, we find that the minimal $\chi^2$ values are smaller than
those for CMSSM points with the same value of $m_{1/2}$,
and that $\chi^2$ is relatively insensitive to $\MA$ but may decrease 
or increase as $\mu$ is varied. The preference for small $m_{1/2}$
is preserved in at least the sparse NUHM sample studied here. However, we 
do find that $m_0$ may differ significantly from its preferred range in 
the CMSSM. Likewise, significantly different values of $\mu$ and $\MA$ are 
also possible. In general, within the NUHM scenarios studied, the prospects 
for observing sparticles at the LHC or the ILC are similar to those in the
CMSSM case, except that in some cases the ${\tilde \tau_1}$ may be
rather heavier than the $\neu{1}$. 

In most of the VCMSSM scenarios with neutralino dark matter (NDM),
looking along the coannihilation strip compatible with WMAP and  
other cosmological data, we find that the
preference for small $m_{1/2}$ noted previously within the CMSSM framework
is repeated (offering good detection prospects for the LHC and the
ILC), and becomes a preference for medium values of 
$\tb$. In addition,  there is a tendency for $\tb$ to increase with
$m_{1/2}$. On the other hand, for $A_0/m_0 = 0$ we find larger values
of $m_{1/2}$ at the minimum $\chi^2$  
(which is significantly larger than for larger values of $A_0/m_0$),
and smaller values of $\tb$ 
which are rather constant with respect to $m_{1/2}$. When $A_0/m_0 = 2$,
we also observe that there are WMAP-compatible VCMSSM models at
$m_{1/2} \sim 140 \gev$ and $m_0 \sim 600 \gev$~\cite{Manuel} with 
$\tb \sim 37$ 
that have even lower $\chi^2$. These occur in the light Higgs funnel, when
$2 \mneu{1} \approx \Mh$, and offer some prospects for detection at the
Tevatron. 

The preference for small $m_{1/2}$ and a medium
range of $\tb$ is also maintained within the VCMSSM with the supplementary
mSUGRA relation $m_{3/2} = m_0$ when the dark matter is composed of
gravitinos (GDM) and the next-to-lightest supersymmetric particle
(NLSP) is the $\Staue$. In this scenario, the global $\chi^2$
that is somewhat smaller than along the WMAP strips in the VCMSSM with
neutralino dark matter. The prospects for sparticle detection at the
LHC and ILC are rather similar to those in the previous VCMSSM NDM
scenarios, but the light Higgs funnel disappears, reducing the
prospects for the Tevatron. We recall that the NLSP is metastable in
such GDM scenarios, suggesting that novel detection strategies should
be explored at the LHC and the ILC~\cite{Moortgat}. 


\section{Current experimental data}

In this Section we review briefly the experimental
data set that has been used for the fits. We focus on parameter points that
yield the correct value of the cold dark matter density, 
$0.094 < \Omega_{\rm CDM} h^2 < 0.129$~\cite{WMAP}, which is, however, not
included in the fit itself. 
The data set furthermore comprises the following
observables: the mass of the $W$~boson, $\MW$, the effective leptonic
weak mixing angle, $\sweff$, the anomalous magnetic moment of the
muon, $(g-2)_\mu$, the radiative $B$-decay
branching ratio $\br(b \to s \ga)$, and the lightest MSSM Higgs boson
mass, $\Mh$.
A detailed description of the first four observables can be found
in~\cite{ehow3,PomssmRep}.  
We limit ourselves here to recalling the current precision of
the experimental results and the theoretical predictions. 
The experimental values of these obervables have not changed
significantly compared to~\cite{ehow3,PomssmRep}, 
and neither have the theoretical calulations. As already commented,
due to the new, lower experimental value of $\mt$, it is necessary to
include the most complete experimental information about $\Mh$ into the 
fit. Accordingly, we give below details about the
inclusion of $\Mh$ and the evaluation of the corresponding $\chi^2$
values obtained from the direct searches for a Standard Model (SM) Higgs 
boson at LEP~\cite{LEPHiggsSM}.

In the following, we refer to the theoretical uncertainties from unknown
higher-order corrections as `intrinsic' theoretical uncertainties and
to the uncertainties induced by the experimental errors of the input
parameters as `parametric' theoretical uncertainties.  
We do not discuss here the theoretical
uncertainties in the renormalization-group
running between the high-scale input parameters
and the weak scale: see \citere{CDMRGEunc} for a recent discussion
in the context of calculations of the cold dark matter density. At present, 
these uncertainties are 
less important than the experimental and theoretical uncertainties in
the precision observables. 

Assuming that the five observables listed above are
uncorrelated, a $\chi^2$ fit has been performed with
\BE
\chi^2 \equiv \sum_{n=1}^{4} \KL \frac{R_n^{\rm exp} - R_n^{\rm theo}}
                                 {\si_n} \KR^2 + \chi^2_{\Mh}.
\label{eq:chi2}
\EE
Here $R_n^{\rm exp}$ denotes the experimental central value of the
$n$th observable ($\MW$, $\sweff$, \mbox{$(g-2)_\mu$} and $\br(b \to s \ga)$),
$R_n^{\rm theo}$ is the corresponding CMSSM prediction and $\si_n$
denotes the combined error, as specified below. $\chi^2_{\Mh}$ denotes the
$\chi^2$ contribution coming from the lightest MSSM Higgs boson mass as
described below.


\subsection{The $W$~boson mass}
\label{subsec:mw}

The $W$~boson mass can be evaluated from
\BE
\MW^2 \KL 1 - \frac{\MW^2}{\MZ^2}\right) = 
\frac{\pi \al}{\sqrt{2} \GF} \left(1 + \De r\KR,
\label{eq:delr}
\EE
where $\al$ is the fine structure constant and $\GF$ the Fermi constant.
The radiative corrections are summarized 
in the quantity $\De r$~\cite{sirlin}.
The prediction for $\MW$ within the Standard Model (SM)
or the MSSM is obtained by
evaluating $\De r$ in these models and solving (\ref{eq:delr}) in an
iterative way.

We include the complete \onel\ result in the
MSSM~\cite{deltarMSSM1lA,deltarMSSM1lB} as well as higher-order QCD
corrections of SM type that are of 
\order{\al\als}~\cite{drSMgfals,deltarSMgfals}
and \order{\al\als^2}~\cite{drSMgfals2,drSMgfals2LF}. Furthermore, we
incorporate 
supersymmetric corrections of \order{\al\als}~\cite{dr2lA} and of
\order{\al_t^2}~\cite{drMSSMal2B,drMSSMal2} to the quantity $\De\rho$.%
\footnote{
A re-evaluation of $\MW$ is currently under way~\cite{MWweber}. Preliminary
results show good agreement with the values used here.
}%

The remaining intrinsic theoretical uncertainty in the prediction for
$\MW$ within the MSSM is still significantly larger than in the SM. It
has been estimated as~\cite{drMSSMal2}
\BE
\De\MW^{\rm intr,current} \lsim 9 \mev~,
\EE
depending on the mass scale of the supersymmetric particles.
The parametric uncertainties are dominated by the experimental error of
the top-quark mass
and the hadronic contribution to the shift in the
fine structure constant. Their current errors induce the following
parametric uncertainties~\cite{lcwsSLACmt,PomssmRep}
\BEA
\de\mt^{\rm current} = 2.9 \gev &\Rightarrow&
\De\MW^{{\rm para},\mt, {\rm current}} \approx 17.5 \mev,  \\[.3em]
\de(\De\al_{\rm had}^{\rm current}) = 36 \times 10^{-5} &\Rightarrow&
\De\MW^{{\rm para},\De\al_{\rm had}, {\rm current}} \approx 6.5 \mev~.
\EEA
The present experimental value of $\MW$ is~\cite{lepewwg,LEPEWWG}
\BE
\MW^{\rm exp,current} = 80.410 \pm 0.032 \gev.
\label{mwexp}
\EE
The experimental and theoretical errors for $\MW$ 
are added in quadrature in our analysis.


\subsection{The effective leptonic weak mixing angle}

The effective leptonic weak mixing angle at the $Z$~boson peak
can be written as
\BE
 \sweff = \frac{1}{4} \, \left( 1 - \re \frac{v_{\rm eff}}{a_{\rm eff}}  
\right) \ ,
\EE
where $v_{\rm eff}$ and $a_{\rm eff}$ 
denote the effective vector and axial couplings
of the $Z$~boson to charged leptons.
Our theoretical prediction for $\sweff$ contains the same class of
 higher-order corrections as described in \refse{subsec:mw}.

In the MSSM, the remaining intrinsic theoretical uncertainty in the
prediction  for $\sweff$ has been estimated as~\cite{drMSSMal2}
\BE
\De\sweff^{\rm intr,current} \lsim 7 \times 10^{-5}, 
\EE
depending on the supersymmetry mass scale.
The current experimental errors of $\mt$ and $\De\al_{\rm had}$
induce the following parametric uncertainties
\BEA
\de\mt^{\rm current} = 2.9 \gev &\Rightarrow&
\De\sweff^{{\rm para},\mt, {\rm current}} \approx 10 \times 10^{-5},  \\[.3em]
\de(\De\al_{\rm had}^{\rm current}) = 36 \times 10^{-5} &\Rightarrow&
\De\sweff^{{\rm para},\De\al_{\rm had}, {\rm current}} \approx 
13 \times 10^{-5} .
\EEA
The experimental value is~\cite{lepewwg,LEPEWWG}
\BE
\sweff^{\rm exp,current} = 0.23153 \pm 0.00016~.
\label{swfit}
\EE
The experimental and theoretical errors for $\sweff$ 
are added in quadrature in our analysis.


\subsection{The anomalous magnetic moment of the muon}

The SM prediction for the anomalous magnetic moment of 
the muon (see~\cite{g-2review,g-2review2}
for reviews) depends on the evaluation of QED contributions (see
\cite{Kinoshita} for a recent update), the
hadronic vacuum polarization and light-by-light (LBL) contributions. The
former have been evaluated in~\cite{DEHZ,g-2HMNT,Jegerlehner,Yndurain}
and the latter in~\cite{LBLwrongsign1,LBLwrongsign2,LBL,LBLnew}. 
The evaluations of the 
hadronic vacuum polarization contributions using $e^+ e^-$ and $\tau$ 
decay data give somewhat different results. 
In view of the additional uncertainties associated with 
the isospin transformation from $\tau$ decay, we use here the latest
estimate based on $e^+e^-$ data~\cite{Hocker:2004xc}:
\BE
\amutheo = 
(11\, 659\, 182.8 \pm 6.3_{\rm had} \pm 3.5_{\rm LBL} \pm 0.3_{\rm QED+EW})
 \times 10^{-10},
\label{eq:amutheo}
\EE
where the source of each error is labelled.
We note that new $e^+e^-$ data sets
have recently been published in~\cite{KLOE,CMD2,SND}, but not yet used in
an updated estimate of $(g - 2)_\mu$. Their inclusion is not expected to alter
substantially the estimate given in (\ref{eq:amutheo}).

The result for the SM prediction is to be compared with
the final result of the Brookhaven $(g-2)_\mu$ experiment 
E821~\cite{g-2exp,g-2exp2}, namely:
\BE
\amuexp = (11\, 659\, 208.0 \pm 5.8) \times 10^{-10},
\label{eq:amuexp}
\EE
leading to an estimated discrepancy
\BE
\amuexp-\amutheo = (25.2 \pm 9.2) \times 10^{-10},
\label{delamu}
\EE
equivalent to a 2.7~$\sigma$ effect. 
While it would be
premature to regard this deviation as a firm evidence for new
physics,
it does indicate a preference for a non-zero supersymmetric contribution.

Concerning the MSSM contribution, the complete one-loop
result was evaluated a decade ago~\cite{g-2MSSMf1l}. 
It indicates that variants of the MSSM with
$\mu < 0$ are already very challenged by the
present data on $\amu$, whether one uses either the $e^+ e^-$ or 
$\tau$ decay data, so we restict our attention in this paper to
models with $\mu > 0$. 
In addition to the full one-loop contributions, the leading QED
two-loop corrections have also been
evaluated~\cite{g-2MSSMlog2l}. Further corrections at the two-loop
level have been obtained recently~\cite{g-2FSf,g-2CNH}, leading to
corrections to the one-loop result that are $\sim 10\%$. These
corrections are taken into account in our analysis according to the
approximate formulae given in~\cite{g-2FSf,g-2CNH}.


\subsection{The decay $b \to s \ga$}

Since this decay occurs at the loop level in the SM, the MSSM 
contribution might {\it a priori} be of similar magnitude. A
recent
theoretical estimate of the SM contribution to the branching ratio
is~\cite{hugr}
\BE
\br( b \to s \ga ) = (3.70 \pm 0.46) \times 10^{-4},
\label{bsga}
\EE
where the calculations have been carried out completely to NLO in the 
\msbar\ renormalization scheme~\cite{ali,bsgKO2,ali2}, and the error is
dominated by  
higher-order QCD uncertainties. We record, however, that the error 
estimate for $\br(b \to s \ga)$ is still under debate, see
also \citeres{hulupo,bsgneubert}.

For comparison, the present experimental 
value estimated by the Heavy Flavour Averaging Group (HFAG)
is~\cite{bsgexp}
\BE
\br( b \to s \ga ) = (3.39^{+ 0.30}_{- 0.27}) \times 10^{-4},
\label{bsgaexp}
\EE
where the error includes an uncertainty due to the decay spectrum, as well 
as the statistical error. The good agreement between (\ref{bsgaexp})
and the SM calculation (\ref{bsga}) imposes important constraints on the 
MSSM. 

Our numerical results have been derived with the 
$\br(b \to s \ga)$ evaluation provided in \citere{bsgMicro}, which
has been checked against other approaches~\cite{bsgKO1,bsgKO2,bsgGH,ali2}.
For the current theoretical uncertainty of the MSSM prediction for 
$\br(b \to s \ga)$ we use the value in (\ref{bsga}). We
add the theory and experimental errors in quadrature.%

We have not included the decay $B_s \to \mu^+ \mu^-$ in our fit, in
the absence of an experimental likelihood function and a suitable 
estimate of the theoretical error. However, it is known that the
present experimental upper limit: 
$\br(B_s \to \mu^+ \mu^-) < 2 \times 10^{-7}$~\cite{Bmumu} may become
important for $\tb > 40$ in the MSSM~\cite{Dedes,ourBmumu}. We mention
below some specific instances where the decay $B_s \to \mu^+ \mu^-$
may already constrain the parameter space studied~\cite{ournewBmumu},
and note that~\cite{ehow3} gives a detailed analysis of its possible
future significance. 


\subsection{The lightest MSSM Higgs boson mass}

\begin{figure}[hbt!]
\begin{center}
\includegraphics[width=.48\textwidth]{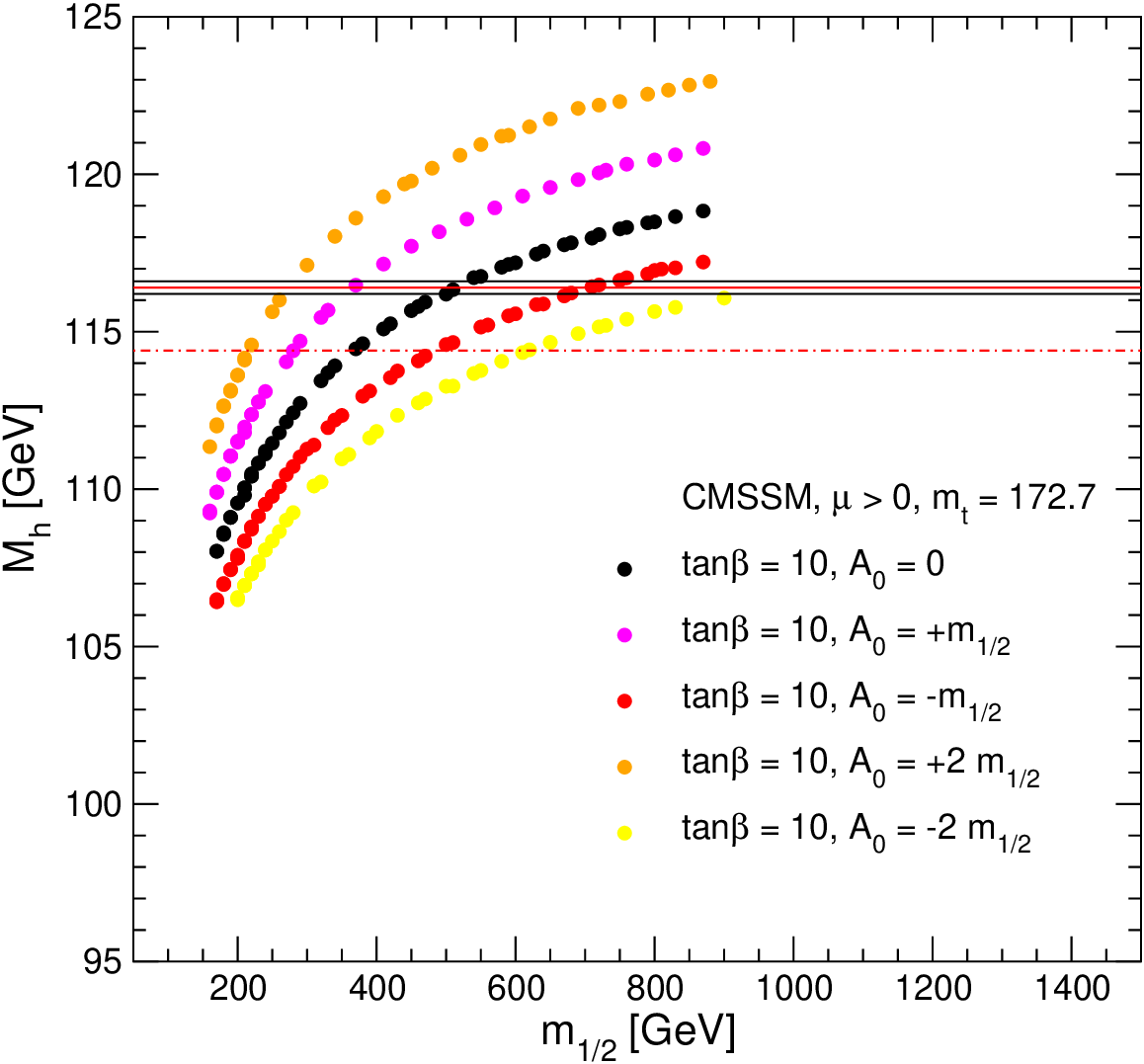}
\includegraphics[width=.48\textwidth]{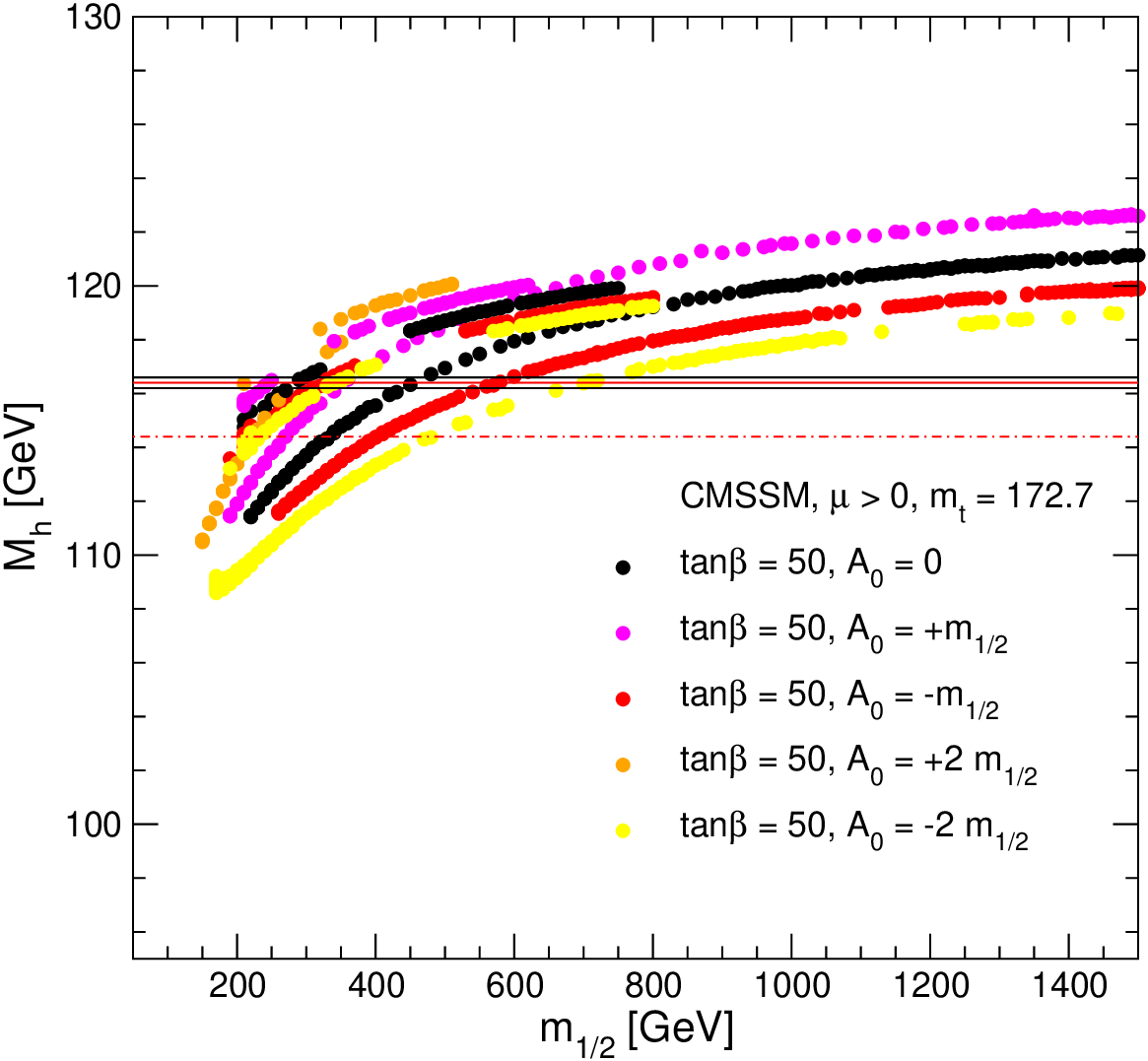}
\caption{
\it The CMSSM predictions for $\Mh$ as functions of $m_{1/2}$ with (a) $\tb = 
10$ and (b) $\tb = 50$ for various $A_0$. 
A hypothetical LHC measurement is shown, namely $\Mh = 116.4 \pm 0.2 \gev$,
as well as the present 95\%~C.L.\ exclusion limit of $114.4 \gev$.
}
\label{fig:Mh}
\end{center}
\end{figure}

The mass of the lightest $\cp$-even MSSM Higgs boson can be predicted in 
terms of
the other CMSSM parameters. At the tree level, the two $\cp$-even Higgs 
boson masses are obtained as functions of $\MZ$, the $\cp$-odd Higgs
boson mass $\MA$, and $\tb$. 
For the theoretical prediction of $\Mh$ we employ the
Feynman-diagrammatic method, using the code 
{\tt FeynHiggs}~\cite{feynhiggs,feynhiggs2},
which includes all numerically relevant known higher-order corrections.
The status of the incorporated results 
can be summarized as follows. For the
one-loop part, the complete result within the MSSM is 
known~\cite{ERZ,mhiggsf1lB,mhiggsf1lC}. 
Computation of the two-loop
effects is quite advanced: see \citere{mhiggsAEC} and
references therein. These include the strong corrections
at \order{\al_t\als} and Yukawa corrections at \order{\al_t^2}
to the dominant one-loop \order{\al_t} term, and the strong
corrections from the bottom/sbottom sector at \order{\al_b\als}. 
In the case of the $b/\Sbot$~sector
corrections, an all-order resummation of the $\Tb$-enhanced terms,
\order{\al_b(\als\tb)^n}, is also known~\cite{deltamb,deltamb1}.
Most recently, the \order{\al_t \al_b} and \order{\al_b^2} corrections
have been derived~\cite{mhiggsEP5}~%
\footnote{
A two-loop effective potential calculation has been presented 
in~\cite{fullEP2l}, but no public code based on this result is 
currently available.
}%
. The current intrinsic error of $\Mh$ due to
unknown higher-order corrections has been estimated to 
be~\cite{mhiggsAEC,mhiggsFDalbals,PomssmRep,mhiggsWN}
\BE
\De\Mh^{\rm intr,current} = 3 \gev~.
\EE
We show in \reffi{fig:Mh} the predictions for $\Mh$ in the CMSSM for  
$\tb = 10$ (left) and $\tb = 50$ (right) along the strips allowed by
WMAP and  other cosmological data~\cite{WMAPstrips}. 
We note that the predicted values of $\Mh$ depend significantly on
$A_0$. Also shown in \reffi{fig:Mh} is the present 95\%~C.L.\ exclusion
limit for a SM-like Higgs boson is $114.4 \gev$~\cite{LEPHiggsSM} and a 
hypothetical LHC measurement at $\Mh = 116.4 \pm 0.2\gev$.

It should be noted that, for the unconstrained MSSM with small values
of $\MA$ and values of 
$\tb$ that are not too small, a significant suppression of the $hZZ$
coupling can occur in the  
MSSM compared to the SM value, in which case the experimental lower
bound on $\Mh$ may be more than 20~GeV below the
SM value~\cite{LEPHiggsMSSM}. However, we have checked that within the 
CMSSM and the other models studied in this paper, the $hZZ$ coupling is
always very close to the SM value. Accordingly, the
bounds from the SM Higgs search at LEP~\cite{LEPHiggsSM}
can be taken over directly (see e.g.\ \citeres{asbs1,ehow1}).
It is clear that low values of $m_{1/2}$, especially for $\tb = 10$,
are challenged by the LEP exclusion bounds. This is essentially because
the leading supersymmetric radiative corrections to $\Mh$ are proportional to 
$\mt^4 \, {\rm ln} (m_{1/2}/\mt)$, so that 
a reduction in $\mt$ must be compensated by 
an increase in $m_{1/2}$ for the same value of~$\Mh$. 


In our previous analysis, we simply applied a cut-off on $\Mh$, considering
only parameter choices for which {\tt FeynHiggs} gave $\Mh > 113.0 \gev$.
However, now that the $\Mh$ constraint assumes greater importance, here we 
use more completely the likelihood information available from LEP.
Accordingly, we evaluate as follows the $\Mh$ contribution to the 
overall $\chi^2$
function~%
\footnote{
We thank P.~Bechtle and K.~Desch for detailed discussions and
explanations.
}%
. Our starting points are the $CL_s(\Mh)$ values provided by the 
final LEP results on the SM Higgs boson search, see Fig.~9 
in~\cite{LEPHiggsSM}~%
\footnote{
We thank A.~Read for providing us with the $CL_s$ values.
}%
. We obtain by inversion from  $CL_s(\Mh)$
the corresponding value of ${\tilde \chi}^2(\Mh)$ determined from~\cite{PDG}
\BE
\edz {\rm erfc}(\sqrt{\edz\tilde\chi^2(\Mh)}) \equiv CL_s(\Mh)~,
\EE
and note the fact that $CL_s(\Mh = 116.4 \gev) = 0.5$ implies that
$\tilde\chi^2(116.4 \gev) = 0$ as is appropriate for a one-sided
limit. Correspondingly we set $\tilde\chi^2(\Mh > 116.4 \gev) = 0$.
The theory uncertainty is included by convolving the likelihood function
associated with
$\tilde\chi^2(\Mh)$ and a Gaussian function, $\tilde\Phi(x)$,
normalized to unity and centred around $\Mh$, whose width is $1.5 \gev$:
\BE
\chi^2(\Mh) = -2 \log\KL
\int_{-\infty}^{\infty} 
    e^{-\tilde\chi^2(x)/2} \; \tilde\Phi(\Mh - x) \, {\rm d}x \KR~.
\EE
In this way, a theoretical uncertainty of up to $3 \gev$ is assigned for 
$\sim 95\%$ of
all $\Mh$ values corresponding to one parameter point. 
The final $\chi^2_{\Mh}$ is then obtained as 
\BEA
\chi^2_{\Mh} = \chi^2(\Mh) - \chi^2(116.4 \gev) &{\rm ~for~}& 
                                                \Mh \le 116.4 \gev~, \\
\chi^2_{\Mh} = 0 &{\rm ~for~}& \Mh > 116.4 \gev ~,
\EEA
and is then combined with the corresponding quantities for the other 
observables we consider, see \refeq{eq:chi2}.


\section{Updated CMSSM analysis}
\label{sec:CMSSMupdate}

As already mentioned, in our previous analysis of the CMSSM~\cite{ehow3} 
we used the range $\mt = 178.0 \pm 4.3 \gev$ that was then preferred
by direct measurements~\cite{oldmt}. 
The preferred range evolved subsequently to $172.7 \pm 2.9 
\gev$~\cite{newmt}. In view of this past evolution and possible future
developments, in this Section we first analyze the current situation
in some detail, 
emphasizing some new aspects related to the lower value of $\mt$, and then
provide a guide to possible future developments.

The effects of the lower $\mt$ value are threefold. First, it drives the
SM prediction of $\MW$ and $\sweff$ slightly further away from the
current experimental value (whereas $(g-2)_\mu$ and $\br(b \to s \ga)$ are
little affected). This increases the favoured magnitude of the
supersymmetric contributions, i.e.,
it effectively lowers the preferred supersymmetric mass scale.
Secondly, the predicted value of the lightest Higgs boson mass in the
MSSM is lowered by the new $\mt$ value, see, e.g., \citere{tbexcl} and
\reffi{fig:Mh}. The effects on the electroweak precision
observables of the downward shift in $\Mh$ are minimal, but
the LEP Higgs bounds~\cite{LEPHiggsSM,LEPHiggsMSSM} now impose a more
important constraint on the MSSM parameter space, notably on $m_{1/2}$.
In our previous analysis, we rejected all parameter points for which {\tt 
FeynHiggs} yielded $\Mh < 113 \gev$. The best fit values in \citere{ehow3} 
corresponded to 
relatively small values of $\Mh$, a feature that is even more pronounced
for the new $\mt$ value. 
Thirdly, the focus-point region of the CMSSM
parameter space now appears at considerably lower $m_0$ than previously,
increasing its importance for the $\chi^2$ analysis.

In view of all these effects, we now update our previous
analysis of the phenomenological constraints on the supersymmetric mass 
scale $m_{1/2}$ in the CMSSM using the new, lower
value~%
\footnote{
See also \citere{LCWS05ehow3}, where a lower bound of 
$\Mh > 111.4 \gev$ has been used.
}%
~of $\mt$
and including a $\chi^2$ contribution from $\Mh$, evaluated as discussed 
in the 
previous Section.
As in \citere{ehow3} we use the experimental information on the cold
dark matter density from WMAP and other observations to reduce the
dimensionality of the CMSSM parameter space. In the parameter region
considered in our analysis we find an acceptable dark matter relic
density along coannihilation strips, in the Higgs funnel region and in
the focus-point region. We comment below on the behaviours of the $\chi^2$
function in each of these regions.

\begin{figure}[htb!]
\begin{center}
\includegraphics[width=.45\textwidth,height=5.4cm]{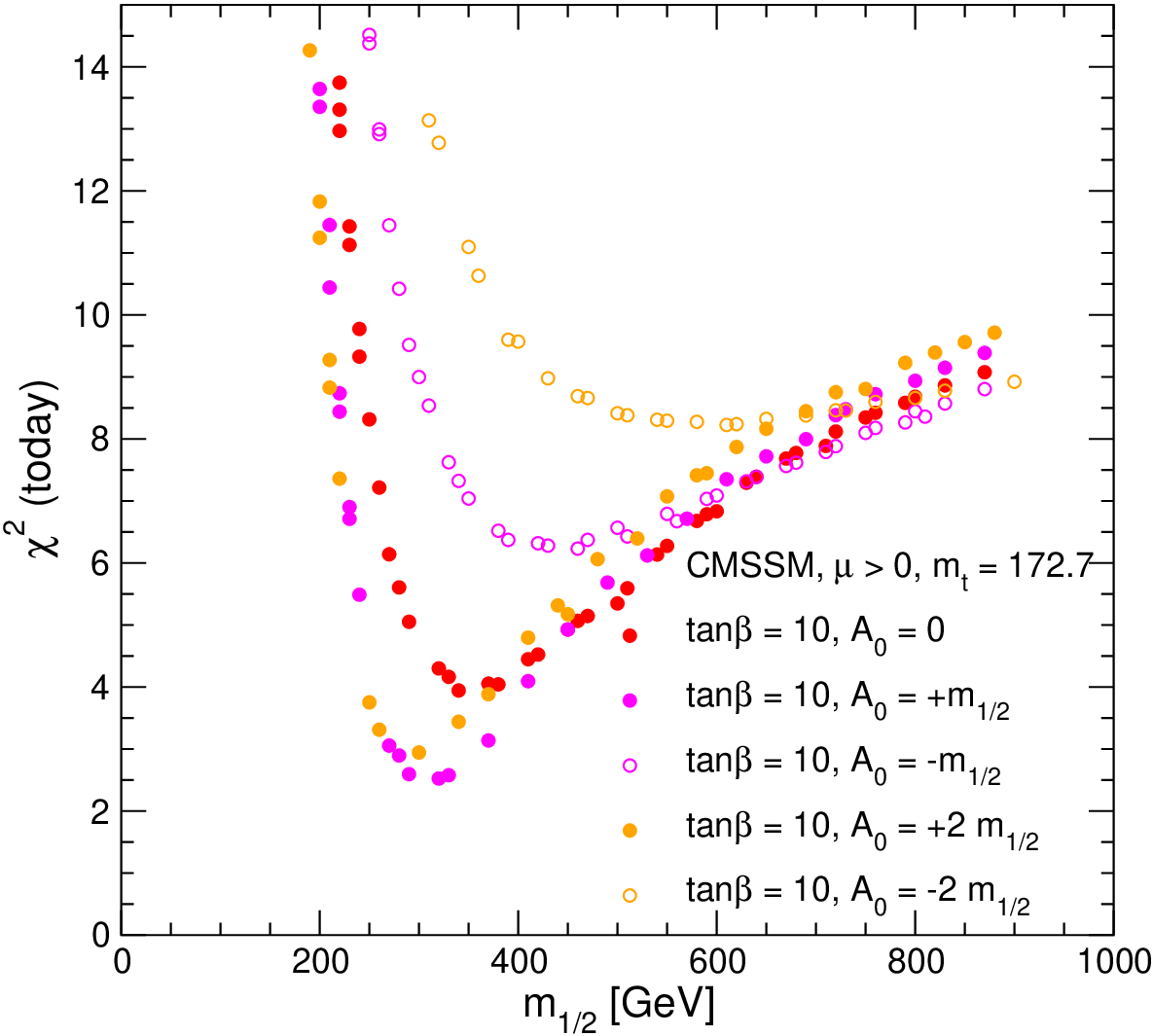}
\includegraphics[width=.45\textwidth,height=5.4cm]{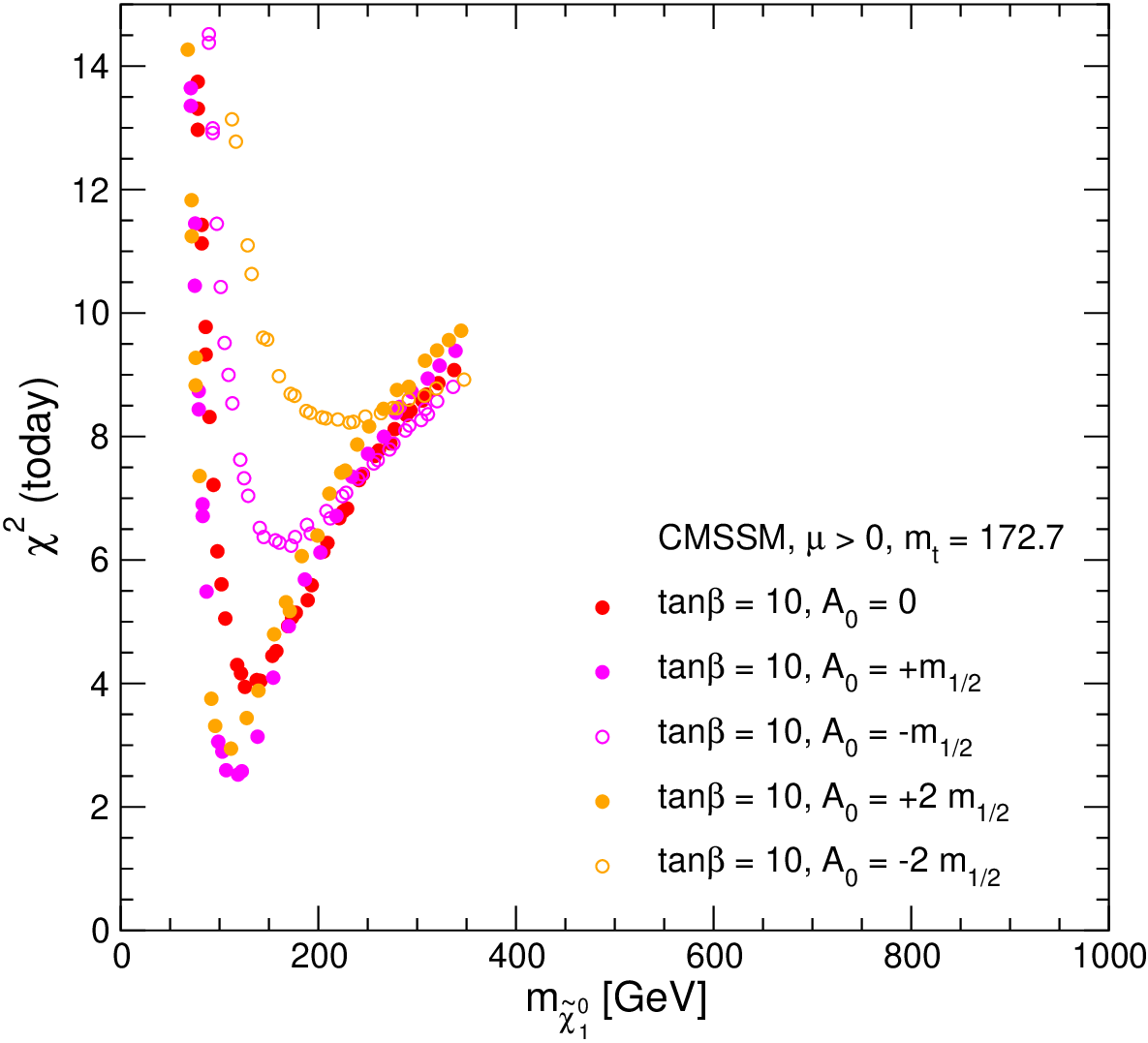}\\[3em]
\includegraphics[width=.45\textwidth,height=5.4cm]{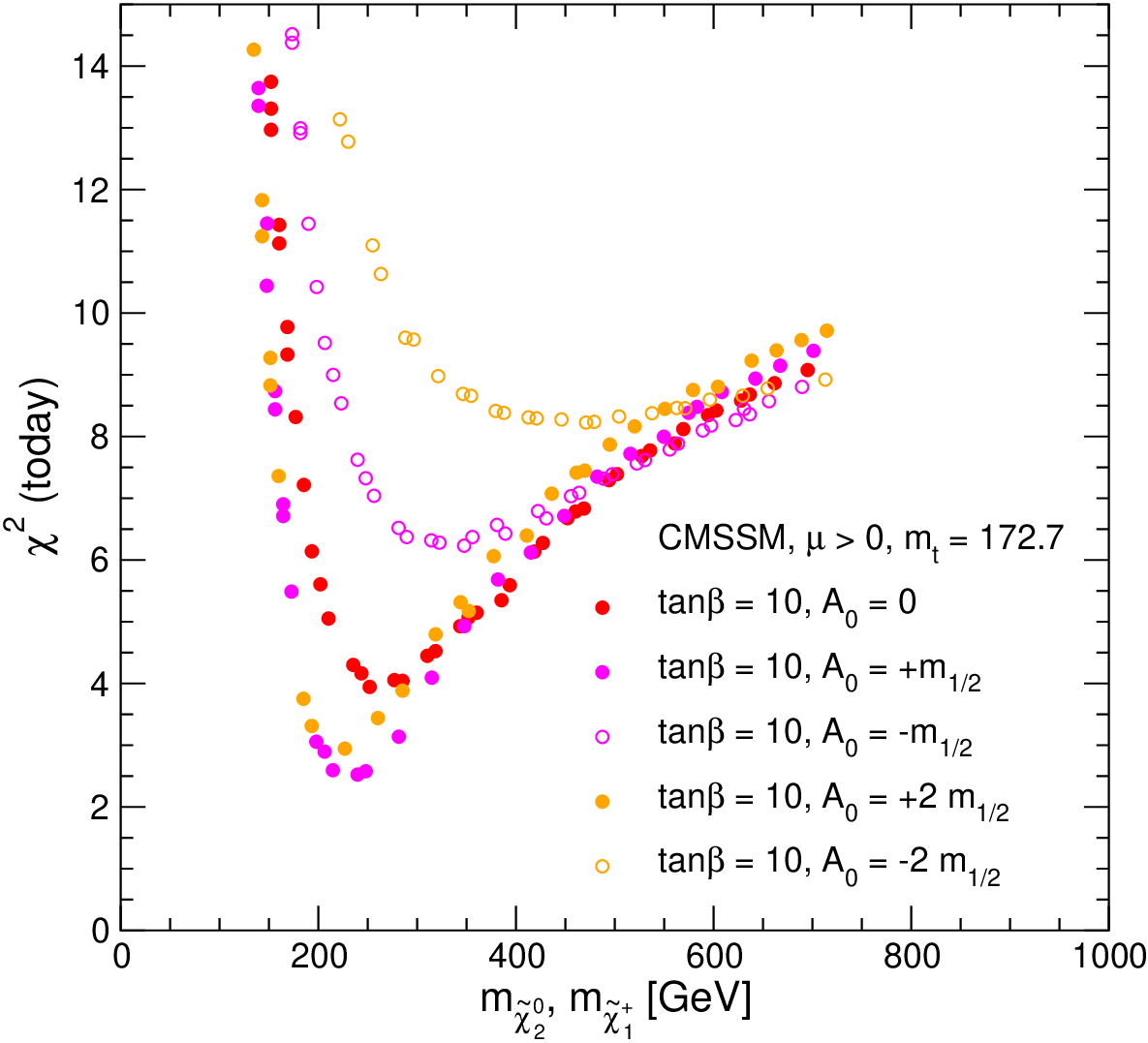}
\includegraphics[width=.45\textwidth,height=5.4cm]{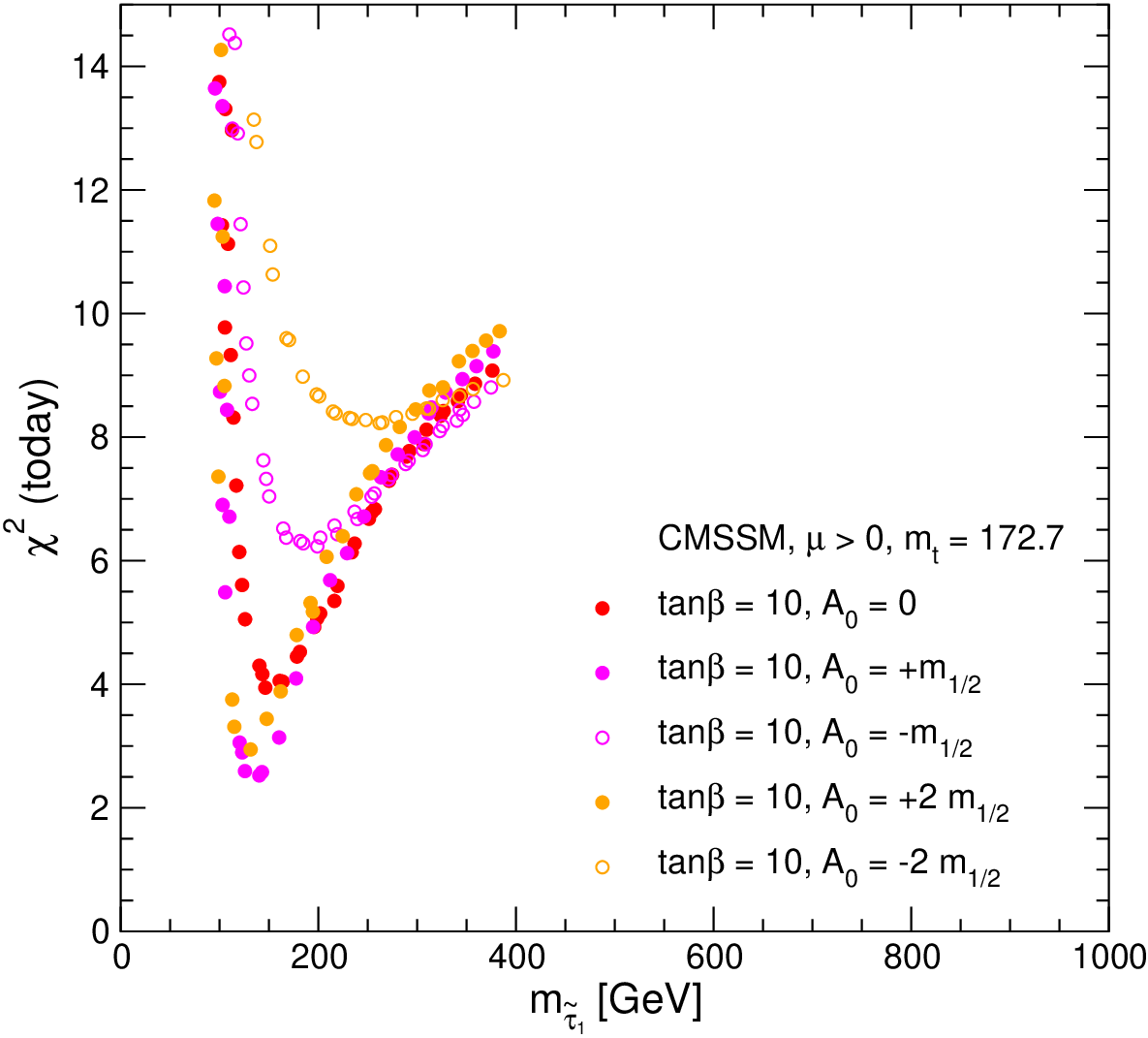}\\[3em]
\includegraphics[width=.45\textwidth,height=5.4cm]{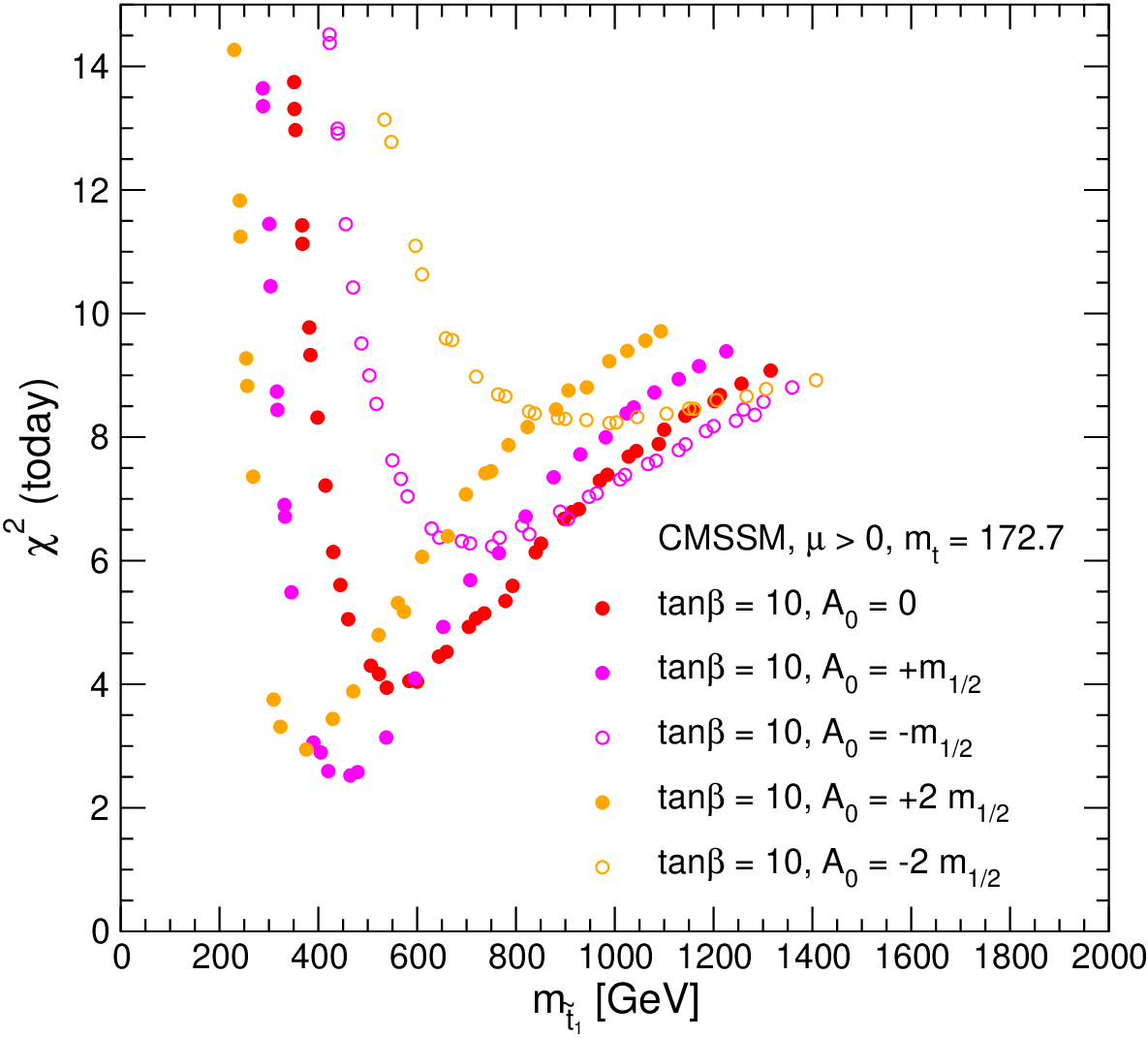}
\includegraphics[width=.45\textwidth,height=5.4cm]{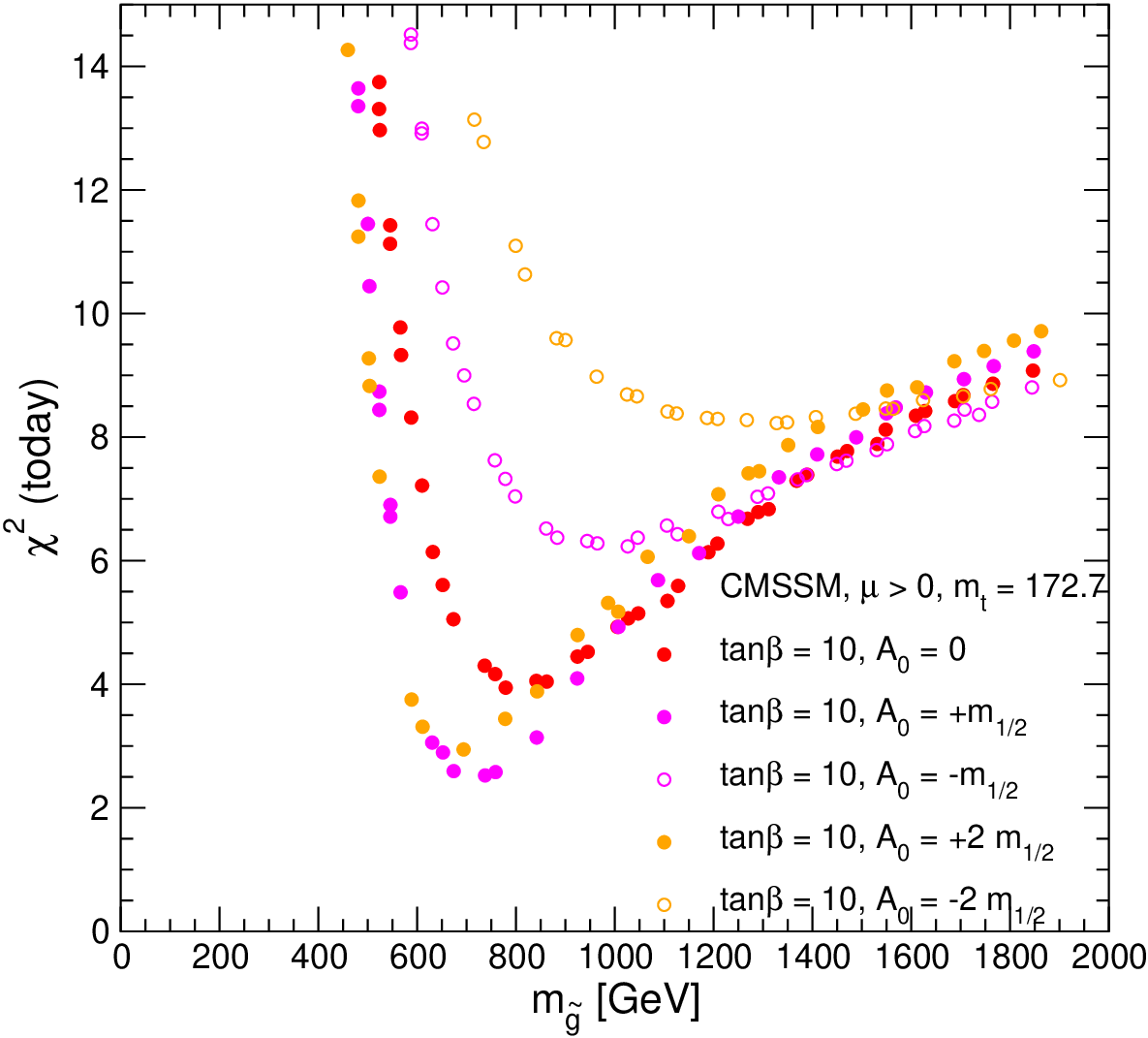}
\begin{picture}(0,0)
\CBox(-315,440)(-228,395){White}{White}
\CBox(-303,450)(-228,440){White}{White}
\CBox(-320,265)(-228,210){White}{White}
\CBox(-110,280)(-018,210){White}{White}
\CBox(-320,079)(-230,020){White}{White}
\CBox(-100,059)(-015,018){White}{White}
\CBox(-082,069)(-013,062){White}{White}
\end{picture}
\caption{
\it The combined likelihood function $\chi^2$ for the electroweak
observables $\MW$, $\sweff$,
$(g - 2)_\mu$, ${\rm BR}(b \to s \ga)$, and $\Mh$
evaluated in the CMSSM for $\tb = 10$,
$\mt = 172.7 \pm 2.9 \gev$ and various discrete values of $A_0$, with
$m_0$ then chosen to yield the central value of the relic neutralino 
density
indicated by WMAP and other observations. We display the $\chi^2$
function for (a) $m_{1/2}$, (b) $\mneu{1}$, 
(c) $\mneu{2}, \mcha{1}$, (d) $\mstaue$, (e) $\mste$ and (f) $\mgl$.
} 
\label{fig:newmt10}
\end{center}
\vspace{-3em}
\end{figure}

As seen in the first panel of \reffi{fig:newmt10}, 
which displays the behaviour of the $\chi^2$ function out to the tips
of typical WMAP coannihilation strips,
the qualitative
feature observed in \citere{ehow3} of a pronounced minimum in $\chi^2$ at
$m_{1/2}$ for $\tb = 10$ is also present for the new value of
$\mt$. However, the $\chi^2$ curve now depends more strongly on the value 
of $A_0$, corresponding to its strong impact on $\Mh$, as seen in
\reffi{fig:Mh}. Values of $A_0/m_{1/2} < -1$ are disfavoured at the 90\%~C.L.,
essentially because of their lower $\Mh$ values, but
$A_0/m_{1/2} = 2$ and~1 give equally good fits and descriptions of the
data. The old best fit point in \citere{ehow3} had $A_0/m_{1/2} = -1$, but
there all $A_0/m_{1/2}$ gave a similarly good description of the experimental
data. The minimum $\chi^2$ 
value is slightly below~3. This is somewhat higher than the result in
\citere{ehow3}, but still represents a good overall fit to the 
experimental data. 
The rise in the minimum value of $\chi^2$, compared to \citere{ehow3},
is essentially a consequence of the lower experimental central value of
$\mt$, and the consequent greater
impact of the LEP constraint on $\Mh$~\cite{LEPHiggsSM,LEPHiggsMSSM}.
In the cases of the observables $\MW$ and $\sweff$, a smaller value of 
$\mt$
induces a preference for a smaller value of $m_{1/2}$, but the opposite is
true for the Higgs mass bound. The rise in the
minimum value of $\chi^2$ reflects the correspondingly increased 
tension between the electroweak precision observables and the $\Mh$
constraint. 

A breakdown of the contributions to $\chi^2$ from the 
different observables can be found for some example points
in Table~\ref{tab:chi2}. 
The best-fit points for $\tb = 10$ and 50 are shown in the first and
third lines, respectively. The second line shows a point near the tip
of the WMAP coannihilation strip for $\tb = 10$, the fourth line shows
a point at the tip of the rapid-annihilation Higgs funnel for 
$\tb = 50$. The fifth till the seventh row show points in 
the focus point region (see below) for $\tb = 50$ with low,
intermediate and high $m_{1/2}$. It is instructive to compare the
contributions to $\chi^2$ at the best-fit points with those at the
coannihilation, Higgs funnel and focus points. 
One can see that, for large $m_{1/2}$ values in all the different regions,
$(g-2)_\mu$ always gives the dominant contribution. However, with the
new lower experimental value of 
$\mt$ also $\MW$ and $\sweff$ give substantial contributions, adding up 
to more than 50\% of the $(g-2)_\mu$ contribution at the coannihilation
and Higgs funnel points. On the other hand, $\Mh$ and 
$\br (b \to s \ga)$ make negligible contributions to $\chi^2$ 
at these points. As seen from the last lines of the Table, the
situation may be different in the focus-point region for low $m_{1/2}$: 
the first example given
yields a reasonably good description of $\MW$, $\sweff$ and even 
$(g-2)_\mu$, while the largest contribution to $\chi^2$ arises from
$\br(b \to s \ga)$~%
\footnote{
We note that, particularly in view of the current uncertainties on
$\mt$ and $\mb$ and the corresponding uncertainties in $\MA$, the
upper limit on the $\br(B_s \to \mu^+ \mu^-)$ currently imposes a weaker
constraint on the CMSSM parameter space than does $b \to s \ga$,
even for $\tb = 50$~\cite{ourBmumu}.
}%
. This smoothly changes to the behavior for large $m_{1/2}$ as
described above also in the focus-point region, as can be seen from
the last two rows in \refta{tab:chi2}. 


\begin{table}[tbh!]
\renewcommand{\arraystretch}{1.5}
\BC
\begin{tabular}{|c|c|c|c||c||c||c|c|c|c|c|}
\hline\hline
$\tb$ & $m_{1/2}$ & $m_0$ & $A_0$ & comment & $\chi^2_{\rm tot}$ & 
$\MW$ & $\sweff$ & $(g-2)_\mu$ & $b \to s \ga$ & $\Mh$ \\ \hline\hline
10 & 320 &  90 &  320 & best fit &
     2.55 & 1.01 & 0.12 & 0.63 & 0.23 & 0.52 \\ \hline
10 & 880 & 270 & 1760 & bad fit & 
     9.71 & 2.29 & 1.28 & 6.14 & 0.01 & 0 \\ \hline\hline
50 &  570 &  390 & -570 & best fit &
     2.79 & 1.44 & 0.31 & 0.08 & 0.91 & 0.04 \\ \hline
50 & 1910 & 1500 & -1910 & bad fit & 
     9.61 & 2.21 & 1.11 & 6.29 & 0.01 & 0 \\ \hline
50 & 250 & 1320 & -250 & focus & 
     7.34 & 0.89 & 0.15 & 1.69 & 3.76 & 0.84 \\ \hline
50 & 330 & 1640 & -330 & focus &
     6.06 & 1.24 & 0.28 & 3.21 & 1.33 & 0 \\ \hline
50 & 800 & 2970 & -800 & focus &
     8.73 & 1.92 & 0.72 & 6.05 & 0.04 & 0 \\
\hline\hline
\end{tabular}
\EC
\renewcommand{\arraystretch}{1}
\caption{\it Breakdown of $\chi^2$ contributions from the different 
precision observables to $\chi^2_{\rm tot}$ for some example points. 
All masses are in GeV. The first and third rows are the best fits for
$\tb = 10$ and 50, the second row is representative of the
coannihilation strip, the fourth row is  
representative of the Higgs funnel region, and the last three rows
are representatives of the focus point-region. 
}
\label{tab:chi2}
\end{table}


The remaining panels of \reffi{fig:newmt10} update our previous
analyses~\cite{ehow3} of the $\chi^2$ functions for various sparticle 
masses within
the CMSSM, namely the lightest neutralino $\neu{1}$, the second-lightest
neutralino $\neu{2}$ and the (almost degenerate) lighter chargino
$\cha{1}$, the lightest slepton which is the lighter stau 
$\Staue$, the lighter stop squark $\Stope$, and the gluino
$\gl$. Reflecting the behaviour of the global $\chi^2$ function
in the first panel of \reffi{fig:newmt10}, the changes in the optimal
values of the sparticle masses are not large.
The 90\% C.L.\ upper bounds on the particle masses are nearly
unchanged compared to the results for $\mt = 178.0 \pm 4.3 \gev$ given in
\citere{ehow3}.


\begin{figure}[htb!]
\begin{center}
\includegraphics[width=.45\textwidth,height=5.5cm]{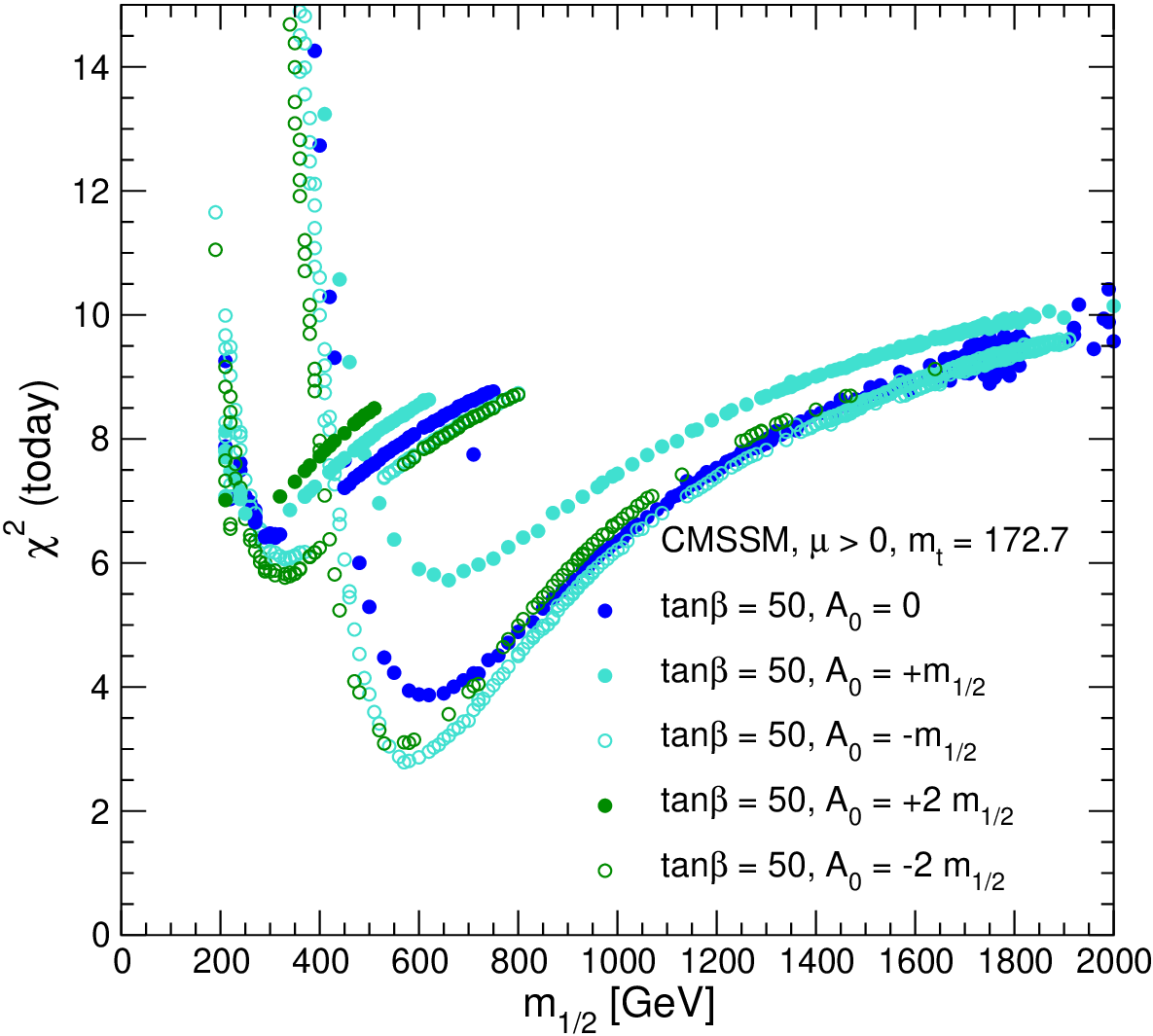}
\includegraphics[width=.45\textwidth,height=5.5cm]{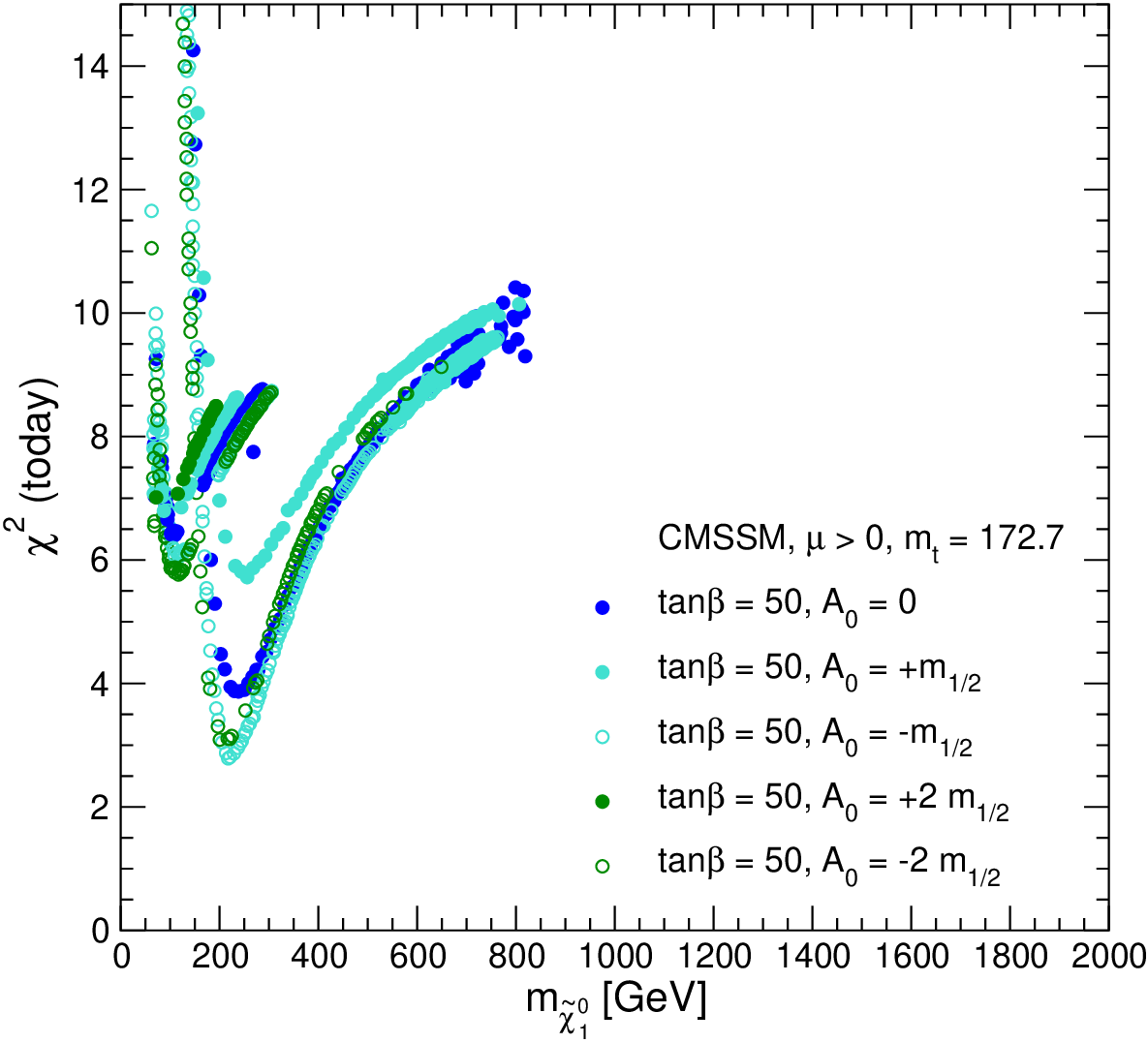}\\[3em]
\includegraphics[width=.45\textwidth,height=5.5cm]{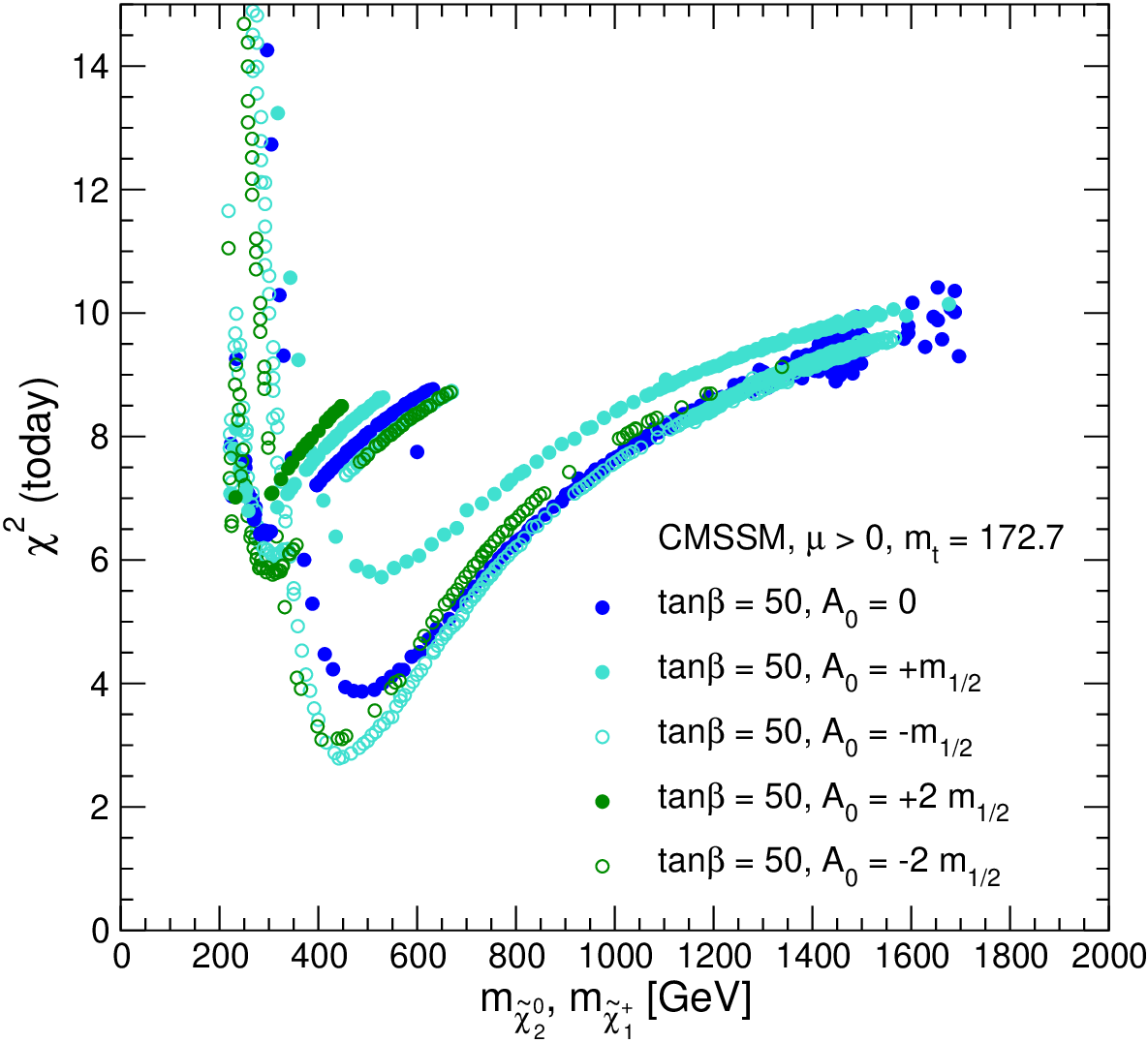}
\includegraphics[width=.45\textwidth,height=5.5cm]{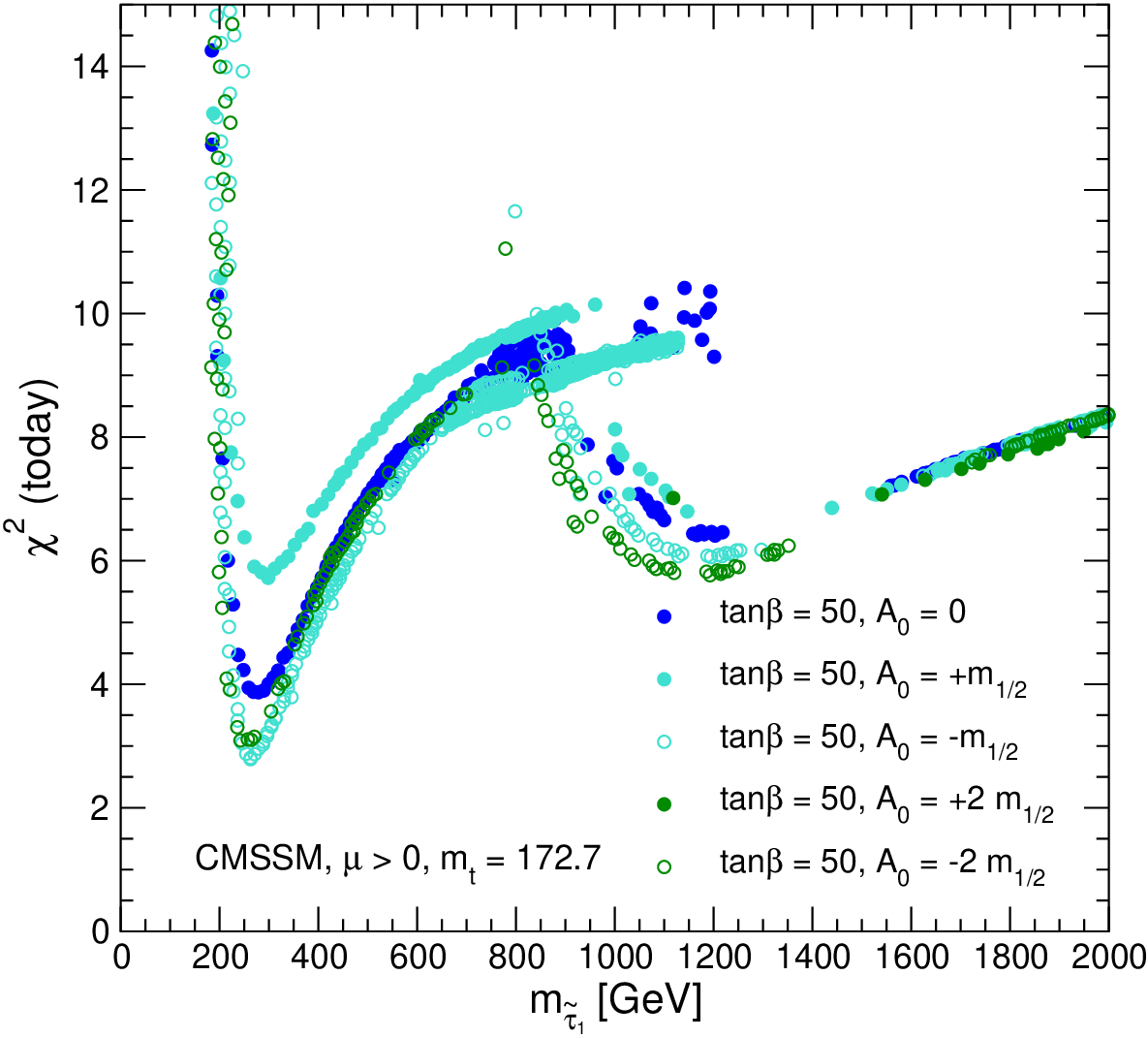}\\[3em]
\includegraphics[width=.45\textwidth,height=5.5cm]{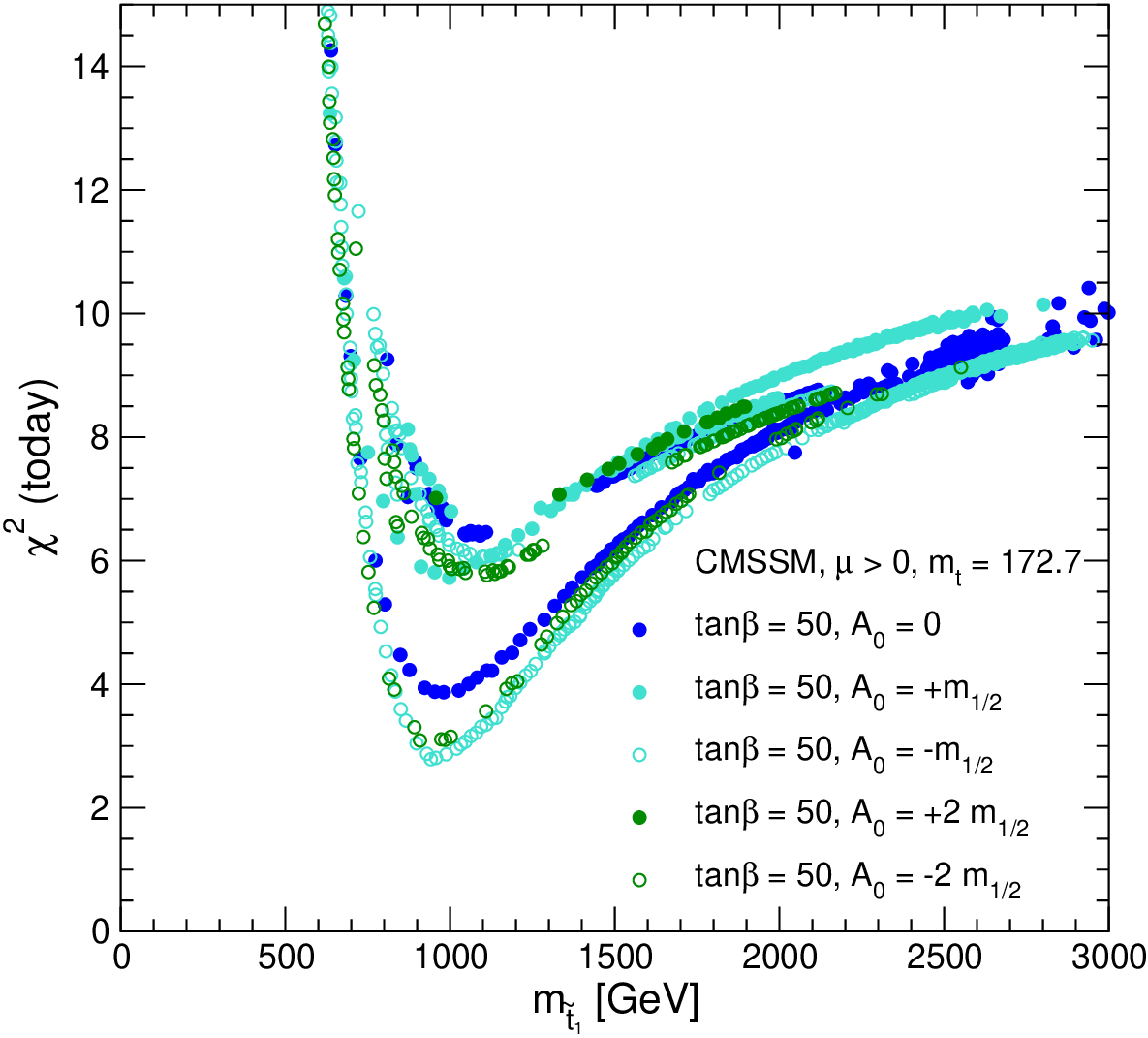}
\includegraphics[width=.45\textwidth,height=5.5cm]{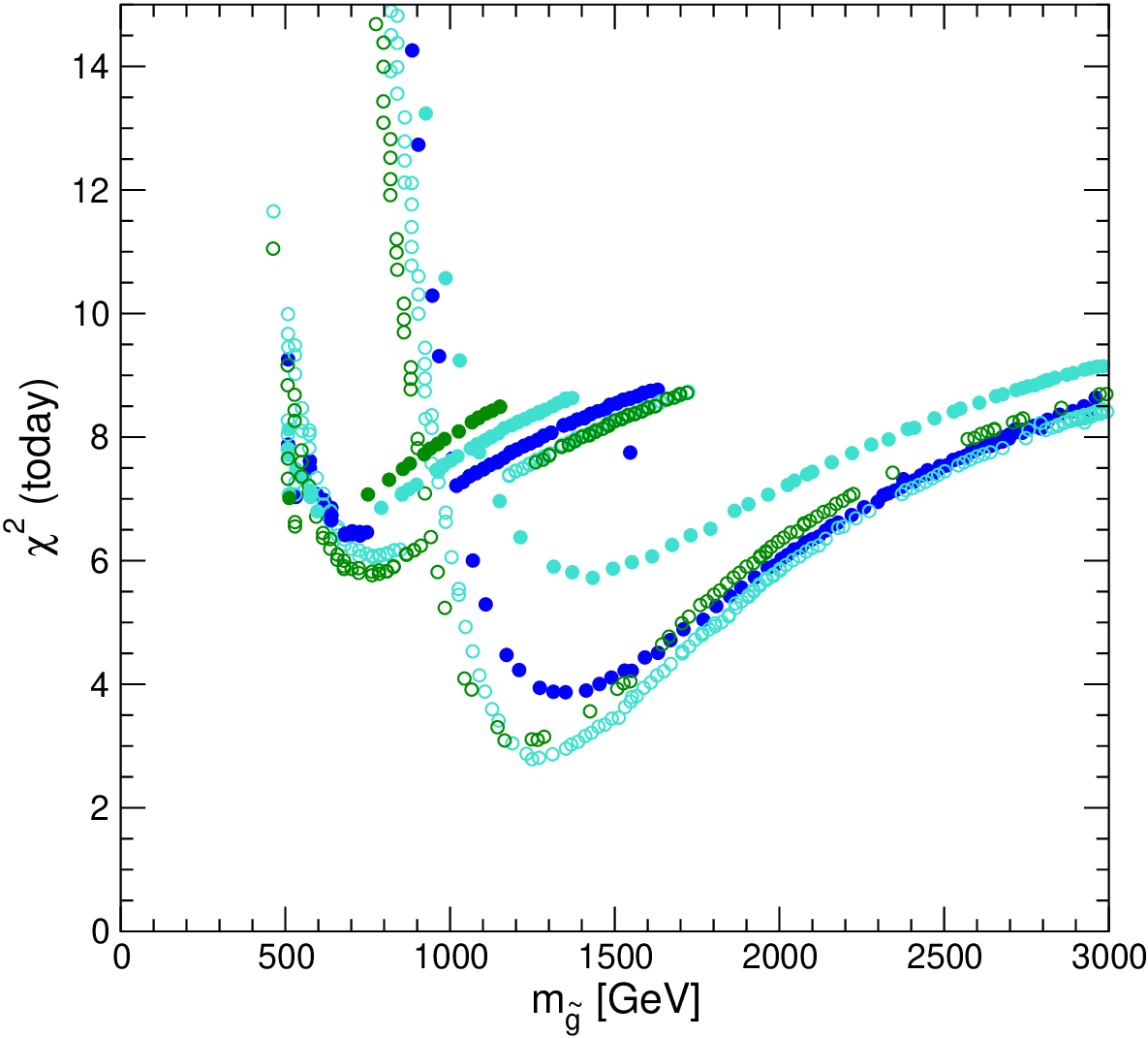}
\begin{picture}(0,0)
\CBox(-318,450)(-228,405){White}{White}
\CBox(-307,462)(-228,450){White}{White}
\CBox(-320,273)(-228,215){White}{White}
\CBox(-095,260)(-018,212){White}{White}
\CBox(-180,222)(-100,213){White}{White}
\CBox(-317,068)(-230,020){White}{White}
\CBox(-301,078)(-230,068){White}{White}
\end{picture}
\caption{
\it
As in Fig.~\protect\ref{fig:newmt10}, but now for $\tb = 50$.
}
\label{fig:newmt50}
\end{center}
\end{figure}

The corresponding results for 
WMAP strips in the coannihilation, Higgs funnel and focus-point regions for
the case $\tb = 50$ are shown in
\reffi{fig:newmt50}.  The spread of points with identical values of
$A_0$ at large $m_{1/2}$ is due to the broadening and bifurcation of
the WMAP strip in the Higgs funnel region, and the higher set of
$\chi^2$ curves originate in the focus-point region, as discussed in
more detail below. We see in panel (a) that the minimum value of
$\chi^2$ for the fit with $m_t = 172.7 \pm 2.9 \gev$ is larger by about a unit
than in our previous analysis with $m_t = 178.0 \pm 4.3 \gev$.  Because of
the rise in $\chi^2$ for the $\tb = 10$ case, however, the minimum values
of $\chi^2$ are now very similar for the two values of $\tb$ shown here. The
dip in the $\chi^2$ function for $\tb = 50$ is somewhat steeper than in
the previous analysis, since the high values of $m_{1/2}$ are slightly
more disfavoured due to their $\MW$ and $\sweff$ values. The best fit
values of $m_{1/2}$ are very similar to their previous values. The
preferred values of the sparticle masses are shown in the remaining panels
of \reffi{fig:newmt50}. Due to the somewhat steeper $\chi^2$ behavior,
the preferred ranges have slightly lower masses than in
\citere{ehow3}. 

We now return to one novel feature as compared to \citere{ehow3}, namely the
appearance of a group of points with 
moderately high $\chi^2$ that have relatively small $m_{1/2} \sim 200 \gev$.
These points have relatively large values of $m_0$, as reflected in the
relatively large values of $\mstaue$ and $\mste$
seen in panels (d) and (e) of \reffi{fig:newmt50}. These points are
located in the focus-point region of the $(m_{1/2}, m_0)$
plane~\cite{focus}, where the LSP has a larger Higgsino content, whose
enhanced annihilation rate brings the relic
density down into the range allowed by WMAP. 
By comparison with our previous analysis, the focus-point region appears at
considerably lower values of $m_0$, because of the reduction in the central
value of $m_t$. This focus-point strip extends to larger values of
$m_0$ and hence $m_{1/2}$ that are not shown. 
The least-disfavoured focus points have a 
$\De\chi^2$ of at least~3.3 (see the discussion of Table~\ref{tab:chi2}
above), and most of them are excluded at the 90\%~C.L. 

Taken at face value, the preferred ranges for the sparticle masses shown
in Figs.~\ref{fig:newmt10} and \ref{fig:newmt50} are quite
encouraging for both the LHC and the ILC. The gluino and squarks lie
comfortably within the early LHC discovery range, and several
electroweakly-interacting sparticles would be accessible to ILC(500)
(the ILC running at $\sqrt{s} = 500 \gev$). The
best-fit CMSSM point is quite similar to the benchmark point
SPS1a~\cite{sps} (which is close to point B of \citere{bench})
which has been shown to offer good
experimental prospects for both the LHC and ILC~\cite{lhcilc}. 
The prospects for sparticle detection are also quite good in the
least-disfavoured part of the focus-point region for $\tb = 50$ shown
in Figs.~\ref{fig:newmt50}, with the exception of the relatively heavy
squarks. 

\begin{figure}[htb!]
\begin{center}
\includegraphics[width=.48\textwidth]{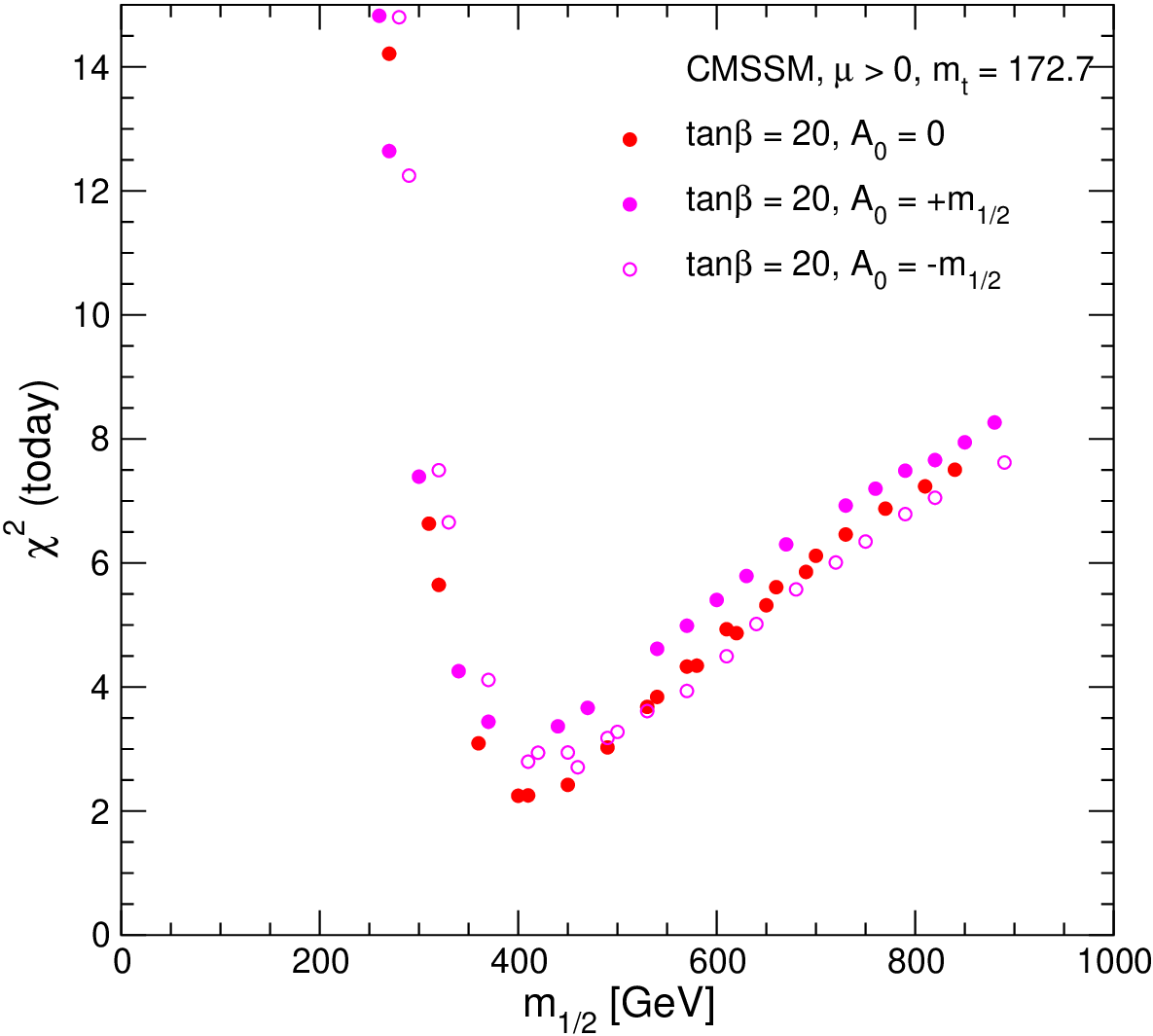}
\includegraphics[width=.48\textwidth]{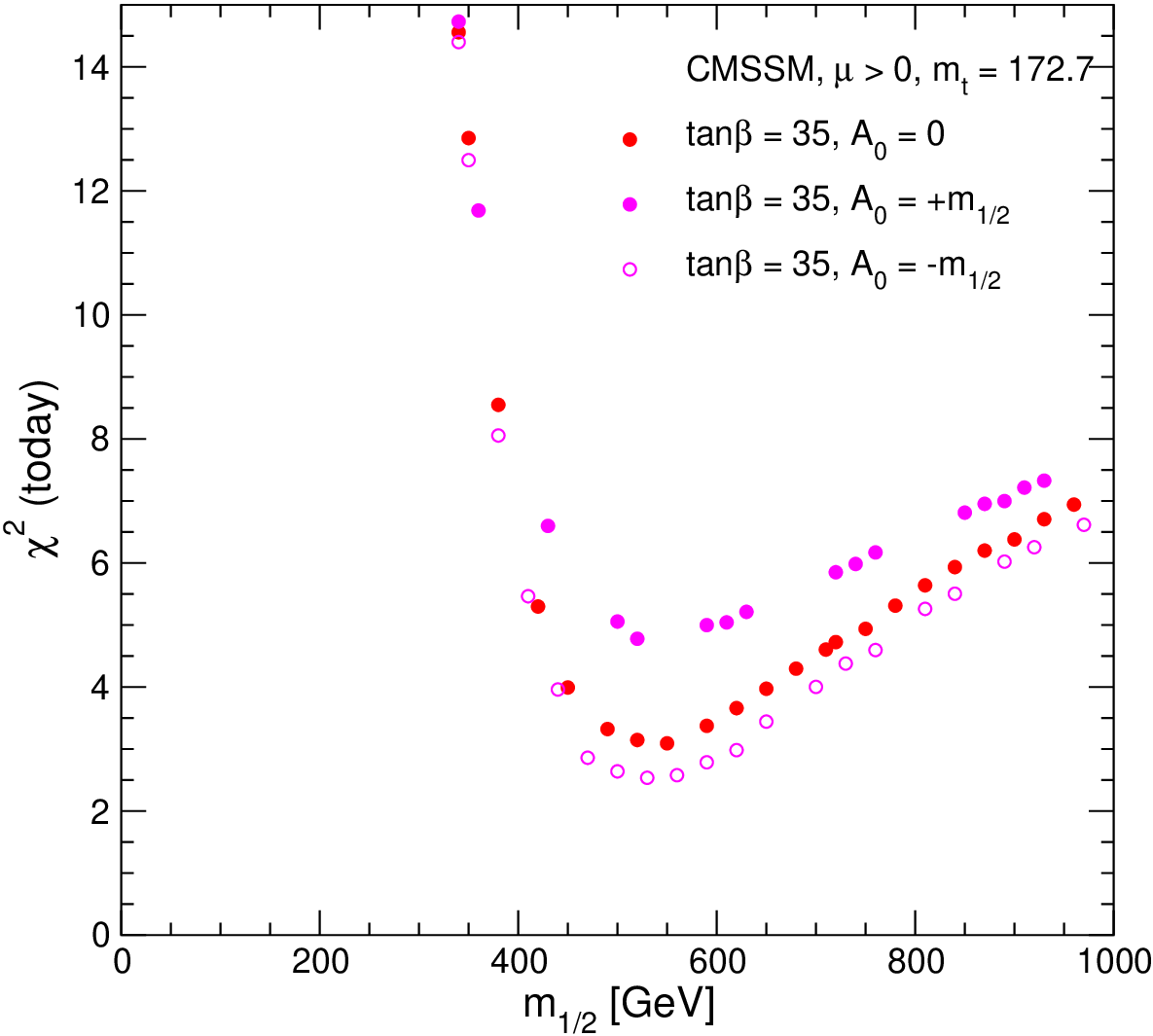}
\caption{
\it The $\chi^2$ functions for $\tb = 20, 35$ and $A_0 = 0, \pm 
m_{1/2}$.
}
\label{fig:2035} 
\end{center}  
\end{figure}

As indicated in \refta{tab:chi2} above,
the minimum values of $\chi^2$ are 2.5 for $\tb = 10$ and 2.8 for 
$\tb = 50$, found for $m_{1/2} \sim 320, 570 \gev$ and 
$A_0 = + m_{1/2}, - m_{1/2}$, respectively, revealing no
preference for either large or small $\tb$~
\footnote{
In our previous
analysis, we found a slight preference for $\tb = 10$ over $\tb = 50$.  
This preference has now been counterbalanced by the
increased pressure exerted by the Higgs mass constraint.
}%
. We display in
\reffi{fig:2035} the $\chi^2$ functions for two intermediate values of
$\tb = 20, 35$, for the cases $A_0 = 0, \pm m_{1/2}$. The minima of
$\chi^2$ are 2.2 and 2.5, respectively, which are not significantly
different from the values when $\tb = 10, 50$. Thus, this analysis reveals
no preference for intermediate values of $\tb$, either. The $\chi^2$
minima are found for $A_0 = 0, - m_{1/2}$, respectively. They appear when
$m_{1/2} \sim 400, 500 \gev$, values intermediate between the locations of
the minima for $\tb = 10, 50$, demonstrating the general stability of this 
analysis.


\begin{figure}[htb!]
\begin{center}
\includegraphics[width=.48\textwidth]{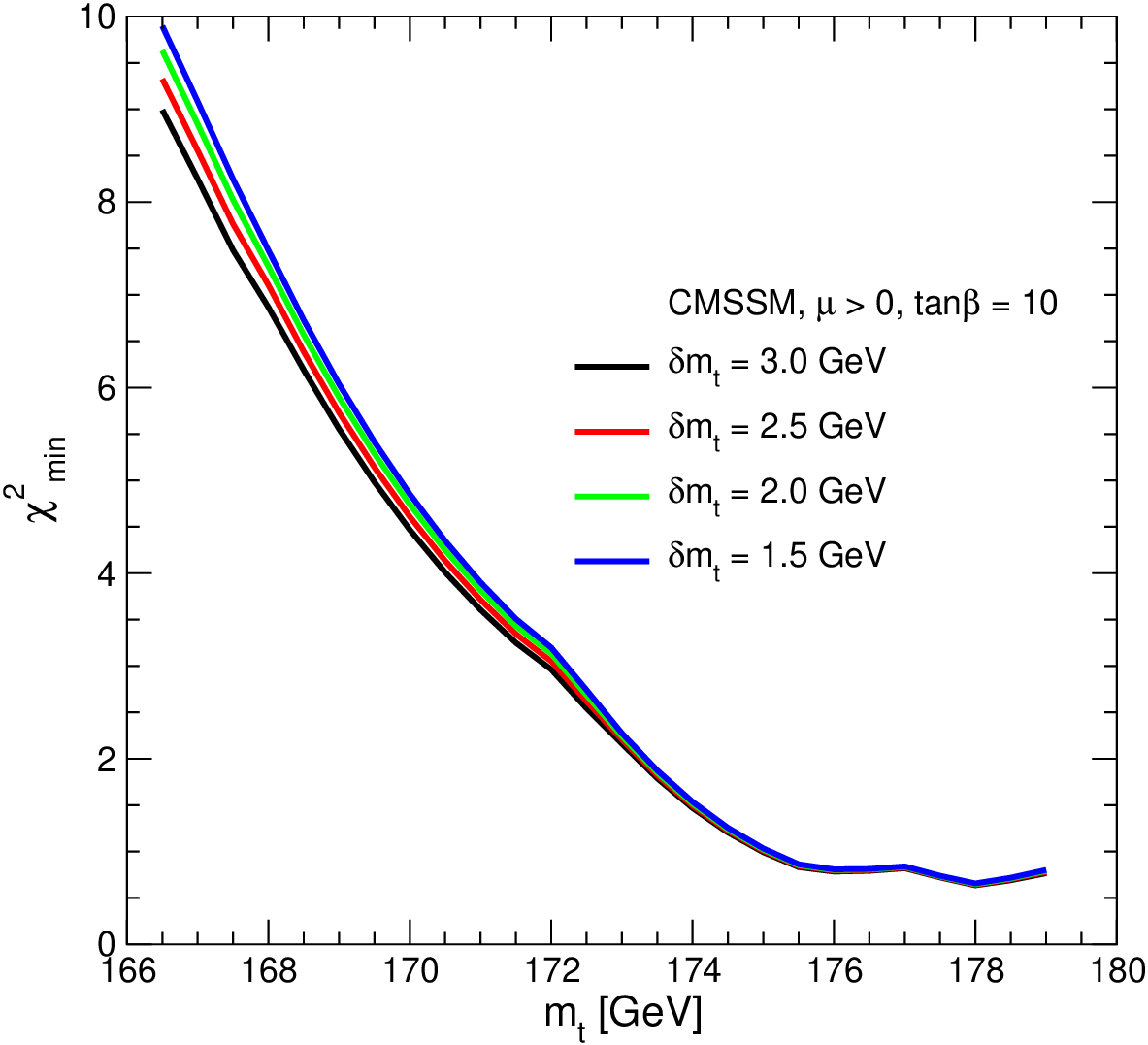}
\includegraphics[width=.48\textwidth]{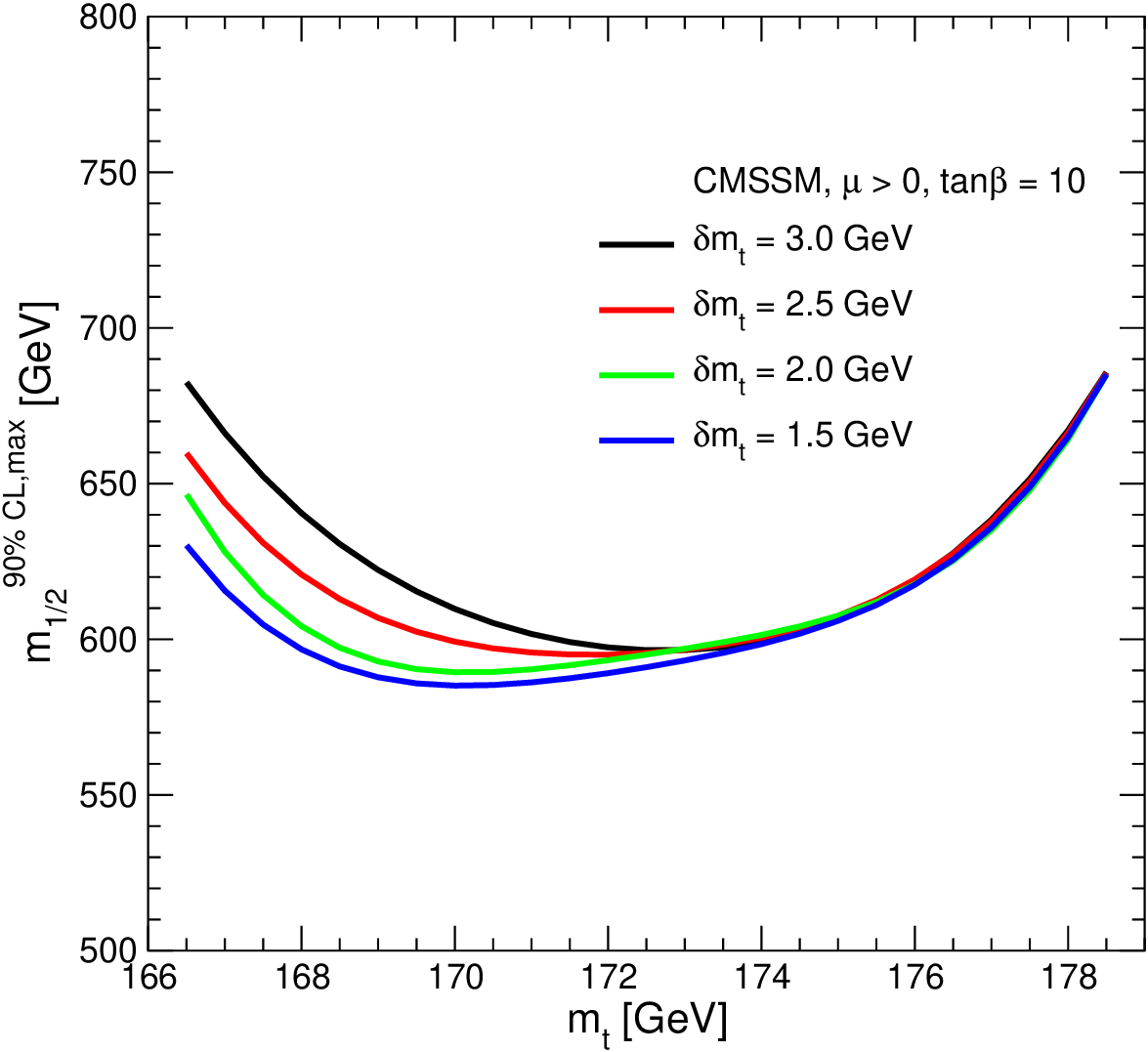}
\begin{picture}(0,0)
\CBox(-350,150)(-250,080){White}{White}
\end{picture}
\caption{
\it The dependence of (a) the minimum value of the $\chi^2$ distribution,
$\chi^2_{\rm min}$, and (b) the 90\% C.L.\ upper
limit for $m_{1/2}$ on $\mt$ and its experimental error $\delta \mt$, 
keeping the experimental values and theoretical predictions 
for the other precision observables
unchanged.
} 
\label{fig:varymt10} 
\end{center}  
\end{figure}  


\begin{table}[htb!]
\renewcommand{\arraystretch}{1.1}
\BC
\begin{tabular}{|c||c|c|c||c||c||c|c|c|c|c|}
\hline\hline
$\mt$ & $m_{1/2}$ & $m_0$ & $A_0$ & $\Mh$ & $\chi^2_{\rm tot}$ & 
$\MW$ & $\sweff$ & $(g-2)_\mu$ & $b \to s \ga$ & $\Mh$ \\ \hline\hline
168 & 270 &  80 &  270 & 111.5 &
      10.10 &  1.79 &  0.14 &  0.01 &  0.60 &  7.57 \\ \hline
168 & 370 & 100 &  370 & 113.5 &
       8.81 &  3.43 &  1.02  & 1.56 &  0.06 &  2.73 \\ \hline
168 & 530 & 140 &  530 & 115.3 &
      10.32 &  4.11 &  1.63 &  3.98 &  0.00 &  0.60 \\ \hline
168 & 800 & 210 &  800 & 116.9 &
      13.09 &  4.87 &  2.45 &  5.77 &  0.00 &  0.00 \\ \hline
\hline
168 & 200 &  80 &  400 & 111.1 &
      17.69 &  0.57 &  0.06 &  1.86 &  6.72 &  8.49 \\ \hline
168 & 300 & 100 &  600 & 114.1 &
       7.11 &  2.90 &  0.68 &  0.50 &  1.19 &  1.83 \\ \hline
168 & 520 & 160 & 1040 & 117.2 & 
      10.07 &  4.20 &  1.71 &  4.08 &  0.08 &  0.00 \\ \hline
168 & 820 & 250 & 1640 & 118.8 & 
      13.70 &  5.08 &  2.66 &  5.95 &  0.01 &  0.00 \\ \hline
\hline
173 & 190 &  70 &  190 & 111.1 & 
      17.20 &  0.03 & 0.36 &  4.56 & 3.78 & 8.49 \\ \hline
173 & 270 &  80 &  270 & 114.2 &  
      2.72 &  0.29 & 0.05 &  0.01 & 0.68 & 1.70 \\ \hline
173 & 330 &  90 &  330 & 115.8 &  
      2.24 &  0.91 & 0.08 &  0.80 & 0.19 & 0.27 \\ \hline
173 & 370 & 100 &  370 & 116.6 &  
      2.95 &  1.12 & 0.18 &  1.56 & 0.08 & 0.00 \\ \hline
173 & 530 & 140 &  530 & 118.8 &  
      6.02 &  1.54 & 0.49 &  3.98 & 0.00 & 0.00 \\ \hline
173 & 800 & 210 &  800 & 120.7 &  
      8.80 &  2.04 & 0.99 &  5.77 & 0.00 & 0.00 \\ \hline
\hline
173 & 170 &  80 &  340 & 112.1 & 
      25.10 &  0.02 & 0.40 &  6.21 &12.57 & 5.91 \\ \hline
173 & 200 &  80 &  400 & 113.7 & 
      12.12 &  0.00 & 0.70 &  1.85 & 7.15 & 2.41 \\ \hline
173 & 300 & 100 &  600 & 117.2 & 
       2.70 &  0.82 & 0.06 &  0.50 & 1.31 & 0.00 \\ \hline
173 & 520 & 160 & 1040 & 120.8 & 
       6.32 &  1.61 & 0.54 &  4.08 & 0.09 & 0.00 \\ \hline
173 & 820 & 250 & 1640 & 122.9 & 
       9.27 &  2.18 & 1.13 &  5.95 & 0.01 & 0.00 \\ \hline
\hline
178 & 210 &  60 &    0 & 112.5 & 
      10.68 & 0.34 & 1.43 & 3.25 & 0.70 & 4.93 \\ \hline
178 & 240 &  60 &    0 & 113.8 &  
      5.38 & 0.41 & 1.52 & 0.76 & 0.27 & 2.41 \\ \hline
178 & 330 &  80 &    0 & 116.7 &  
      0.76 & 0.01 & 0.17 & 0.58 & 0.00 & 0.00 \\ \hline
178 & 450 & 110 &    0 & 119.0 &  
      2.89 & 0.11 & 0.00 & 2.76 & 0.02 & 0.00 \\ \hline
178 & 600 & 140 &    0 & 120.9 &  
      4.75 & 0.22 & 0.02 & 4.48 & 0.03 & 0.00 \\ \hline
178 & 800 & 190 &    0 & 122.4 &  
      6.19 & 0.36 & 0.13 & 5.67 & 0.02 & 0.00 \\ \hline
\hline
178 & 190 &  70 &  190 & 113.6 & 
      13.26 & 0.43 & 1.51 & 4.56 & 4.03 & 2.73 \\ \hline
178 & 270 &  80 &  270 & 117.1 &  
       1.53 & 0.08 & 0.68 & 0.01 & 0.77 & 0.00 \\ \hline
178 & 330 &  90 &  330 & 119.0 & 
       1.14 & 0.02 & 0.10 & 0.80 & 0.23 & 0.00 \\ \hline
178 & 370 & 100 &  370 & 119.9 & 
       1.76 & 0.06 & 0.03 & 1.56 & 0.10 & 0.00 \\ \hline
178 & 530 & 140 &  530 & 122.4 & 
       4.20 & 0.20 & 0.01 & 3.98 & 0.00 & 0.00 \\ \hline
178 & 800 & 210 &  800 & 124.7 & 
       6.35 & 0.41 & 0.17 & 5.77 & 0.00 & 0.00 \\ 
\hline\hline
\end{tabular}
\EC
\renewcommand{\arraystretch}{1}
\caption{\it Breakdown of $\chi^2$ contributions from the different 
precision observables to $\chi^2_{\rm tot}$ for some example points
with $\mt = 168, 173, 178 \gev$, $\de\mt = 2.5 \gev$ and $\tb = 10$.
All masses shown are in GeV. The fifth column shows the $\Mh$ value for
the corresponding point, and the last column shows the $\chi^2$
contribution of this $\Mh$ value. The values of $A_0$ were selected so as to
minimize  $\chi^2_{\rm tot}$ for the corresponding value of $m_{1/2}$. 
}
\label{tab:chi2mt}
\vspace{-3em}
\end{table}

In view of the possible future evolution of both the central value of 
$\mt$ and its experimental uncertainty $\delta \mt$, we have analyzed the 
behaviour of
the global $\chi^2$ function for $166 \gev < \mt < 179 \gev$ and
$1.5 \gev < \delta \mt < 3.0 \gev$ for the case of $\tb = 10$ 
(assuming that the experimental results and theoretical predictions for 
the precision observables are otherwise unchanged),
as seen in the left panel of \reffi{fig:varymt10}. 
We see that the minimum value of $\chi^2$ is
almost independent of the uncertainty $\delta \mt$, but increases
noticeably as the assumed central value of $\mt$ decreases. This effect
is not strong when $\mt$ decreases from $178.0 \gev$ to $172.7 \gev$,
but does become significant for $\mt < 170 \gev$. This effect is not
independent of the known preference of the ensemble of precision
electroweak data for $\mt \sim 175 \gev$ within the
SM~\cite{lepewwg,LEPEWWG}, to 
which the observables $\MW$ and $\sweff$ used here make important
contributions. On the other hand, as already commented, within the 
CMSSM there is the additional effect that the best fit values of $m_{1/2}$ 
for very low
$\mt$ result in $\Mh$ values that are excluded by the LEP Higgs
searches~\cite{LEPHiggsSM,LEPHiggsMSSM} and have a very large
$\chi^2_{\Mh}$, resulting in an increase of the
lowest possible $\chi^2$~value for a given top-quark mass value. 
This effect also increases the value of  $m_{1/2}$ where the $\chi^2$
function is minimized.
This is analyzed in more detail in Table~\ref{tab:chi2mt}, where we show
the breakdown of the different contributions to $\chi^2$ for $\mt = 168, 173, 178 \gev$
for $\de\mt = 2.5 \gev$ and $\tb = 10$. The $A_0$ values are chosen so as to
minimize $\chi^2_{\rm tot}$ for each choice of $m_{1/2}$. For $\mt = 168 \gev$,
$\chi^2_{\rm tot}$ exhibits only a shallow and relatively high
minimum, and $\Mh$ and $\br(b \to s \ga)$ give the
largest contribution for low $m_{1/2}$, shifting smoothly to large contributions from $\MW$, $\sweff$ and
$(g-2)_\mu$ for larger $m_{1/2}$.
For $\mt = 173 \gev$, a more pronounced minimum of $\chi^2_{\rm tot}$
appears for relatively small $m_{1/2}$ values. For lower $m_{1/2}$,
again $\Mh$ and $\br(b \to s \ga)$ give large contributions, whereas for
higher values this shifts again to $\MW$, $\sweff$ and $(g-2)_\mu$,
after passing through a minimum with a very good fit quality where no
single contribution exceeds unity. The same trend, just slightly more
pronounced, can be observed for $\mt = 178 \gev$. 
Finally, in the right panel of 
\reffi{fig:varymt10} we demonstrate that the 90\% C.L.\ upper
limit on $m_{1/2}$ shows only a small variation, less than 10\% for
$\mt$ in the preferred range above $170 \gev$~%
\footnote{
The plot has been
obtained by putting a smooth polynomial through the otherwise slightly
irregular points.
}%
. 
Finally we note that the upper limit 
on $m_{1/2}$ is essentially independent of $\delta \mt$~%
\footnote{
Note added: this analysis demonstrates, in particular, that
incorporating the latest global fit value 
$m_t = 172.5 \pm 2.3$~GeV~\cite{newestmt} would have a negligible
effect on our $\chi^2$ analysis.
}%
.  

It is striking that the preference noted earlier for relatively low values
of $m_{1/2}$ remains almost unaltered after the 
change in $\mt$ and the change in the
treatment of the LEP lower limit on $\Mh$. There seems to be little chance
at present of evading the preference for small $m_{1/2}$ hinted by the
present measurements of $\MW$, $\sweff$, $\br(b \to s \ga)$ and 
$(g - 2)_\mu$, at least within the CMSSM framework. It should be noted
that the preference for a relatively low SUSY scale is correlated with 
the top mass value lying in the interval 
$170 \gev \lsim \mt \lsim 180 \gev$.


\section{NUHM Analysis}

In the NUHM, one may parametrize the soft supersymmetry-breaking
contributions to the squared masses of the two Higgs multiplets, $m_{1,2}^2$,
as follows:
\begin{equation}
m_i^2 \; = \; m_0^2 (1 + \delta_i)~,
\label{NUHMdelta}
\end{equation}
where $m_0^2$ is the (supposedly) universal soft supersymmetry-breaking
squared mass for the squarks and sleptons.
As already mentioned, the increase of the dimensionality of
the NUHM parameter space compared to the CMSSM, 
due to the appearance of the two new parameters $\delta_{1,2}$,
makes a systematic survey quite involved. Here, as
illustrations of what may happen in the NUHM, we analyze some specific
parameter planes that generalize certain specific CMSSM points. We
note that certain combinations of input parameter choices
lead to soft SUSY-breaking Higgs mass squares 
which are negative at the GUT scale.
When either $m_{1}^2 + \mu^2 < 0$ or  $m_{2}^2 + \mu^2 < 0$,
the point is excluded, so as to ensure vacuum stability at the GUT 
scale~\cite{NUHM}.

\begin{figure}[htb!]
\vspace{2em}
\begin{center}
\includegraphics[width=.48\textwidth]{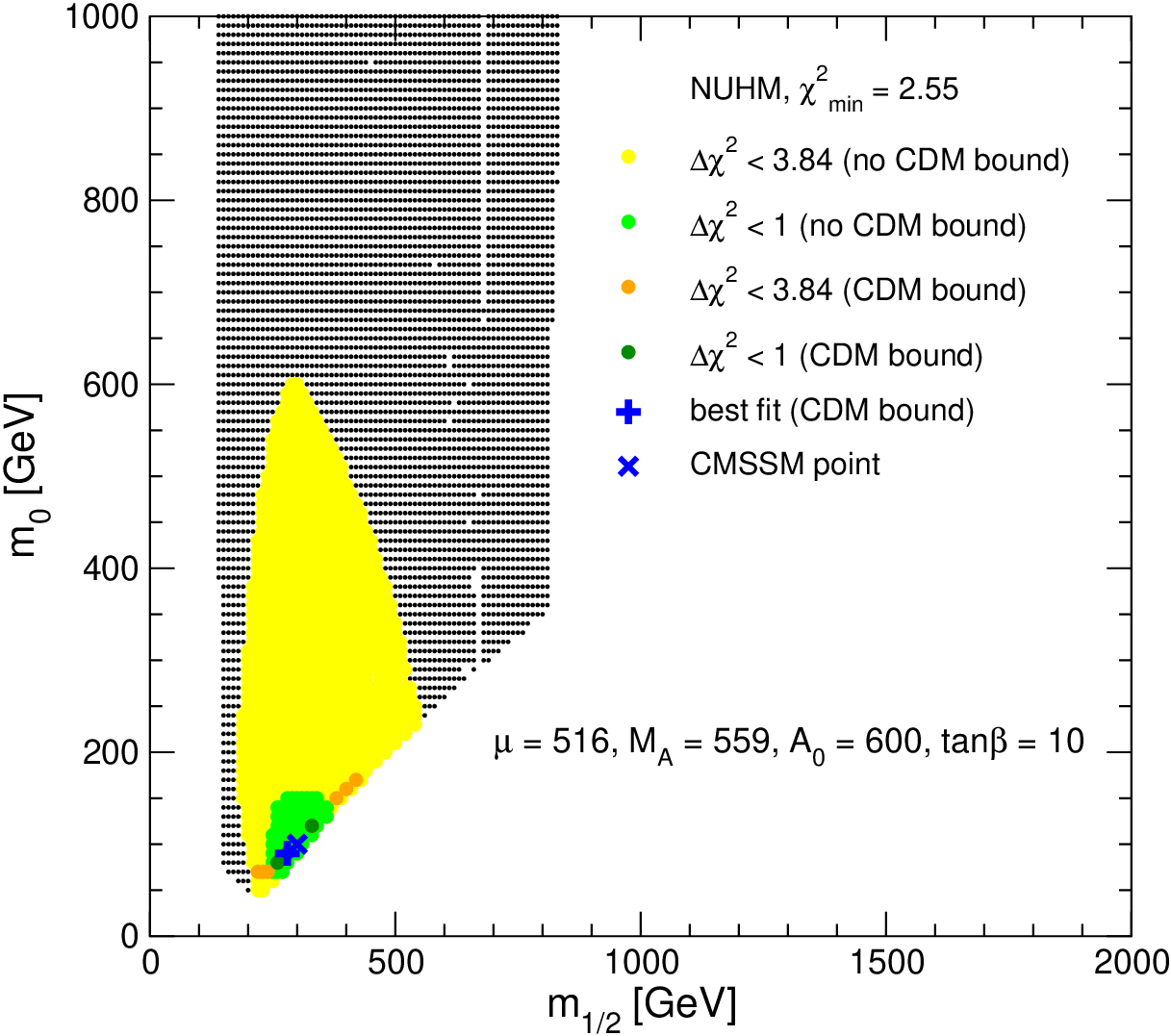}
\includegraphics[width=.48\textwidth]{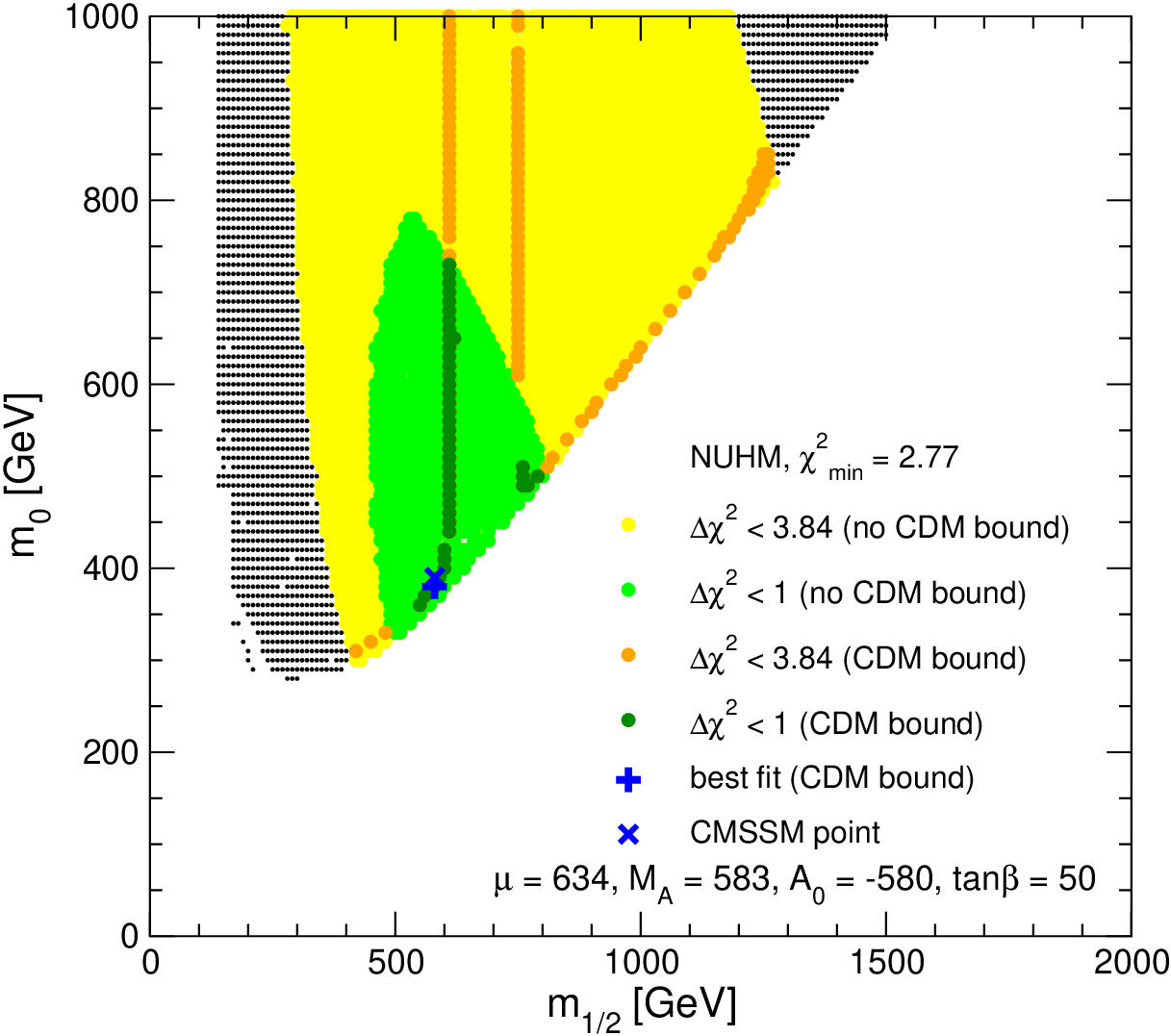}\\[2em]
\includegraphics[width=.48\textwidth]{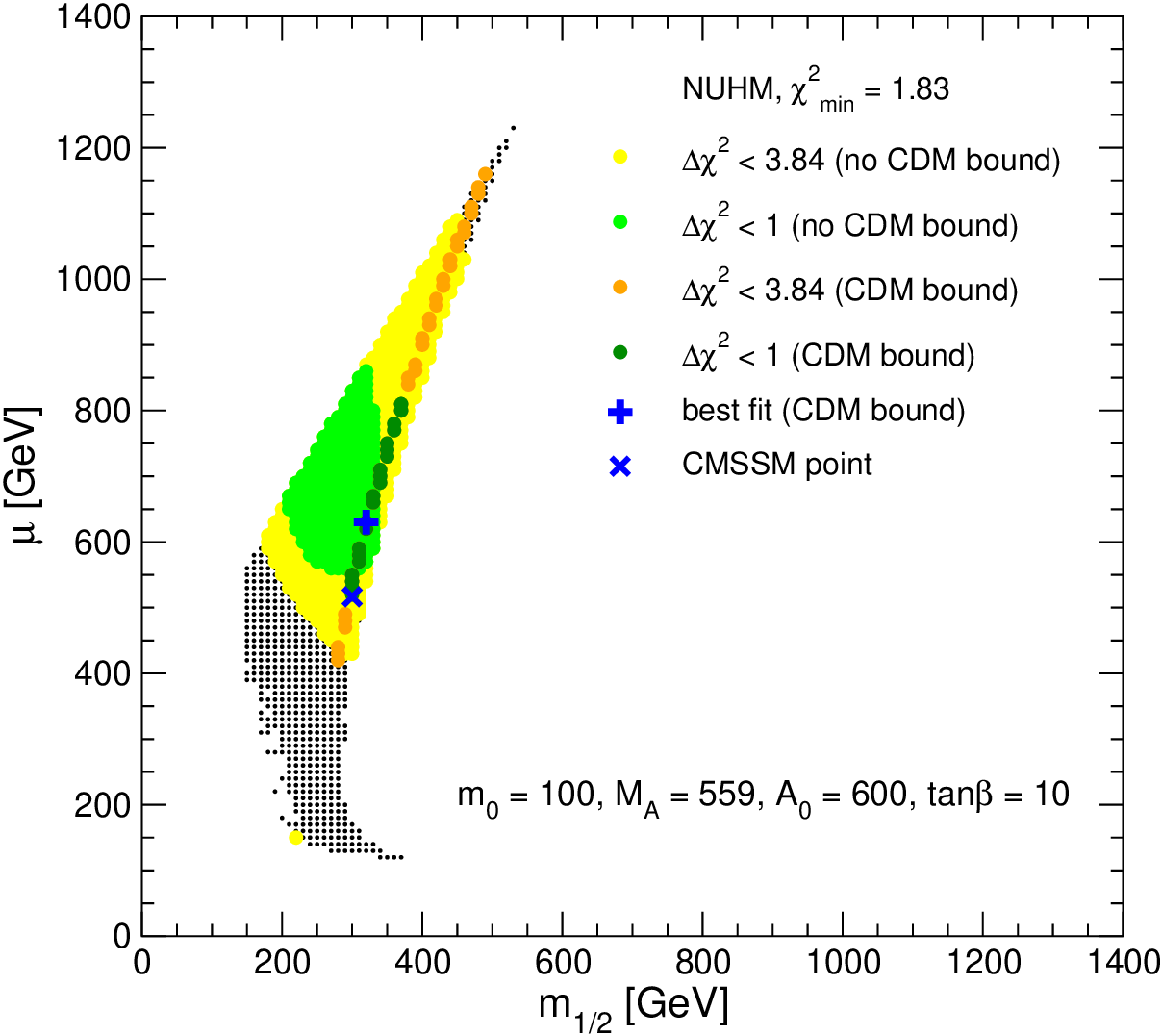}
\includegraphics[width=.48\textwidth]{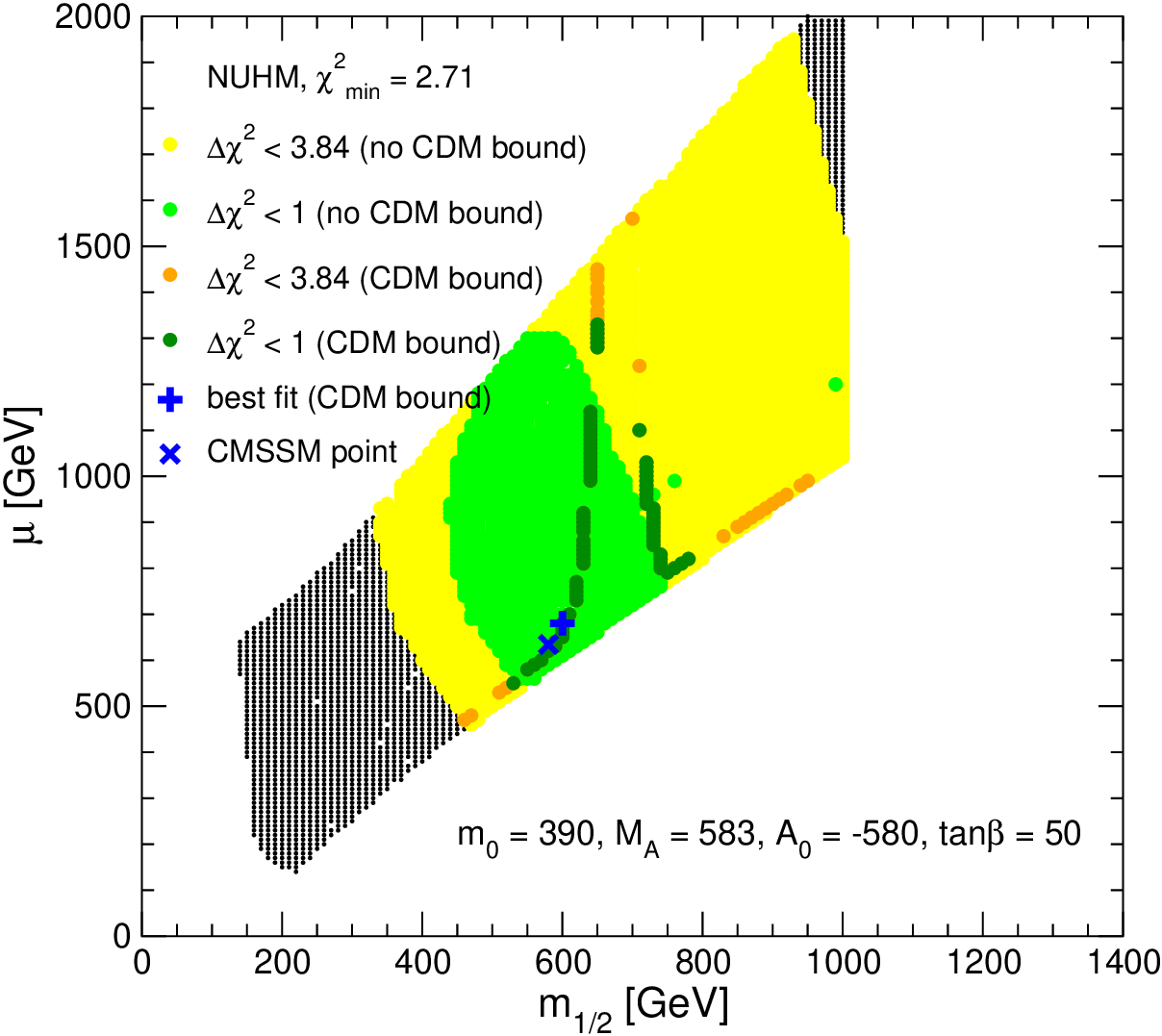}
\vspace{1em}
\begin{picture}(0,0)
\CBox(-340,400)(-250,300){White}{White}
\CBox(-350,175)(-250,080){White}{White}
\CBox(-110,327)(-020,260){White}{White}
\end{picture}
\caption{\it
Sample NUHM scenarios shown in the \plane{m_{1/2}}{m_0} (top row) and 
\plane{m_{1/2}}{\mu} (bottom row). The CMSSM points shown in the 
left (right) column correspond to the best fit points for $\tb = 10 \;(50)$. 
The other parameters are given in the plots. 
The green [medium grey] (yellow [light grey]) regions have 
$\De \chi^2 < 1 (3.84)$, whilst the
black regions have larger $\De \chi^2$. The strips where the 
neutralino relic density respects the WMAP limits have darker
shadings. The blue plus sign marks the best-fit NUHM point that
respects the relic density bounds, and the CMSSM point is marked with
a blue cross.  
} 
\label{fig:NUHMother}
\end{center}
\end{figure}

\begin{figure}[htb!]
\vspace{4em}
\begin{center}
\includegraphics[width=.48\textwidth]{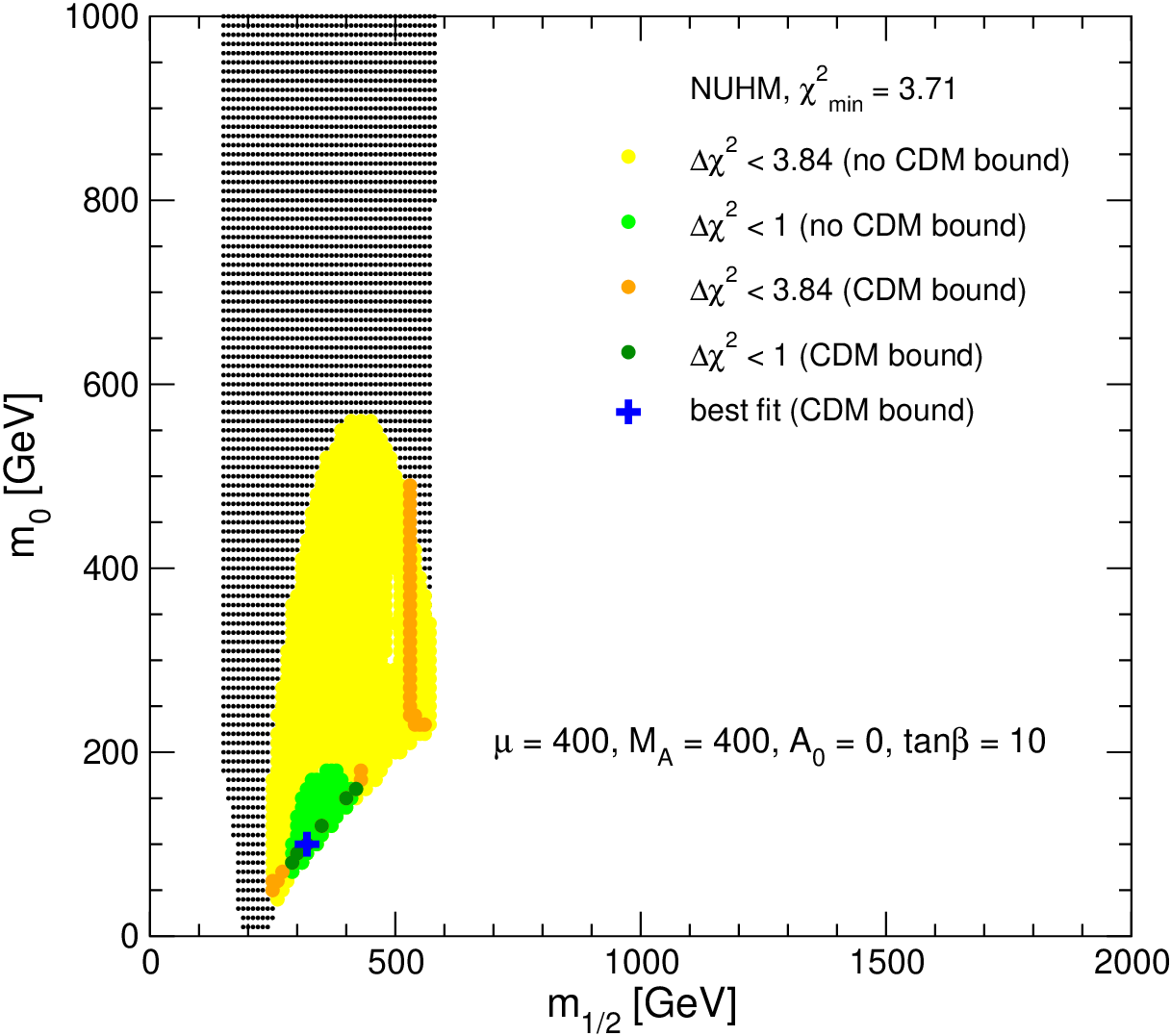}
\includegraphics[width=.48\textwidth]{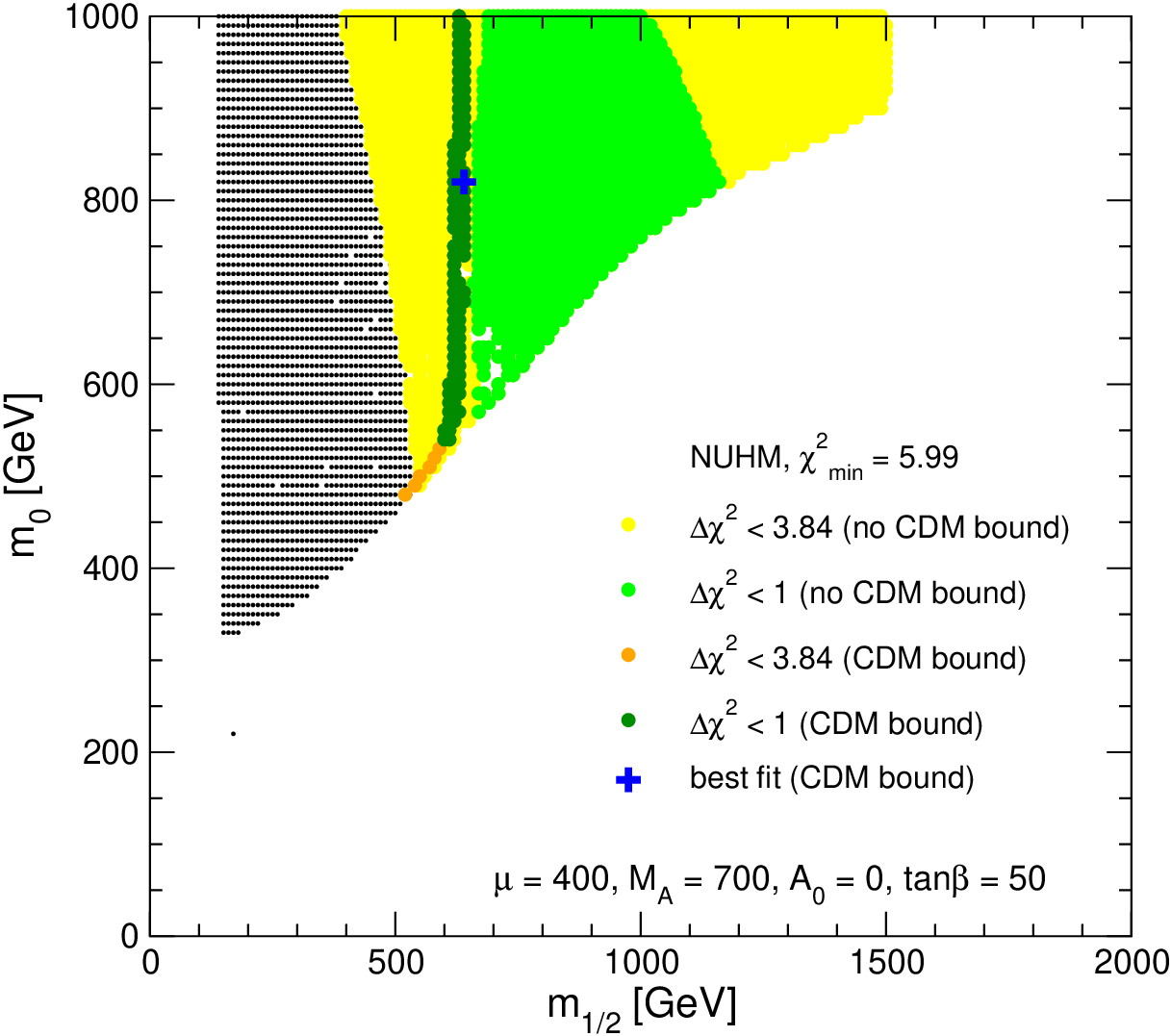}\\[3em]
\includegraphics[width=.48\textwidth]{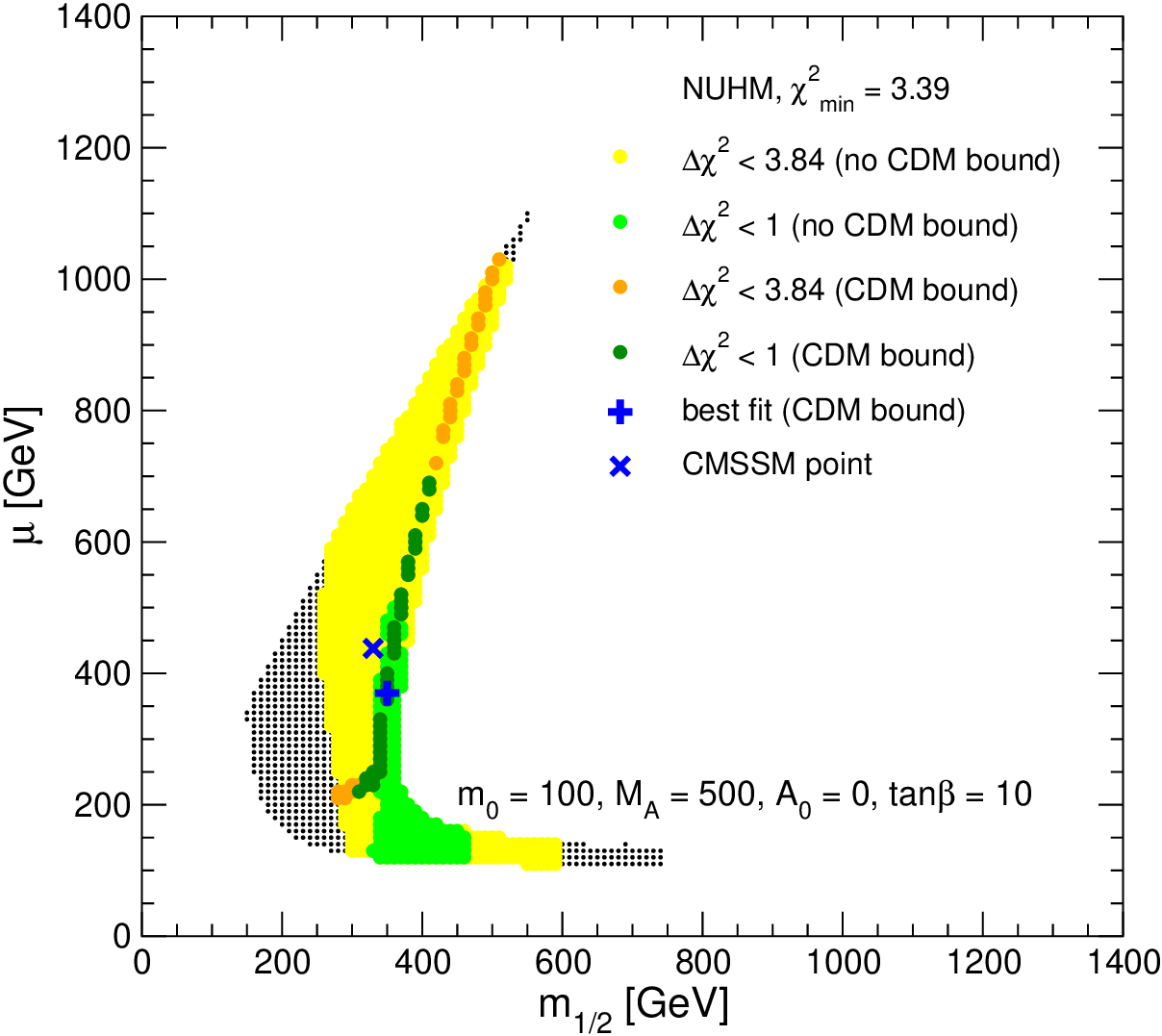}
\includegraphics[width=.48\textwidth]{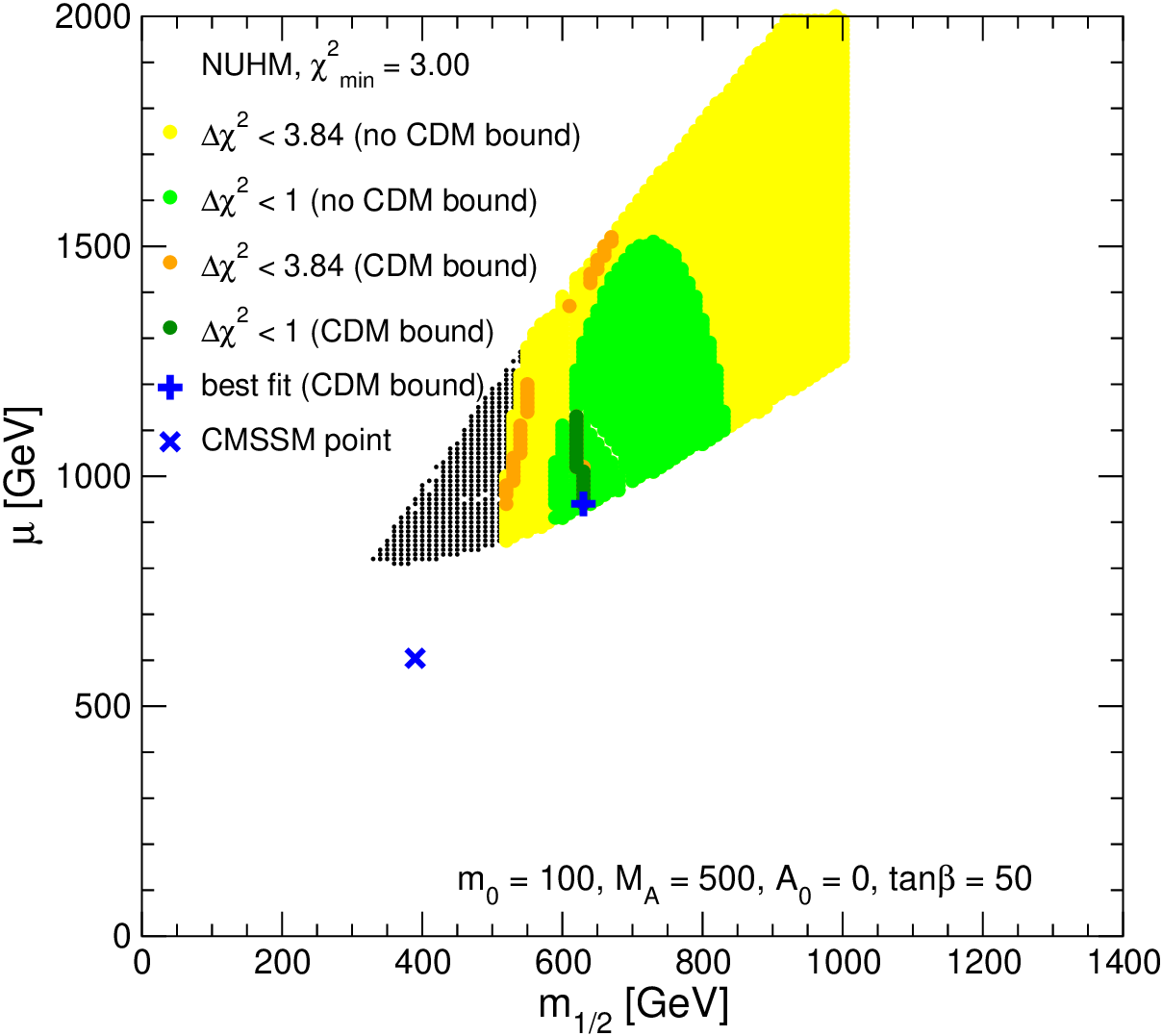}
\vspace{2em}
\begin{picture}(0,0)
\CBox(-340,410)(-245,300){White}{White}
\CBox(-350,175)(-250,080){White}{White}
\CBox(-110,340)(-020,280){White}{White}
\end{picture}
\caption{\it
Additional sample NUHM scenarios shown in the \plane{m_{1/2}}{m_0} (top 
row) and \plane{m_{1/2}}{\mu} (bottom row). 
The colour coding is the same as in \reffi{fig:NUHMother}.
} 
\label{fig:NUHManother}
\end{center}
\vspace{3em}
\end{figure}

\begin{figure}[htb!]
\vspace{4em}
\begin{center}
\includegraphics[width=.48\textwidth]{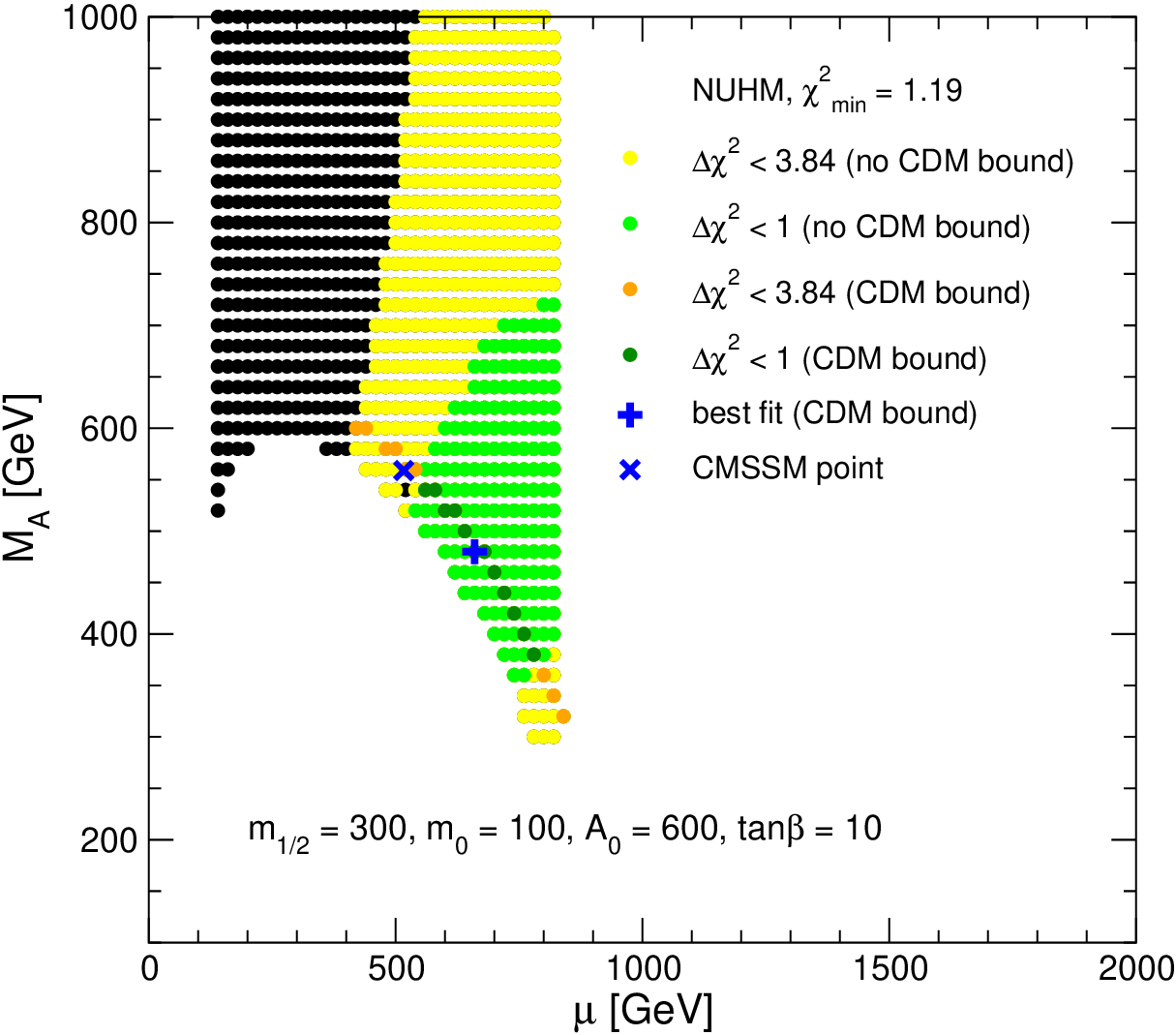}
\includegraphics[width=.48\textwidth]{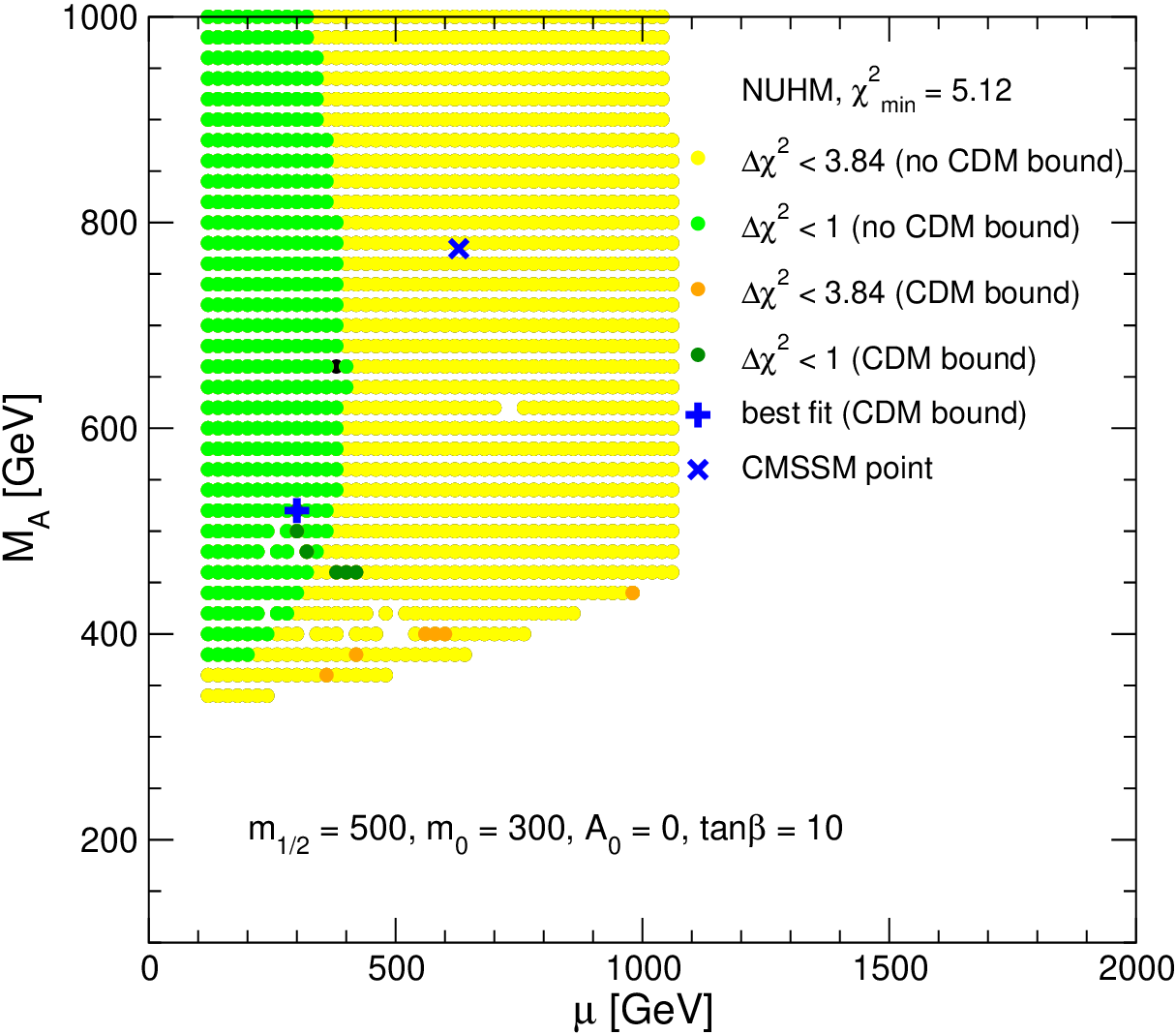}\\[3em]
\includegraphics[width=.48\textwidth]{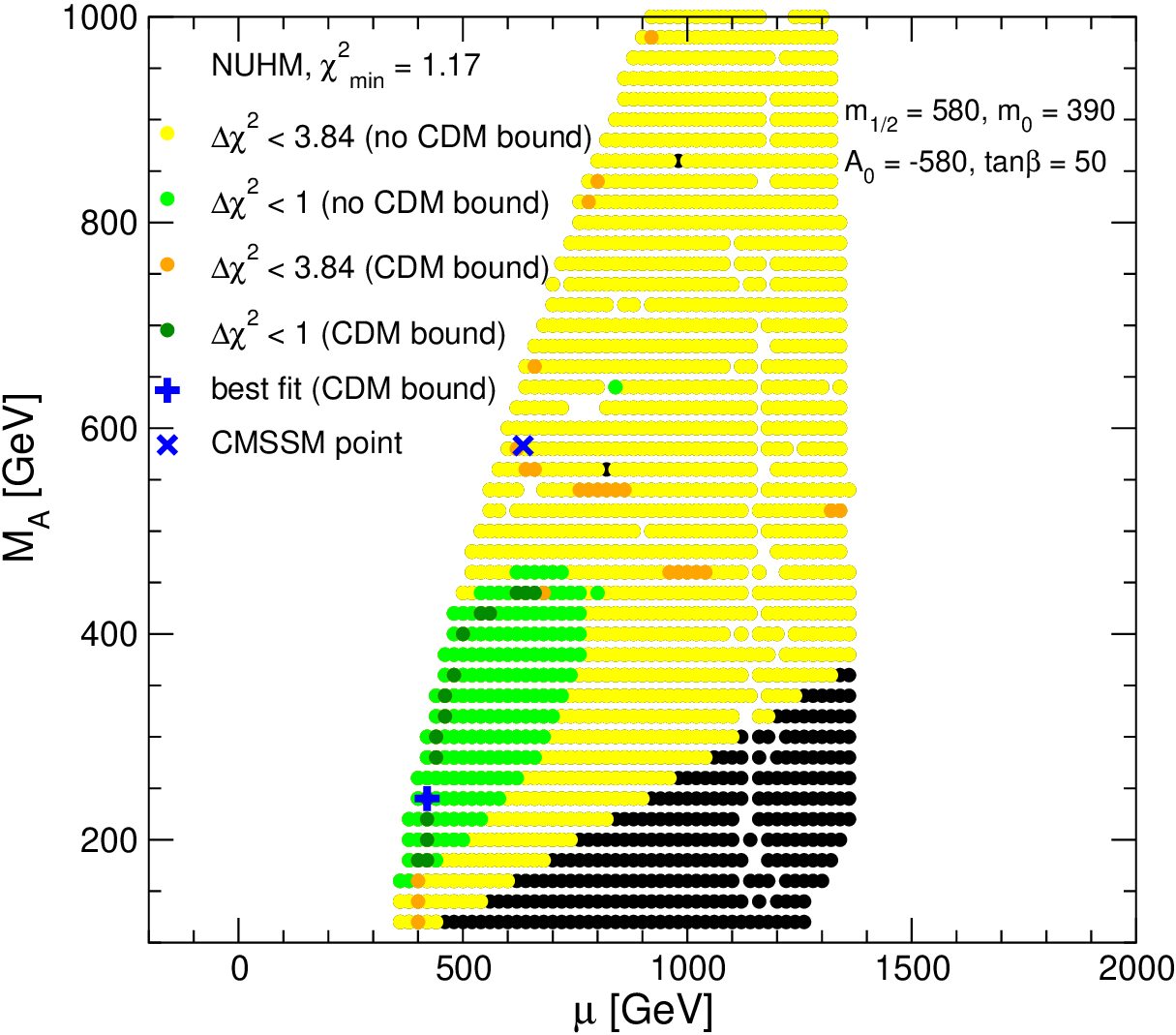}
\includegraphics[width=.48\textwidth]{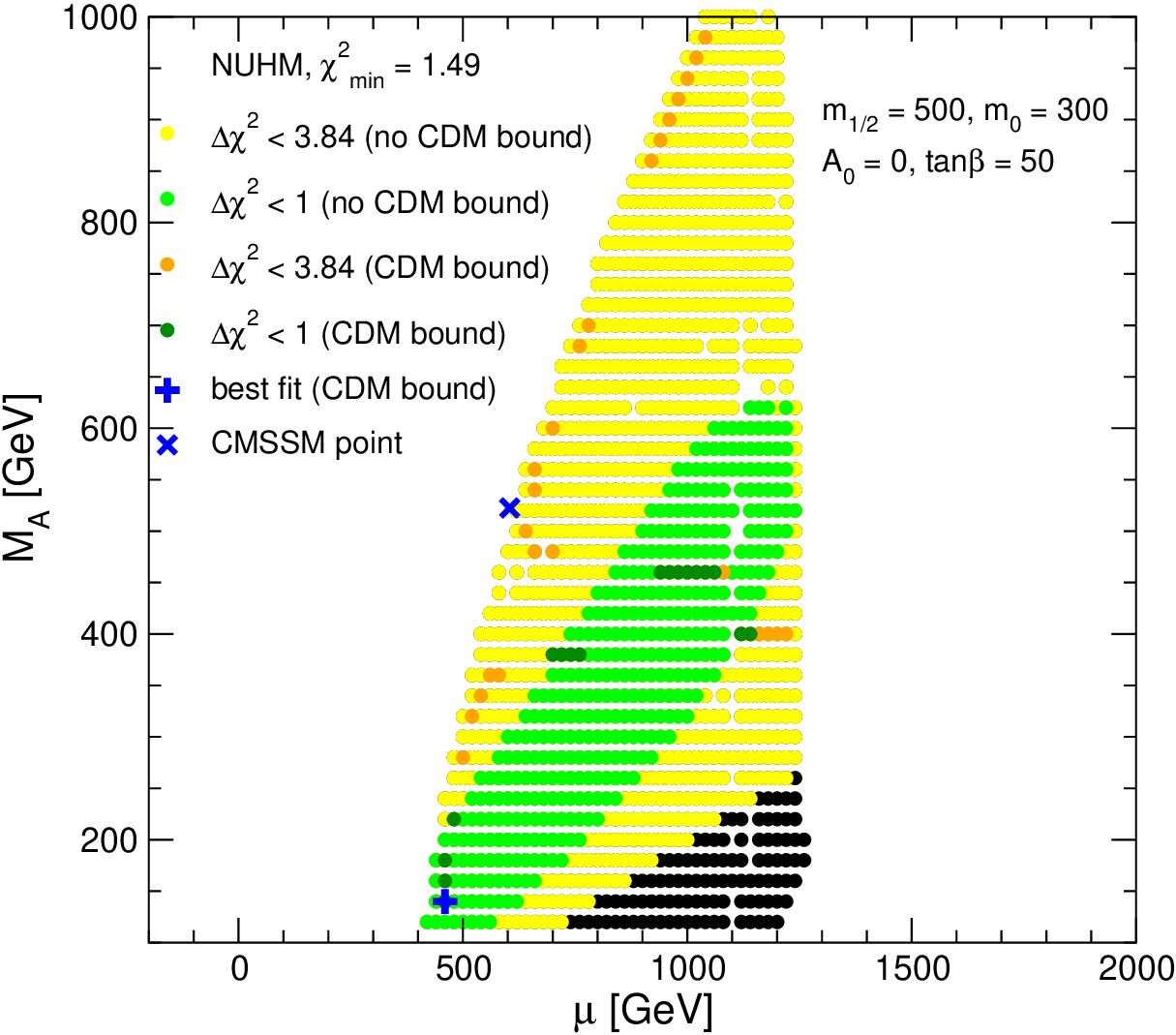}
\vspace{1em}
\begin{picture}(0,0)
\CBox(-340,405)(-245,300){White}{White}
\CBox(-095,410)(-013,330){White}{White}
\CBox(-427,180)(-344,168){White}{White}
\CBox(-427,165)(-350,153){White}{White}
\CBox(-427,151)(-352,141){White}{White}
\CBox(-427,139)(-356,129){White}{White}
\CBox(-427,127)(-360,117){White}{White}
\CBox(-427,116)(-370,106){White}{White}
\end{picture}
\caption{\it 
Sample NUHM $(\mu, \MA)$ planes for different choices of 
$(m_{1/2}, m_0, A_0, \tb)$:  
(a) $(300 \gev, 100 \gev, 600 \gev, 10)$, 
(b) $(500 \gev, 300 \gev, 0, 10)$, 
(c) $(580 \gev, 390 \gev$, $- 580 \gev, 50)$, and 
(d) $(500 \gev, 300 \gev, 0, 50)$. 
The colour coding is as in \reffi{fig:NUHMother}.
} 
\label{fig:NUHMmumA}
\end{center}
\vspace{2em}
\end{figure}

\begin{figure}[htb!]
\vspace{3em}
\begin{center}
\includegraphics[width=.48\textwidth]{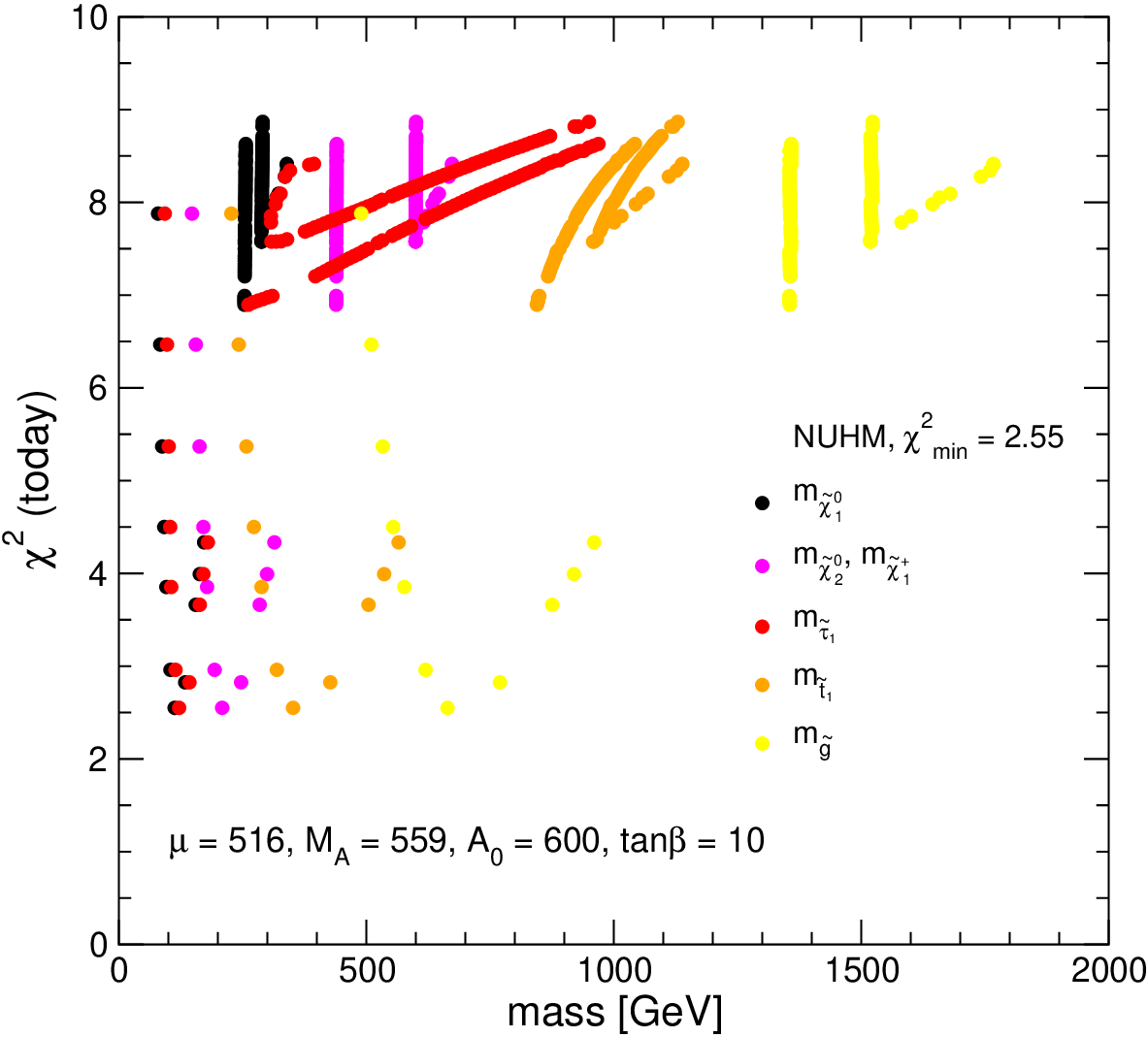}
\includegraphics[width=.48\textwidth]{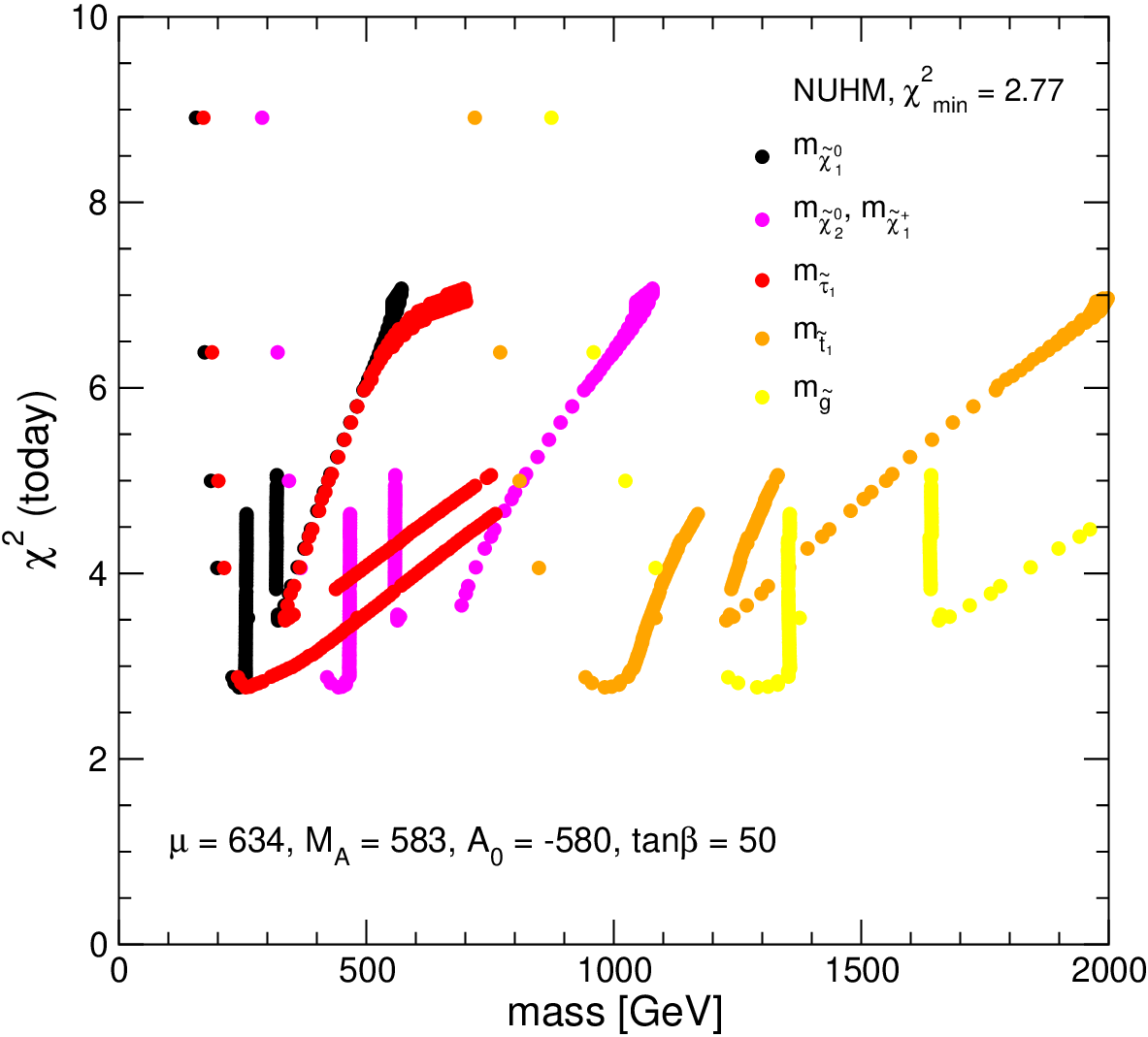}\\[3em]
\includegraphics[width=.48\textwidth]{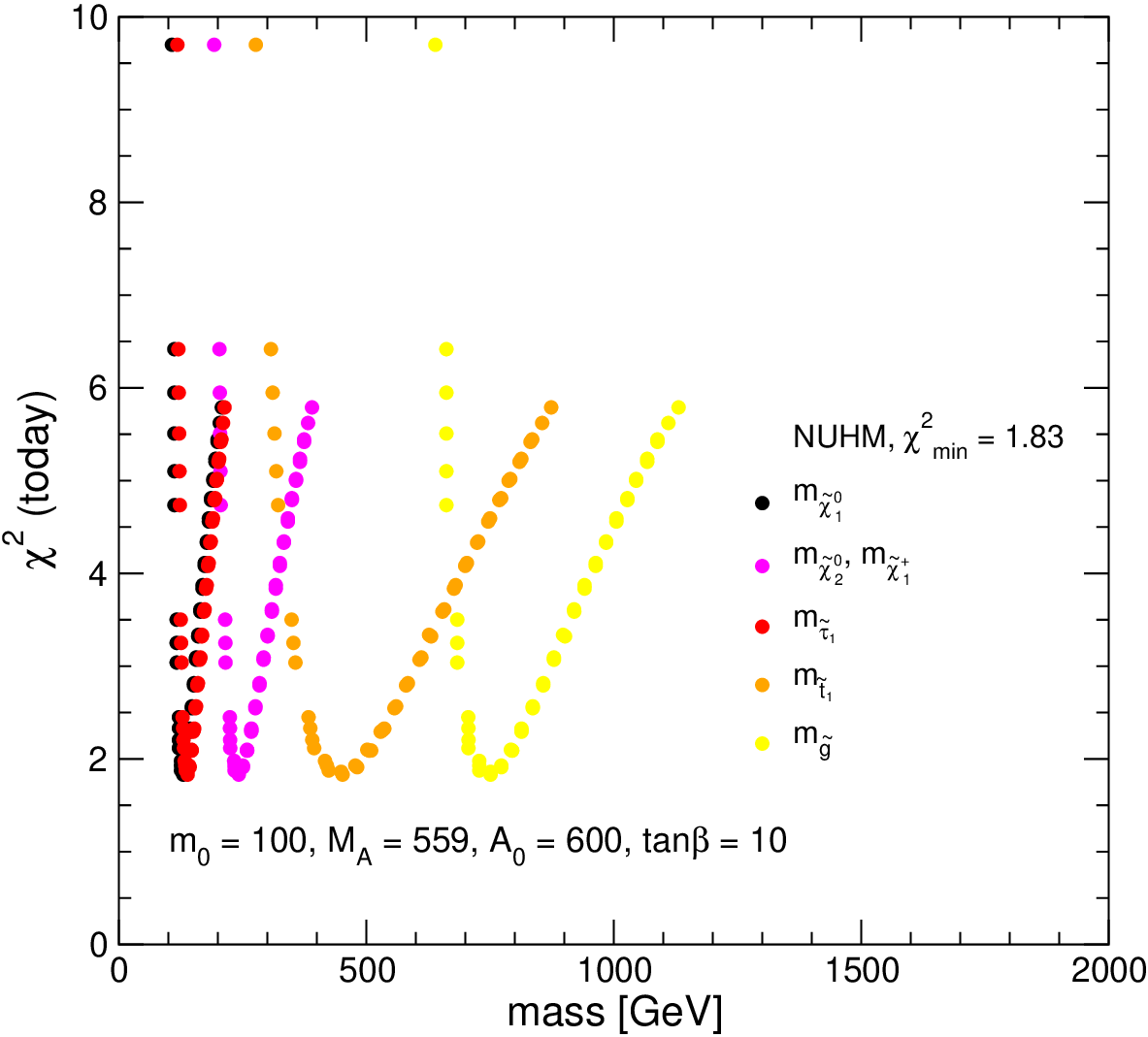}
\includegraphics[width=.48\textwidth]{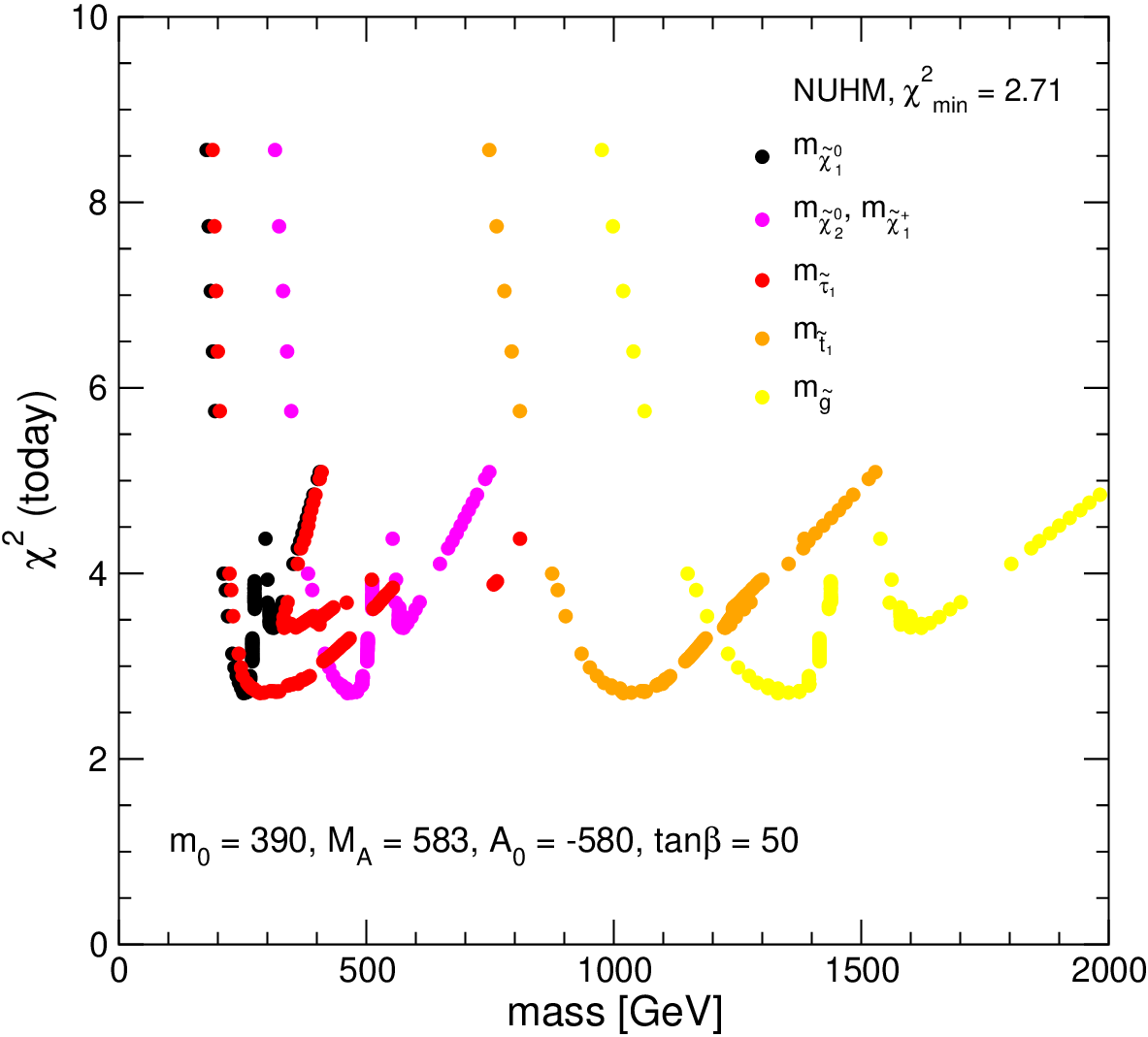}
\vspace{2em}
\begin{picture}(0,0)
\CBox(-080,417)(-040,360){White}{White}
\CBox(-320,110)(-260,050){White}{White}
\CBox(-085,180)(-040,115){White}{White}
\end{picture}
\caption{\it
The likelihood $\chi^2$ along the WMAP strips in the sample NUHM 
scenarios shown in Fig.~\protect\ref{fig:NUHMother}, as a function of the 
masses of the $\neu{1}$, $\neu{2}/\cha{1}$, $\Staue$, $\Stope$ and
$\gl$. The branchings in the $\chi^2$ curves reflect the corresponding
branchings in the WMAP strips in Fig.~\protect\ref{fig:NUHMother}.} 
\label{fig:NUHMmasses}
\end{center}
\vspace{3em}
\end{figure}

Since it is the value of $m_{1/2}$ that affects most importantly the masses of
the sparticles that might be observable at the LHC or ILC, our primary
objective is to  investigate whether the introduction of extra NUHM
parameters affects significantly the preference for small $m_{1/2}$
found previously within the CMSSM: see
\reffis{fig:NUHMother} and \ref{fig:NUHManother}.
After satisfying ourselves on this point, we subsequently explore the possible
dependences on $\MA$ and $\mu$: see \reffi{fig:NUHMmumA}.
In order to present our results we use parameter planes
with generic points that do not necessarily satisfy the CDM constraint.
Exhibiting full parameter planes rather than just the regions where 
the neutralino relic density respects the WMAP limits (we
indicate these strips in the plots) 
provides a better understanding of the dependences of the $\chi^2$
function on the different NUHM variables. It also provides a context
for understanding the branchings of the $\chi^2$ function visible in
\reffi{fig:NUHMmasses},
which are due to the  bifurcations of the WMAP strips in the
parameter planes. We also note that, in NUHM 
models with a light gravitino where the CDM constraint does not apply, generic 
regions of these parameter planes may be consistent with cosmology.

In view of our primary objective,
\reffi{fig:NUHMother} shows two examples of $(m_{1/2}, m_0)$ planes for
fixed values of $\mu, \MA, A_0$ and $\tb$ (top row) and two examples of
$(m_{1/2}, \mu)$ planes for fixed values of $m_0, \MA, A_0$ and $\tb$
(bottom row).  In both the two top panels, the left boundaries of the
shaded regions are provided by the LEP lower limit on the chargino mass,
the upper bounds on $m_{1/2}$ are provided by the GUT stability
constraint, and the lower edges of the shaded regions are provided by the
stau LSP constraint. The colour codings are as follows. In each panel, the
best fit NUHM point that respects the WMAP constraints on the relic
neutralino density is marked by a (blue) plus sign, and the (blue) cross
indicates the CMSSM values of $(m_{1/2}, m_0)$ [or $(m_{1/2}, \mu)$] for
the chosen values of the other parameters. The green (medium grey) regions
have $\De \chi^2 < 1$ relative to the minimum when the WMAP/CDM
constraint is {\it not} employed. Hence, some points in this region may
have a lower $\chi^2$ than our best fit point when the CDM constraint {\it
is} employed. 

In all four panels of \reffi{fig:NUHMother}, our best CDM
fit, denoted by the plus sign, is within 1 sigma of the overall minimum
$\chi^2$, and hence lies within the green region. The yellow (light grey)
regions have $\De \chi^2 < 3.84$, and the black points have larger
values of $\De \chi^2$ relative to the absolute minimum.  Traversing
the regions with $\De \chi^2 < 1, 3.84$, there are thin, darker shaded
strips where the relic neutralino density lies within the range favoured
by WMAP. That is, in these regions, $\chi^2$ is within 1 or 3.84 of the
minimum $\chi^2$ when the WMAP/CDM bound is included.  The blue cross must
always lie within these regions. Our sampling procedure causes these WMAP
strips to appear intermittent. In the top right panel of
\reffi{fig:NUHMother}, we note two vertical tramlines, which are due to
rapid annihilation via the direct-channel $A$ pole. Since $\MA$ is fixed
in each of these panels, there is always a value of $m_{1/2}$ such that 
$2 \mneu{1} \approx \MA$, in principle even for $\tb = 10$. We note
that the analogous tramlines are invisible in panel (a), because they
have a $\De \chi^2 > 4$ and thus would be located in the black shaded
region.

In the lower two panels, large values of $\mu$ are excluded due to the GUT
constraint, large values of $m_{1/2}$ are excluded by the stau LSP
constraint, and low values of $\mu$ and $m_{1/2}$ are exuded by the
chargino mass limit. In the lower left panel, large values of $m_{1/2}$
are excluded because the ${\tilde \tau}_1$ becomes the LSP, whereas in the
right panel our computation was limited to $m_{1/2} < 1000 \gev$, thus
producing the right-hand boundary. Within the regions allowed by these
constraints, the same colour codings are used. In the lower right panel,
one sees clearly the effect of the pseudoscalar funnel at $m_{1/2}
\approx 680 \gev$.  In the lower left panel, this possibility is excluded
by the GUT stability constraint.


The planes in \reffi{fig:NUHMother} have been defined such that the
CMSSM points marked by (blue) crosses in the different panels of
\reffi{fig:NUHMother} lie at the minima of the CMSSM $\chi^2$ functions
shown in \reffis{fig:newmt10} and \ref{fig:newmt50}. 
They enable us to study whether the CMSSM preference for relatively small
$m_{1/2}$ may be perturbed by generalizing to the larger NUHM parameter space.
In each case, we
see that the CMSSM point lies 
close to the best NUHM fit, whose $\chi^2$ is lower by just 0.00, 0.02,
0.72 and 0.08, respectively.  We also note that the ranges of $m_{1/2}$
favoured at this level are quite close to the CMSSM values.  Thus, in
these cases, {\it the introduction of two extra parameters in the NUHM does not
modify the preference for relatively small values of $m_{1/2}$ observed
previously in the CMSSM}. In the top left panel for $\tb = 10$, we see that
the preferred range of $m_0$ is also very close to the CMSSM value. On the
other hand, we see in the top right panel that rather larger values of
$m_0$ would be allowed for $\tb = 50$ at the $\De \chi^2 < 1$ level.
This is due to the insensitivity of the annihilation cross section to
$m_0$ in the funnel due to rapid annihilation via the pseudoscalar Higgs
boson $A$.  We also see in the bottom two panels that quite wide ranges of
$\mu$ would be allowed for either value of $\tb$~%
\footnote{
In all panels of \reffi{fig:NUHMother}, the assumed values of $\MA$
are sufficiently large that $B_s \to \mu^+ \mu^-$ currently does not
impose any useful constraint~\cite{ournewBmumu}.
}%
.


\reffi{fig:NUHManother} displays four analogous NUHM planes, specified
this time by values of $\mu, \MA$ and $A_0$ in the top row and $m_0, \MA,
A_0$ in the bottom row that do {\it not} correspond to minima of the
$\chi^2$ function for the CMSSM with the corresponding values of $\tb$. 
These examples were studied in detail in \cite{NUHM},
and enable us to explore whether there may be good NUHM fits that are
not closely related to the best CMSSM fits. 
In the top panels, the left boundaries are due to the chargino constraint,
and the bottom boundaries are due to the stau LSP constraint. In the left
panel, the right boundary is due to GUT stability, but in the right panel
it is due to a sampling limit. In the bottom left panel the GUT, stau and
chargino constraints operate similarly as in \reffi{fig:NUHMother}, and
the tail at low $\mu$ and large $m_{1/2}$ is truncated by the GUT
stability constraint. In the bottom right panel, the top boundary is due
to GUT stability, the bottom boundary to the stau, and the boundary at
large $m_{1/2}$ is another sampling limitation~%
\footnote{
We note that, in this
example, the CMSSM point is excluded by the stau LSP constraint.
}%
. Within the allowed regions of \reffi{fig:NUHManother}, the colour
codings are the same as in 
\reffi{fig:NUHMother}. The best fit CDM point lies within the 
$\De \chi^2 < 1$ green regions in the top left and bottom right panels, 
whereas
in the upper right panel the best fit point has $\De \chi^2$ slightly
larger than 1, and its $\De \chi^2$ is even greater in the bottom left
panel.

In the $(m_{1/2}, m_0)$ planes shown in the top row, we see that {\it the
ranges of $m_{1/2}$ favoured at the $\De \chi^2 < 1$ level are again
limited to values close to the best-fit CDM values}. The range of $m_0$
for a given $\De \chi^2$ is somewhat restricted for $\tb = 10$ (top
left), but is again considerably larger for $\tb = 50$ (top right). As for
the $(m_{1/2}, \mu)$ planes in the bottom row, we see in the left panel
for $\tb = 10$ that {\it the range of $m_{1/2}$ is again restricted at the
$\De \chi^2 < 1$ level}, whereas the range of $\mu$ is almost completely
unrestricted.  A similar conclusion holds in the bottom right panel for
$\tb = 50$, though here the range of $m_{1/2}$ is somewhat broader~%
\footnote{
In all panels of \reffi{fig:NUHManother}, the assumed values of
$\MA$ are again sufficiently large that $B_s \to \mu^+ \mu^-$
currently does not impose any useful constraint~\cite{ournewBmumu}.
}%
.

Having established that the CMSSM preference for small values of $m_{1/2}$
is generally preserved in the NUHM, whereas different values of $m_0$ and
$\mu$ are not necessarily disfavoured, we now study further the
sensitivity to $\mu$ and $\MA$ via the four examples of $(\mu, \MA)$
planes shown in \reffi{fig:NUHMmumA}. In each case, we have made specific
choices of $A_0$, $\tb$,  $m_{1/2}$ and $m_0$. In the two panels on
the left, these 
correspond to the best CMSSM fit along the corresponding WMAP strip. 
The examples on the right were studied in \cite{NUHM}. In
each case, we restrict our attention to the regions of the plane that have
no vacuum instability below the GUT scale. This constraint provides the
near-vertical right-hand edges of the coloured regions, whereas the other
boundaries are due to various phenomenological constraints. The
near-vertical boundaries at small $\mu$ in the top panels are due to the
LEP chargino exclusion, and those in the bottom panels are due to the stau
LSP constraint. The boundary at low $\MA$ in the top left panel is also
due to the stau LSP constraint, whereas that in the top right panel is
again the GUT stability constraint.


Within the allowed regions of \reffi{fig:NUHMmumA}, the colour codings are
the same as in \reffi{fig:NUHMother}. We see that in the top left panel
the WMAP strip runs parallel to the lower boundary defined by the stau LSP
constraint. The best fit NUHM point has $\chi^2 = 1.19$, which is somewhat
less than two units smaller than for the CMSSM point. This is hardly
significant, and suggests that the absolute minimum of the NUHM $\chi^2$
lies at a similar value of $m_{1/2}$. As seen from the location and shape
of the green region with $\De \chi^2 < 1$, the fit is relatively
insensitive to the magnitudes of $\mu$ and $\MA$, as long as they are
roughly proportional, but small values of $\mu / \MA$ are disfavoured.
In contrast, for the larger value of $m_{1/2}$ shown in the top right
panel of \reffi{fig:NUHMmumA}, we see that low values
of $\mu$ are preferred. However, the minimum value of $\chi^2 = 5.12$ in
the NUHM is not much lower than in the CMSSM, even though it occurs for
significantly smaller values of both $\mu$ and $\MA$~%
\footnote{
We recall
that, in this case, the NUHM WMAP strip has two near-horizontal branches
straddling the $\MA = 2 \mneu{1}$ contour, with the upper branch heading to
large $\MA$ at small $\mu$, features not seen clearly in this panel
because of the coarse parameter sampling.
}%
.

Turning now to the bottom left panel of \reffi{fig:NUHMmumA} for $\tb = 50$
and $A_0 = 580 \gev$, with $m_{1/2}$ and $m_0$ again chosen so as to minimize
$\chi^2$ (i.e., to reproduce the corresponding best-fit point), we
note several features familiar from the two previous panels. 
The WMAP strip clings close to the left boundary of the allowed region,
apart from an intermittent funnel straddling the $\MA = 2 \mneu{1}$ line.  
The minimum of $\chi^2 = 1.13$ for the NUHM is somewhat smaller than for
the CMSSM. The $\De \chi^2 < 1$ NUHM region is a lobe extending away
from the origin at small $\mu$ and $\MA$. Similar features are seen in the
bottom right panel for $\tb = 50$ and $A_0 = 0$, except that the 
$\De \chi^2 < 1$ lobe extends up to rather larger values of $\mu$ and
$\MA$~%
\footnote{
We note, however, that the lower ranges of $\MA \lappeq 300 \gev$ in
the two bottom panels of \reffi{fig:NUHMmumA} are likely to be
excluded by the current upper limit on 
$\br(B_s \to \mu^+ \mu^-)$~\cite{ournewBmumu}, 
once the experimental likelihood is made available and combined with
the corresponding theoretical errors. 
}%
. 

These examples show that, although the absolute values of $\mu$ and $\MA$
are typically relatively unconstrained in the NUHM~%
\footnote{
The prospects for an indirect determination of $\MA$ and
$\mu$ using future Higgs-sector measurements have been discussed in
\cite{ehow2}.
}%
, their values tend to
be correlated, often with a restricted range for their ratio: $\MA/ \mu
\sim 1.4, \le 1, \sim 2$ at the $\De \chi^2 < 1$ level
in the first three panels of
\reffi{fig:NUHMmumA}. On the other hand, the correlation in the fourth
panel takes the form $\MA \sim \frac{1}{2}(\mu - 400 \gev$).

\begin{figure}[htb!]
\begin{center}
\includegraphics[width=.48\textwidth]{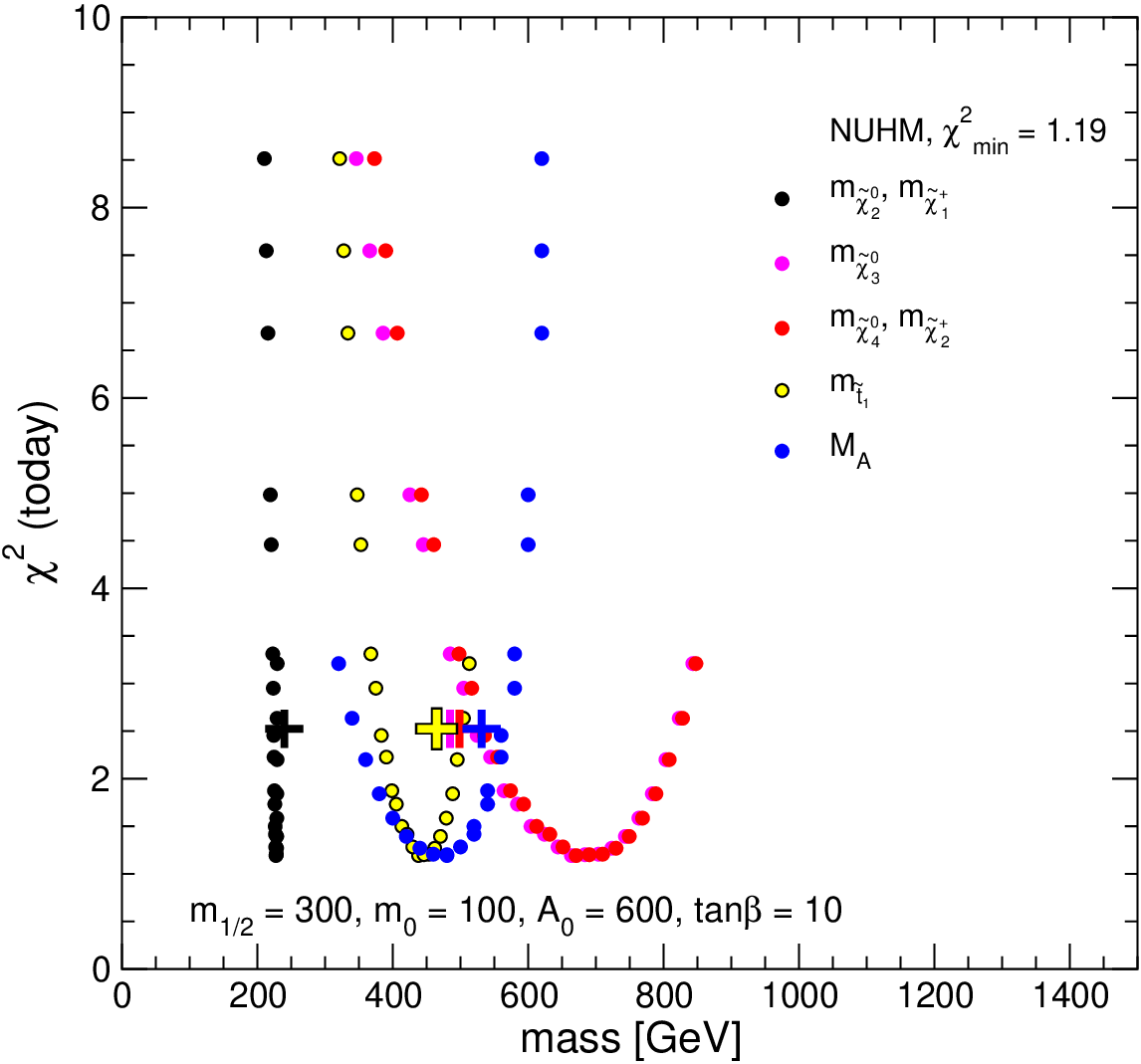}
\includegraphics[width=.48\textwidth]{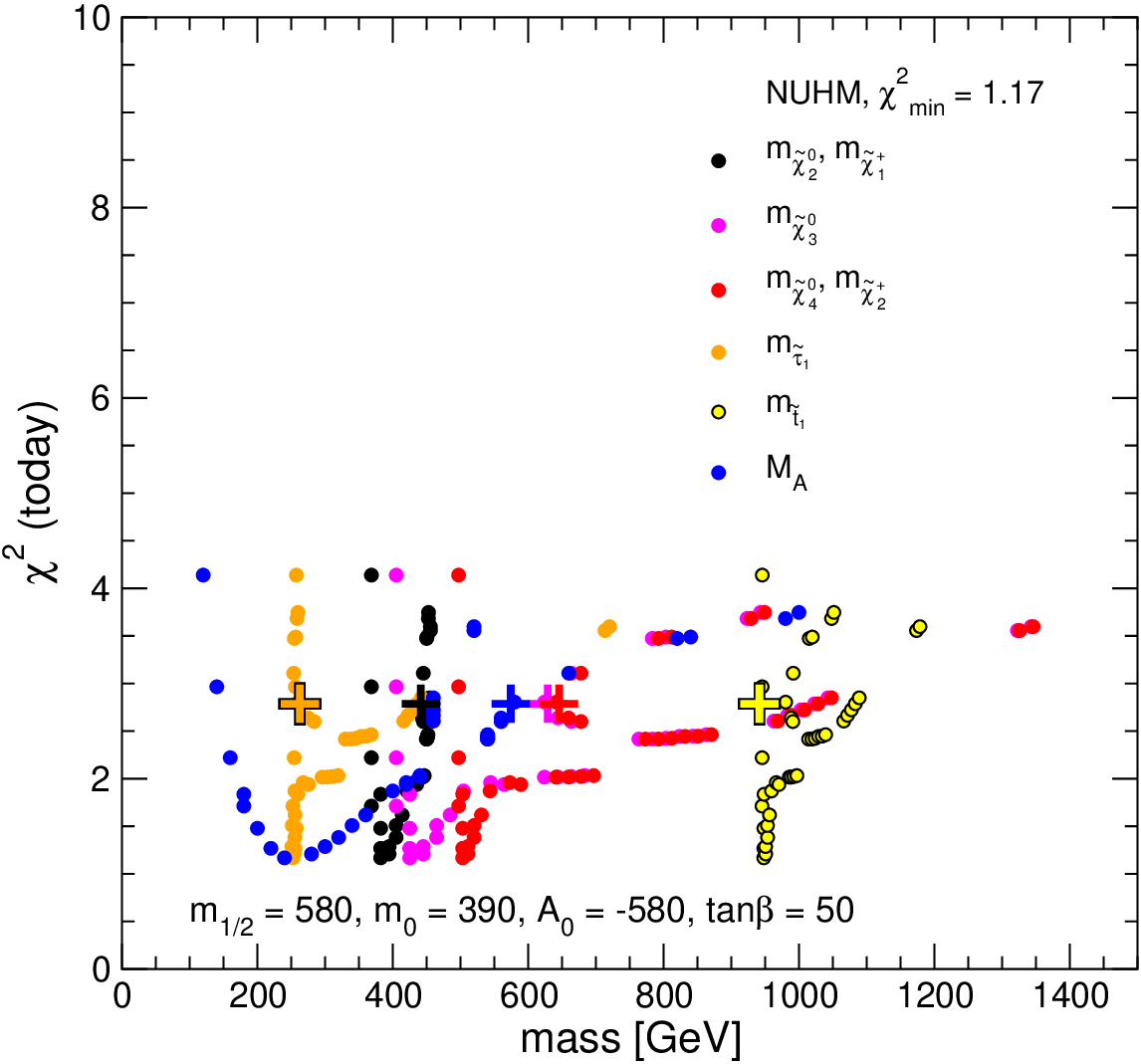}
\vspace{2em}
\caption{\it
The likelihood $\chi^2$ along the WMAP strips in the sample NUHM 
scenarios shown in the left panels of Fig.~\protect\ref{fig:NUHMmumA},
as a function of the masses of the $\neu{2}/\cha{1}$ (black), 
$\neu{3}$ (pink), $\neu{4}/\cha{2}$ (red), $\Staue$ (orange)
[omitted from the left panel],  $\Staue$ (yellow with black border) 
and $A$ boson (blue). The branchings in the $\chi^2$ curves in the
right panel reflect the corresponding branchings in the WMAP strips in
the bottom left panel of Fig.~\protect\ref{fig:NUHMmumA}.
The crosses indicate the corresponding best fit points in the CMSSM.
}  
\label{fig:NUHMmasses2}
\end{center}
\end{figure}

To conclude this Section, we make some remarks about the preferred masses
of sparticles and their possible detectability within the NUHM framework,
in the light of the above $\chi^2$ analysis. {\it Since the ranges of $m_{1/2}$
favoured within the CMSSM are also favoured in the NUHM, one should expect
that the LHC prospects for detecting the gluino and several other
sparticles may also be quite good in the NUHM}. On the other hand, the
greater uncertainties in $m_0, \mu$ and $\MA$ suggest that the prospects
for sparticle studies at the ILC may be more variable within the NUHM.
These remarks are borne out by \reffis{fig:NUHMmasses} and 
\ref{fig:NUHMmasses2}, which display
$\chi^2$ functions for various sparticle masses in a selection of NUHM 
scenarios. 
\reffi{fig:NUHMmasses} presents masses in the four NUHM scenarios
shown in \reffi{fig:NUHMother}, in which the CMSSM points
correspond to the best-fit points from \refse{sec:CMSSMupdate}, and 
Fig.~\ref{fig:NUHMmasses2} presents masses in two of the scenarios shown 
in \reffi{fig:NUHMmumA}. 

In each panel of \reffi{fig:NUHMmasses}, we display the $\chi^2$
functions for the masses of the $\neu{1}$, $\neu{2}/\cha{1}$,
$\Staue$, $\Stope$ and $\gl$, for NUHM parameters along the WMAP strips
in the corresponding panels of \reffi{fig:NUHMother}. Since there are
several branches of the WMAP strips in some cases, the $\chi^2$ functions
are sometimes multivalued.
In the top left panel of \reffi{fig:NUHMmasses}, we see well-defined
preferred values for the sparticle masses, with the gluino and stop masses
falling comfortably within reach of the LHC, and the $\neu{1}, \Staue$
and possibly also the $\neu{2}$ and $\cha{1}$ within reach of
the ILC(500). When $\De \chi^2 \sim 4.5$, new branches of the $\chi^2$
function appear, corresponding to a branching of the WMAP strip around a
rapid-annihilation funnel when $\MA = 559 \gev \sim 2 \mneu{1}$.  
This funnel is not visible in \reffi{fig:NUHMother}, but would appear
in the black-spotted region of large $\De \chi^2$. The ILC(1000) would
have a good chance to see even the lighter stop.
Turning to the top
right panel of \reffi{fig:NUHMmasses}, we see that the branching due to
the rapid-annihilation funnel appears at much lower $\De \chi^2$,
reflecting the closeness of the funnel to the best-fit point in the top
right panel of \reffi{fig:NUHMother}. In this case, whereas the
${\tilde g}$ should be observable at the LHC, the ${\tilde t}_1$ might
well be problematic~%
\footnote{
We recall that it is thought to be observable
at the LHC if it weighs less than about 1~TeV.
}%
. The $\neu{1}$ would be
kinematically accessible at the ILC(500), but the ${\tilde \tau}_1$ might
well be too heavy: the rises in the branches of its $\chi^2$ function at
larger masses reflect the extension of the WMAP strip to large $m_0$ that
is seen in the corresponding panel of \reffi{fig:NUHMother}. In this
particular scenario, the $\neu{2}$ and $\cha{1}$ would probably not be
observable at the ILC(500). The ILC(1000) on the other hand, would
have a high potential to detect them. The bottom left panel of
\reffi{fig:NUHMmasses} has the most canonical $\chi^2$ functions: the
gluino and stop would very probably lie within reach of the LHC and the
$\Staue$ within reach of the ILC(500), whereas the $\neu{2}$ and
$\cha{1}$ might be more problematic. Again the ILC(1000) offers much
better opportunities here, possibly even for the lighter
stop. Finally, the prospective 
observabilities in the bottom right scenario would be rather similar to
those in the top right scenario:  we again see that, as one moves away
from the coannihilation strip, the $\Staue$ may become much
heavier than the $\neu{1}$, and too heavy to observe at the
ILC(500). The ILC(1000) should, on the other hand, offers very good
prospects. 


\reffi{fig:NUHMmasses2} presents a similar analysis of sparticle masses 
of the two favoured scenarios in \reffi{fig:NUHMmumA}, namely in the 
two left-hand panels. In these cases, we show the variations of the 
$\chi^2$ functions for the different masses as one follows the WMAP strip 
to larger values of $\MA$. In the left panel of \reffi{fig:NUHMmasses2}, 
we display the masses of the $\neu{2}$ and $\cha{1}$ (which are 
nearly equal) in black, the mass of the $\neu{3}$ in pink, the masses of 
the $\neu{4}$ and $\cha{2}$ (which are nearly equal) in red, the 
mass of the $\Stope$ in yellow (with black border), and $\MA$ in
blue. In each case, the  
$+$ sign of the same colour represents the best fit in the CMSSM for the 
same values of $m_{1/2}, m_0, A_0$ and $\tb$. The fact that the 
minima of the NUHM lie somewhat below the CMSSM points reflect the
fact that 
the NUHM offers a slightly better fit, but the difference is not significant. 
In this case, the preferred masses of the $\neu{2}$ and $\cha{1}$ 
are almost identical to the best-fit CMSSM values, and the same would be true 
for the $\neu{1}$ and $\gl$, which are not shown. The masses of 
the $\neu{3}$, $\Stope$ and $A$ are also very similar to their CMSSM
values, but the $\neu{4}$ and $\cha{2}$ may be significantly heavier.
In addition to the above sparticle masses, the right panel also includes 
the mass of the $\Staue$ in orange. In this case, whereas the masses 
of the $\neu{1}$ (not shown), $\neu{2}/\cha{1}$ and 
$\gl$ (not shown) preferred in the NUHM are similar to 
their values at the best-fit CMSSM point, this is not true for the other 
sparticles shown. The $A$ boson may be
considerably lighter, the $\neu{3}$, $\neu{4}$ and the $\cha{2}$ may
be either lighter or 
heavier, and the $\Staue$ and $\Stope$ might be significantly heavier for
points along the Higgs funnel visible in
\reffi{fig:NUHMmumA}. Thus, in this case the prospects for
detecting some sparticles at the LHC or ILC may differ substantially
in the NUHM from the CMSSM. 


To summarize: these examples demonstrate that, although the preferred value 
of the overall sparticle mass scale set by $m_{1/2}$ may be quite similar 
in the NUHM to its CMSSM value, the masses of some sparticles in the
NUHM may differ significantly from the corresponding CMSSM values.


\section{VCMSSM Analysis}

As an alternative to the above NUHM generalization of the CMSSM, we now
examine particular CMSSM models with the additional constraint $B_0 = A_0
- m_0$ motivated by minimal supergravity models, namely the VCMSSM
framework introduced earlier. We still assume that the gravitino is too 
heavy to be the LSP. The extra constraint reduces the
dimensionality of the VCMSSM parameter space, as compared with the CMSSM,
facilitating its exploration. In the CMSSM case, the electroweak vacuum
conditions can be used to fix $|\mu|$ and $\MA$ as functions of
$m_{1/2}, m_0$ and $A_0$ for a large range of fixed values of $\tb$. On
the other hand, in the VCMSSM case the expression for $B_0$ in terms of
$A_0$ and $m_0$ effectively yields a relation between $|\mu|$ and $\MA$
that is satisfied typically for only one value of $\tb$, for any fixed
set of $m_{1/2}, m_0$ and $A_0$ values~\cite{tbfixed,VCMSSM}.

As already mentioned, motivated by $(g - 2)_\mu$ and (to a lesser extent)  
$\br(b \to s \ga)$, we restrict our attention here to the case
$\mu > 0$. As is well known, other phenomenological constraints tend to
favour $\tb \gsim 5$, see e.g.\ \citeres{tbexcl,asbs2}. This condition
is generally obeyed along the WMAP 
coannihilation strip for neutralino dark matter in the VCMSSM if one
assumes $A_0 \ge 0$, in which case the resultant value of $\tb$ tends to
increase with $m_{1/2}$ and $m_0$ along the WMAP strip. We have studied
the choices $A_0/m_0 = 0, 0.75, 3 - \sqrt{3}$ and 2.  In this Section we
restrict our attention to these cases, and in the next Section
we compare the VCMSSM results with the corresponding gravitino dark matter
scenarios.

\begin{figure}[htb!]
\begin{center}
\includegraphics[width=.45\textwidth,height=5.4cm]{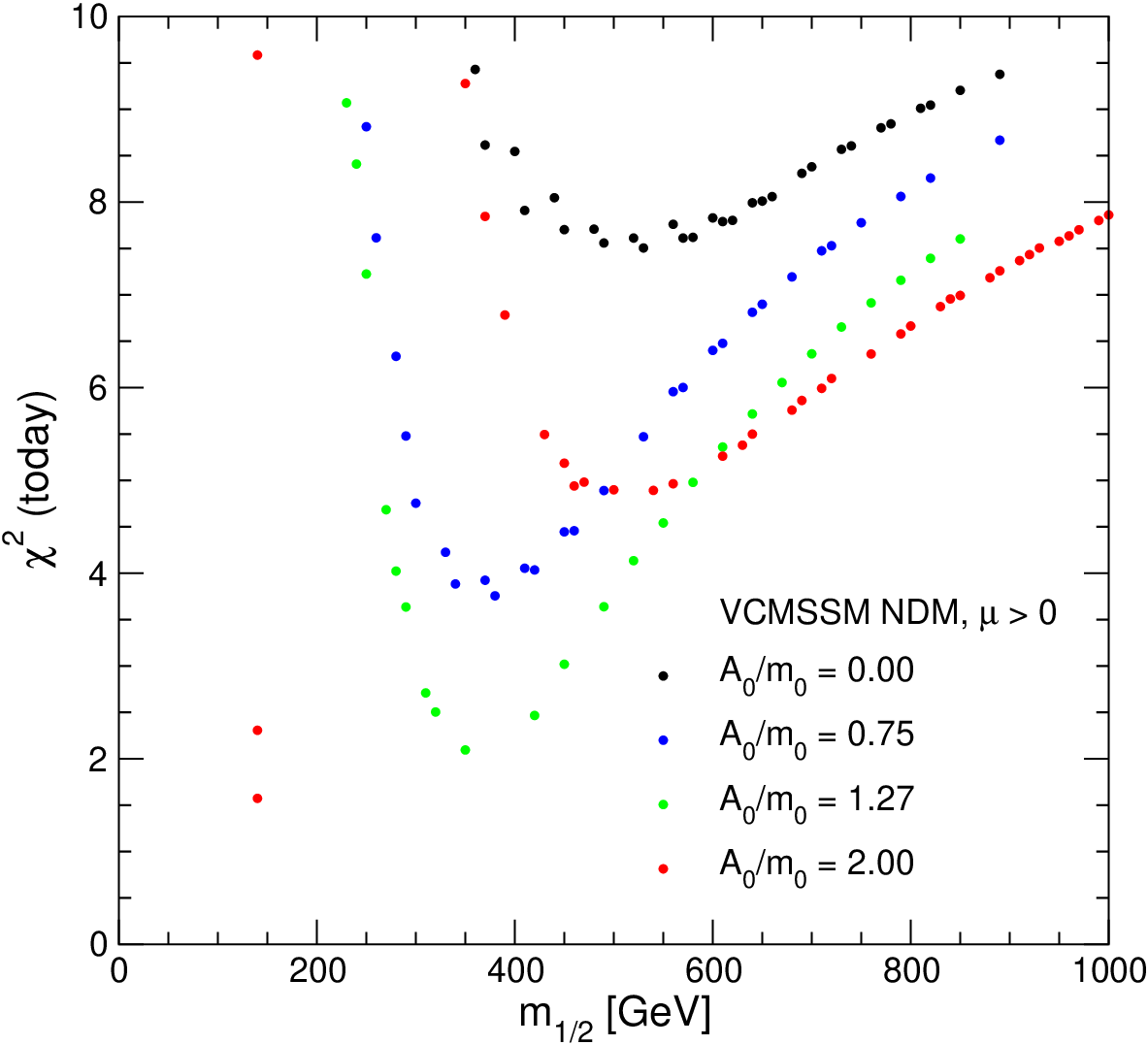}
\includegraphics[width=.45\textwidth,height=5.4cm]{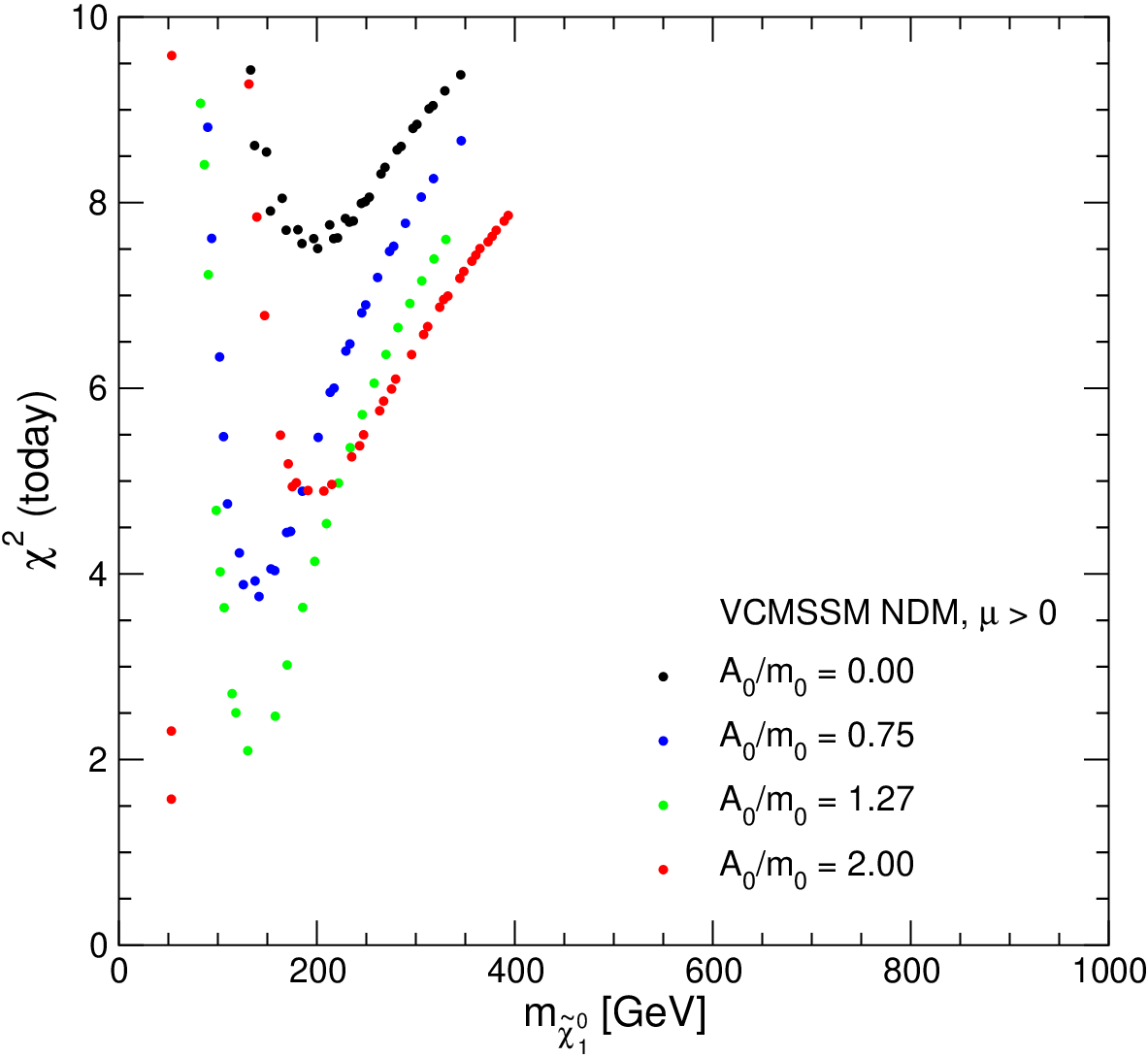}\\[3em]
\includegraphics[width=.45\textwidth,height=5.4cm]{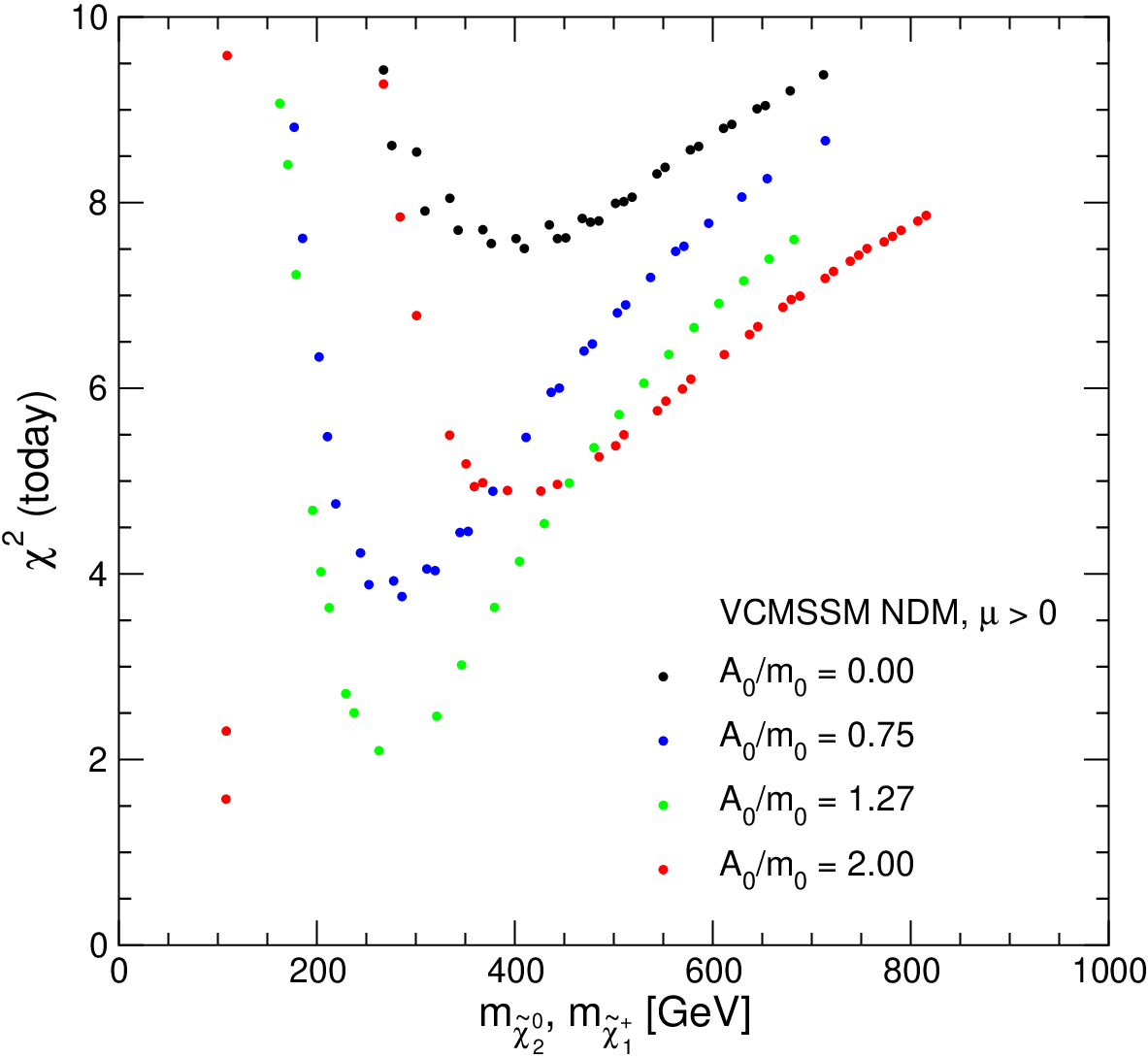}
\includegraphics[width=.45\textwidth,height=5.4cm]{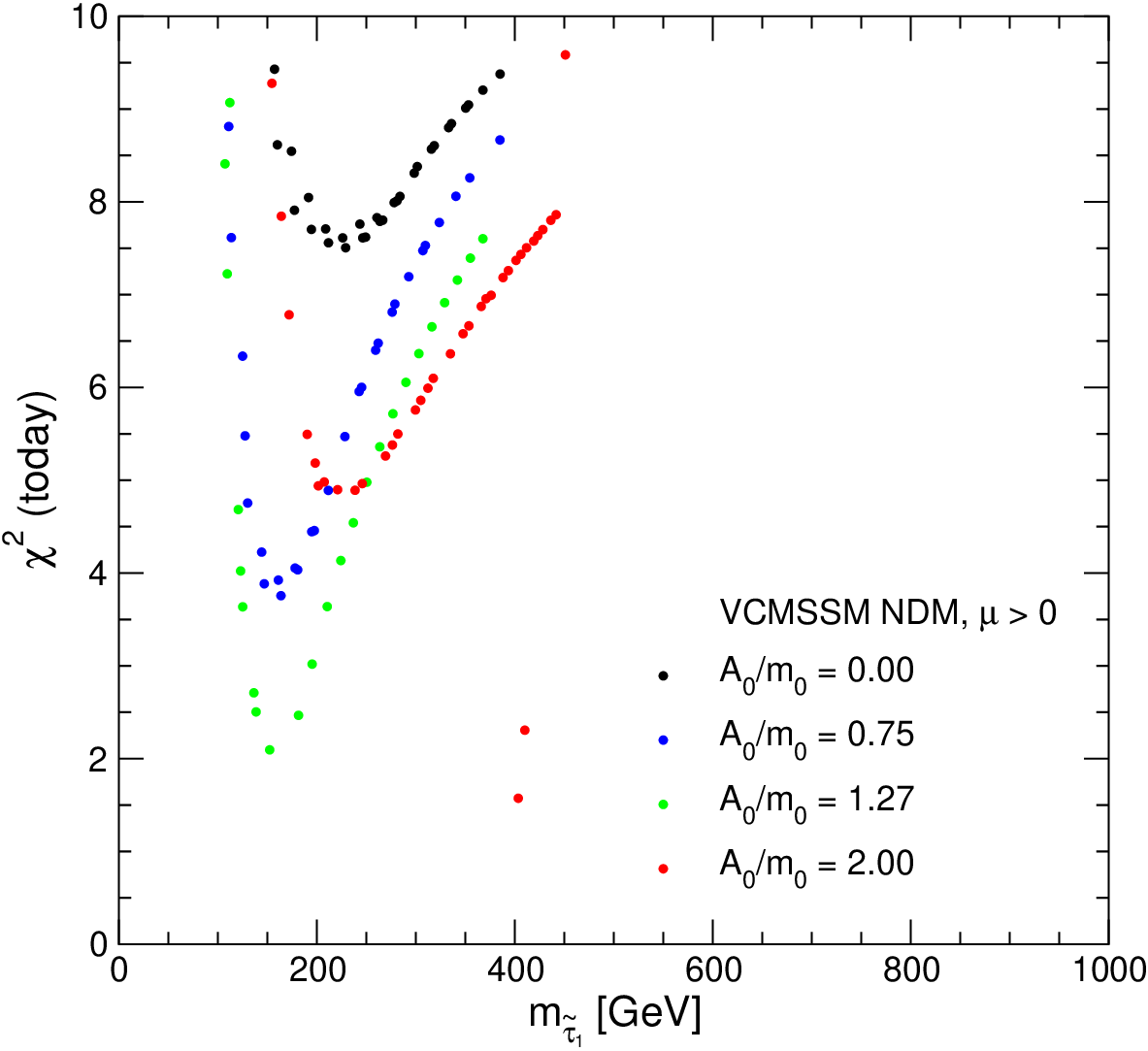}\\[3em]
\includegraphics[width=.45\textwidth,height=5.4cm]{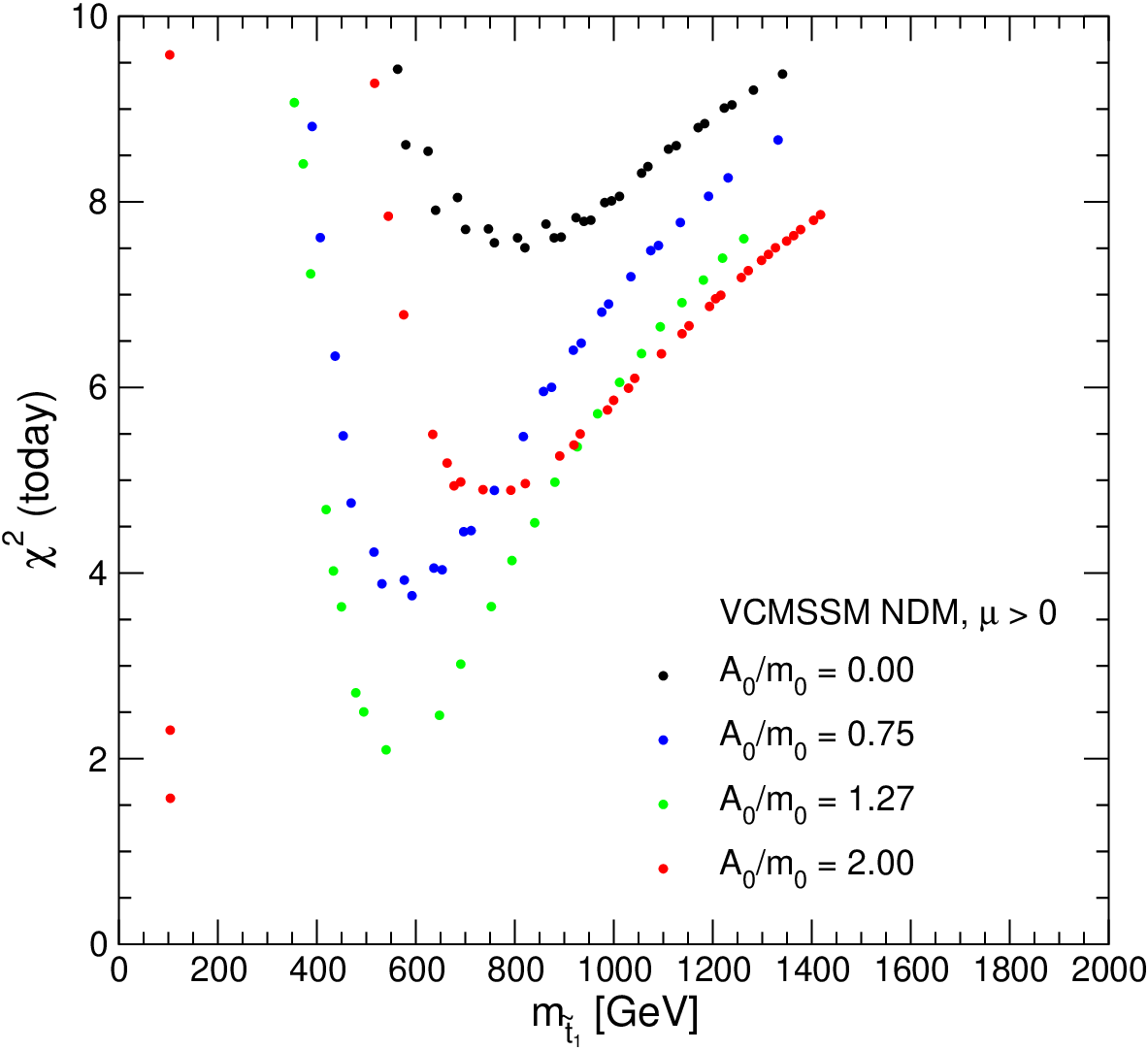}
\includegraphics[width=.45\textwidth,height=5.4cm]{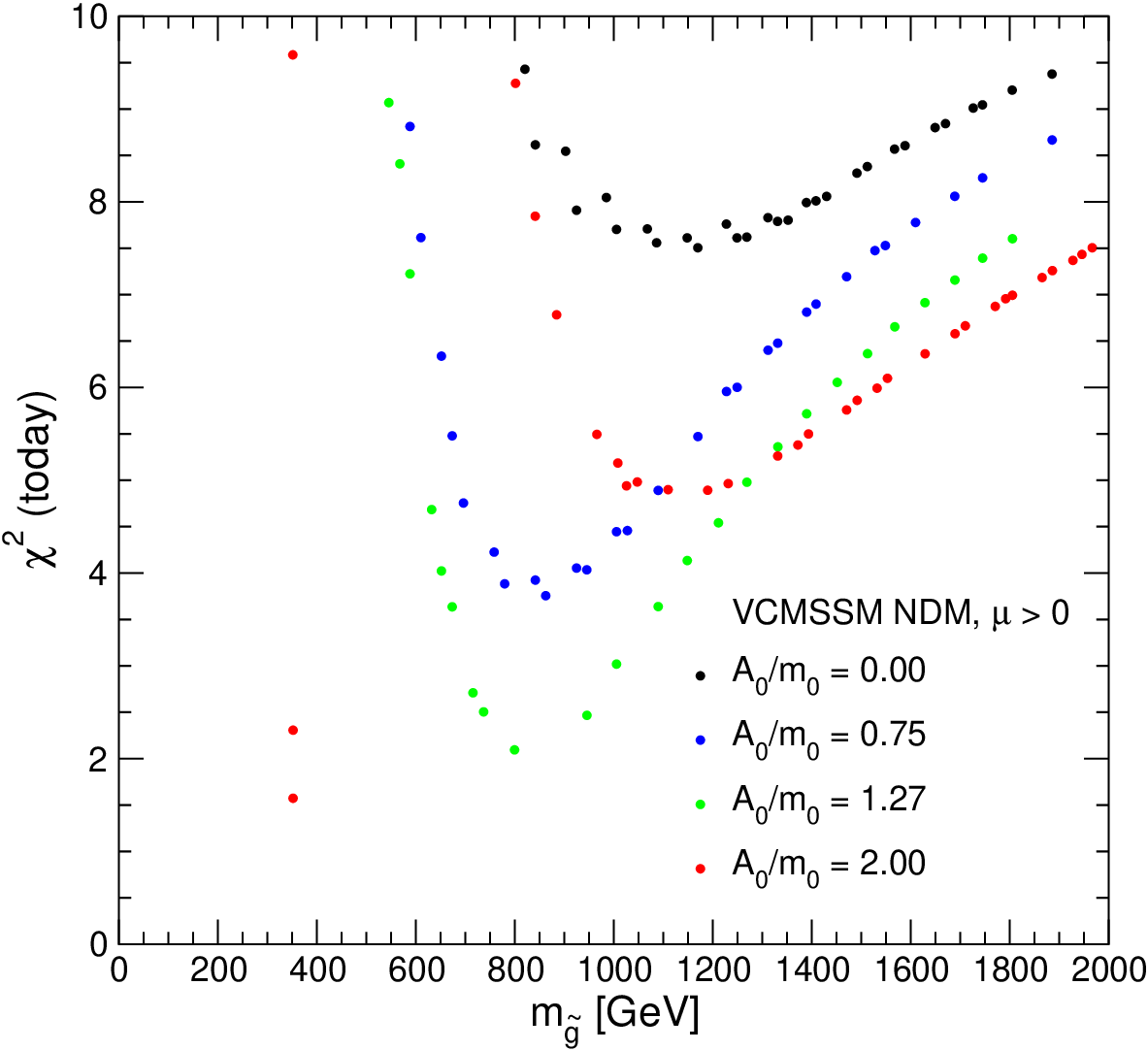}
\begin{picture}(0,0)
\CBox(-095,470)(-015,400){White}{White}
\CBox(-310,270)(-230,210){White}{White}
\CBox(-095,270)(-015,210){White}{White}
\CBox(-310,070)(-230,022){White}{White}
\CBox(-090,070)(-015,022){White}{White}
\end{picture}
\caption{
\it The combined likelihood $\chi^2$ along the WMAP strips for
NDM scenarios with $A_0/m_0 = 0, 0.75, 3 - \sqrt{3}$ and $2$,
as a function (a) of $m_{1/2}$, (b) of $\mneu{1}$, (c) of 
$\mneu{2}$ and $\mcha{1}$, (d) of $\mstaue$, (e) of 
$\mste$ and (f) $\mgl$.}
\label{fig:VCMSSMNDM}
\end{center} 
\end{figure}

Since in the CMSSM the value of $\chi^2$ tends first to decrease and then
to increase with $m_{1/2}$, but does not vary strongly with $\tb$, we
would expect the $\chi^2$ function to exhibit a similar dependence on
$m_{1/2}$ also in the VCMSSM scenario.  This effect is indeed seen in the
first panel of \reffi{fig:VCMSSMNDM}: there are well-defined local
minima at $m_{1/2} \sim 400$ to 600~GeV, as $A_0/m_0$ varies from 0 to 2.  
However, for the latter value of $A_0 / m_0$, we notice some isolated
(red) points with $m_{1/2} \sim 140 \gev$ and much lower $\chi^2 \sim 2$.%
\footnote{
Similar points appear in the CMSSM, but at values of
$A_0/m_{1/2}$ much larger than those considered in~\cite{ehow3} and
here.
}%
~At these points, which barely survive the LEP chargino limit,
rapid annihilation through a direct-channel light-Higgs pole brings the
neutralino relic density down into the WMAP
range~\cite{Manuel}. The remaining panels of \reffi{fig:VCMSSMNDM}
display the 
$\chi^2$ functions for the masses of the $\neu{1}, \Staue,
\neu{2}, \cha{1}, \Stope$ and $\gl$. Their qualitative
features are similar to those shown earlier for the CMSSM, with the
addition of the exceptional low-mass rapid-annihilation points. In these
VCMSSM NDM scenarios, the LHC has good prospects for the $\gl$ and
$\Stope$ and the ILC(500) has good prospects for the $\neu{1}$ and
$\Staue$, whereas the prospects for the $\neu{2}$ and $\cha{1}$
would be dimmer, except at the isolated rapid-annihilation points. 
The ILC(1000), on the other hand, would have a good chance to detect
the $\neu{2}$ and the $\cha{1}$, depending somewhat on $A_0/m_0$.
These points might also be accessible to the Tevatron, in particular via 
searches for gluinos.

We find no analogous focus-point regions in the VCMSSM.
When $A_0/m_0$ is large,  the RGE evolution of $\mu$
does not reduce it, even when $m_0$ is very large~%
\footnote{
This is true also in the CMSSM.
}%
. For smaller $A_0/m_0$, the value of
$\tan \beta$ fixed by the electroweak vacuum conditions in the VCMSSM becomes
small: $\tan \beta < 5$ when $m_0$ is large.  In this case, as in the CMSSM,
the focus-point region is not reached.

\begin{figure}[htb!]
\begin{center}
\includegraphics[width=.35\textwidth]{ehow4.MW11.cl.eps}\hspace{2em}
\includegraphics[width=.35\textwidth]{ehow4.SW11.cl.eps}\\[1em]
\includegraphics[width=.35\textwidth]{ehow4.BSG11.cl.eps}\hspace{2em}
\includegraphics[width=.35\textwidth]{ehow4.AMU11.cl.eps}\\[1em]
\includegraphics[width=.35\textwidth]{ehow4.Mh11.cl.eps}
\includegraphics[width=.35\textwidth]{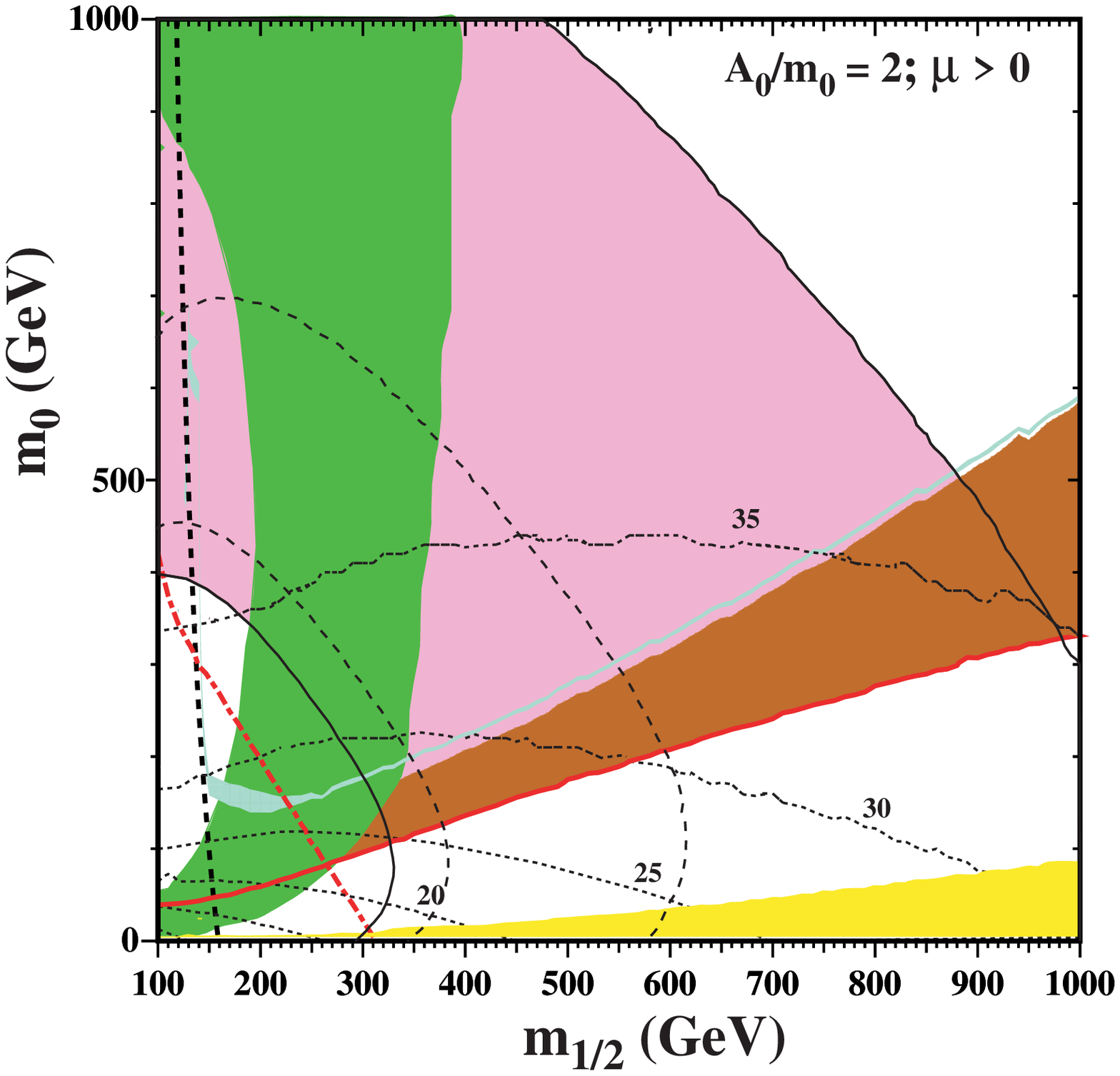}
\begin{picture}(0,0)
\CBox(-085,430)(-000,400){White}{White}
\CBox(-270,235)(-190,210){White}{White}
\CBox(-085,310)(-000,285){White}{White}
\CBox(-270,070)(-180,040){White}{White}
\end{picture}
\caption{\it
Results as functions of $m_{1/2}$ 
along the VCMSSM WMAP strip
for neutralino dark matter when $A_0/m_0 = 2$, for (a) $\MW$, (b) $\sweff$, (c)
${\rm BR}(b \to s \ga)$, (d) $(g - 2)_\mu$ and (e) $\Mh$.
Note the NDM points at low $m_{1/2} \sim 140 \gev$ that simultaneously
fit very well ${\rm BR}(b \to s \ga)$ and $(g - 2)_\mu$. In panel (f) we 
display phenomenological constraints in the $(m_{1/2}, m_0)$ plane for the 
VCMSSM with $A_0/m_0 = 2$, including both the NDM WMAP strip (blue) and 
the GDM wedge (yellow). The regions {\it disfavoured} by $b \to s
\gamma$ and {\it favoured} by $(g - 2)_\mu$ are shaded green and pink
(darker and lighter grey), respectively, the LEP Higgs constraint is a
near-vertical (red) dashed line, and the (blue) dotted lines are contours of 
$\tb$, as fixed by the VCMSSM vacuum conditions.} 
\label{fig:VCMSSMobs}
\end{center}
\vspace{-3em}
\end{figure}

In order to understand better the variation of $\chi^2$ with $m_{1/2}$ in
\reffi{fig:VCMSSMNDM}, and in particular to understand its relatively low
value at the rapid light-Higgs annihilation points with $m_{1/2} \sim
140 \gev$~\cite{Manuel}, we display separately in \reffi{fig:VCMSSMobs} the
dependences of (a) $\MW$, (b) $\sweff$, (c) ${\rm BR}(b \to s \ga)$, (d)
$(g - 2)_\mu$ and (e) $\Mh$ on $m_{1/2}$ for the case $A_0/m_0 = 2$~%
\footnote{
The values of $\tb$ in the VCMSSM are too small for $B_s \to \mu^+
\mu^-$ currently to make any significant contribution to the $\chi^2$
function~\cite{ournewBmumu}.
}%
. Along the VCMSSM WMAP strip, we 
see that $\MW$ prefers a very low value of $m_{1/2}$, with the
rapid-annihilation points slightly disfavoured, whereas $\sweff$ prefers a
range of somewhat larger values of $m_{1/2}$, with the rapid-annihilation
points slightly favoured. However, we then see that both ${\rm BR}(b \to s
\ga)$ and $(g - 2)_\mu$ independently strongly disfavour $m_{1/2} \sim
200 \gev$, whereas the rapid-annihilation points fit these measurements
very well. The same tendency is observed for $\Mh$. 


These behaviours can be understood by referring to panel (f) of
\reffi{fig:VCMSSMobs}, where the regions {\it disfavoured} by $b \to s
\gamma$ and {\it favoured} by $(g - 2)_\mu$ are shaded green and pink
(darker and lighter grey), respectively. The shaded  $(g - 2)_\mu$
region represents a 2-$\sigma$ deviatation based on (\ref{delamu}), while
the dashed lines represent the region favoured at the 1-$\sigma$ level.  
The LEP Higgs constraint is a
diagonal (red) dot-dashed line, while the near-vertical black dashed
line shows the LEP constraint on the chargino mass. The pale (blue)
shaded strip is favoured by WMAP for NDM. Below this strip, there is a
red shaded region in which the  LSP is the $\Staue$ and therefore 
excluded.  Below the $\Staue$ LSP region, the gravitino is the
LSP~\cite{GDM}. In the unshaded portion of the GDM region, the
next-to-lightest supersymmetric particle  (NLSP) will decay
into a gravitino with unacceptable effects on the abundances of the
light elements and is excluded by BBN \cite{CEFO,GDM,eov,others}. 
The pale (yellow)  shaded wedge is favoured for
gravitino dark matter as this region is allowed by BBN constraints. 
Finally, the black dotted curves labeled 20, 25, 30 and 35 correspond to the
values of $\tb$ required by the VCMSSM vacuum conditions.
We see that the rapid-annihilation tail of the WMAP
strip rises at low $m_{1/2}$ into a region allowed by $b \to s \gamma$,
favoured by $(g - 2)_\mu$ and tolerated by $\Mh$. It is the synchronized
non-monotonic behaviour of these last three observables that explains the
similar non-monotonic behaviour of $\chi^2$ along the NDM WMAP strip in
\reffi{fig:VCMSSMNDM} and the low value of $\chi^2$ for the isolated
rapid-annihilation point at $m_{1/2} \sim 140 \gev$~\cite{Manuel}. This is in 
fact the best overall fit point in this VCMSSM scenario, as seen in
\reffi{fig:VCMSSMNDM}.

\begin{figure}[htb!]
\vspace{1em}
\begin{center}
\includegraphics[width=.48\textwidth]{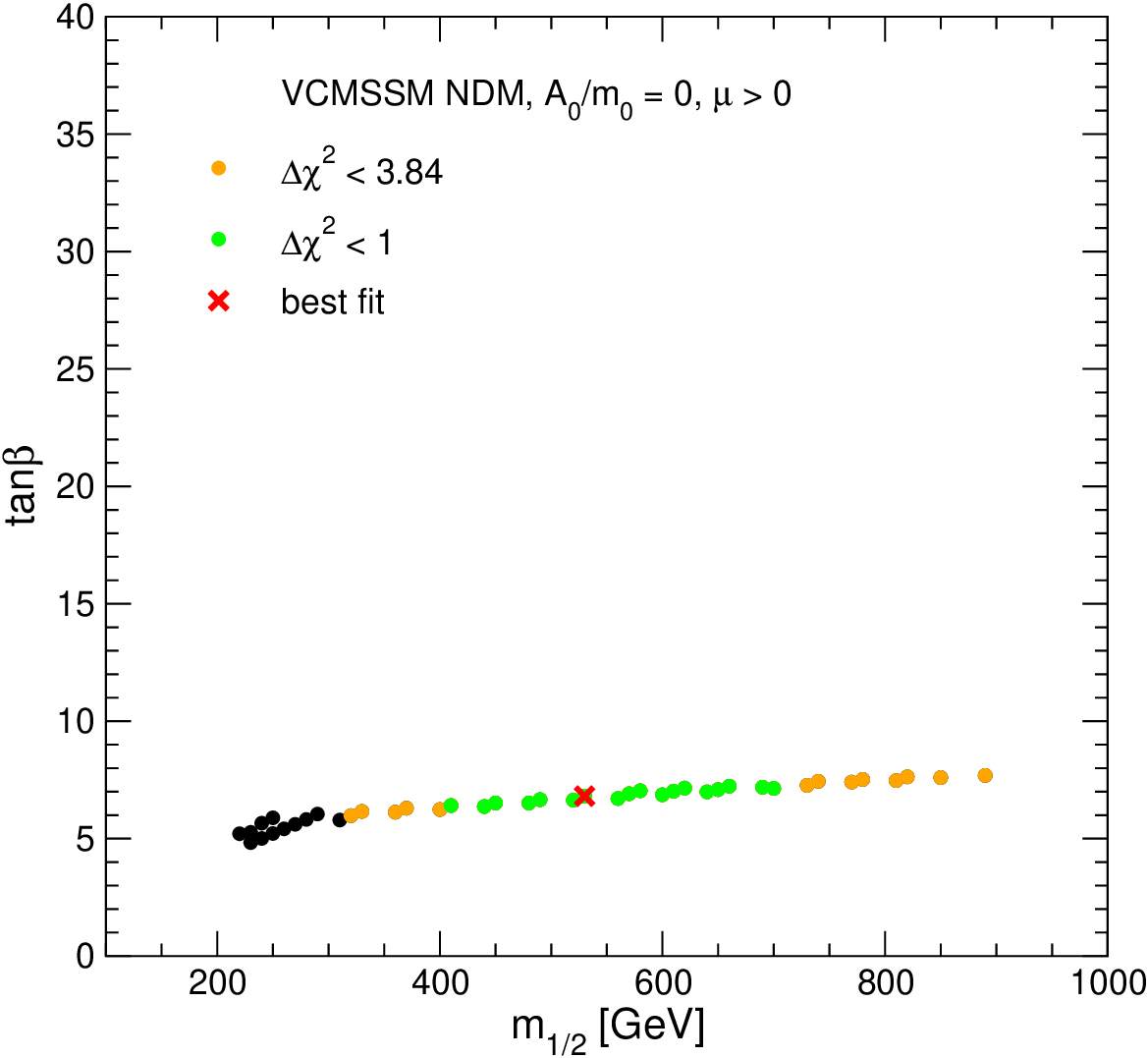}
\includegraphics[width=.48\textwidth]{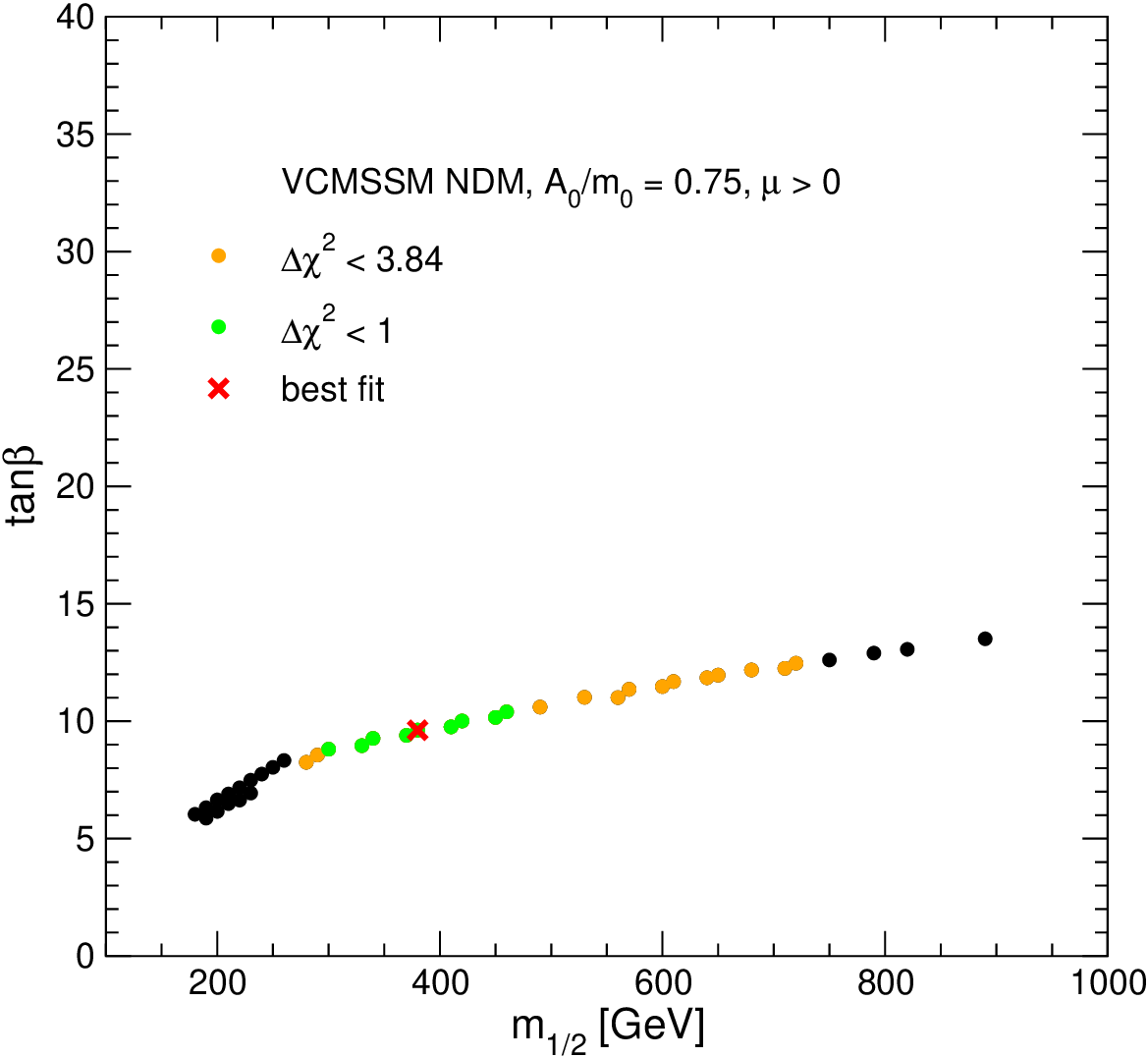}\\[3em]
\includegraphics[width=.48\textwidth]{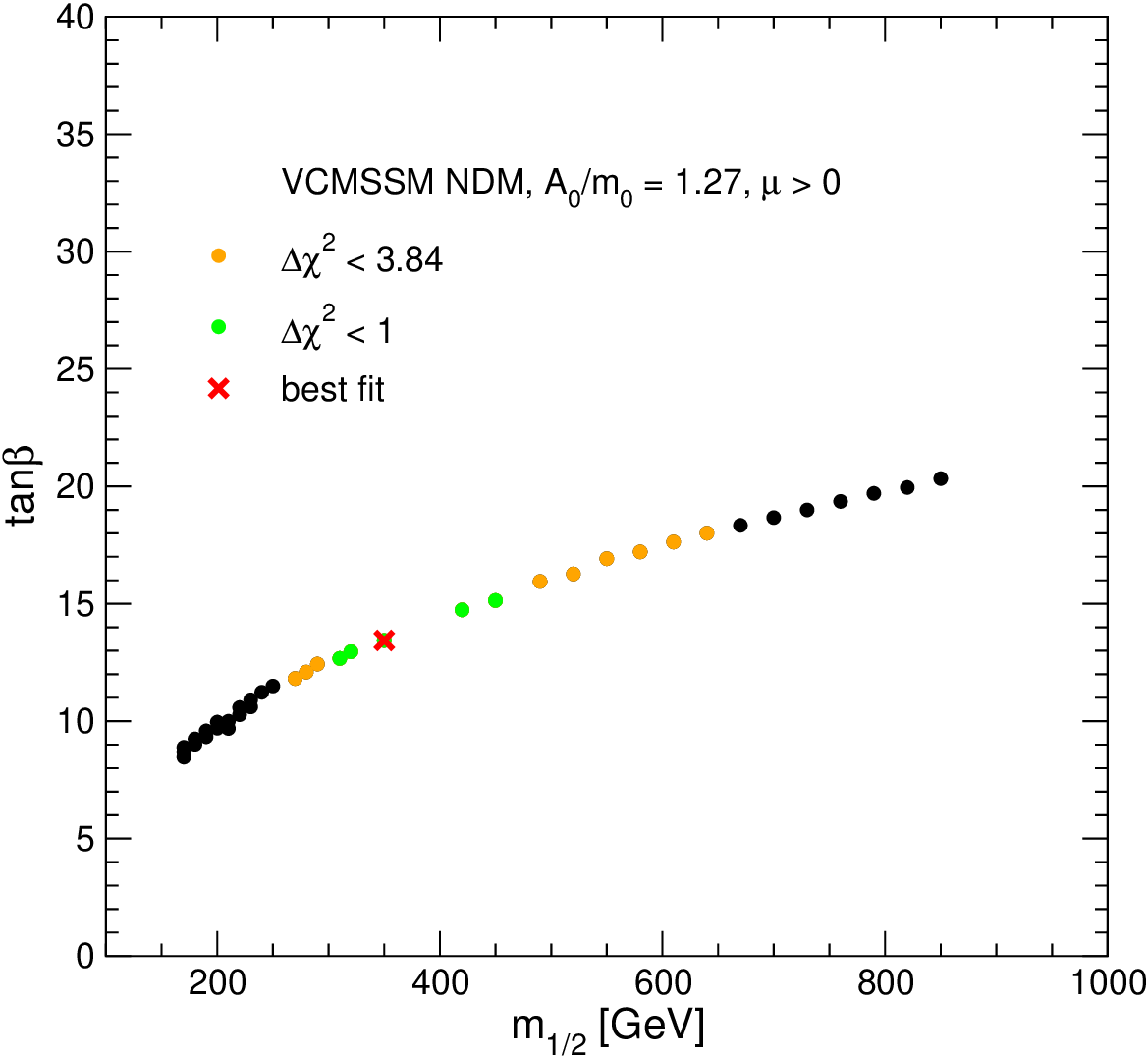}
\includegraphics[width=.48\textwidth]{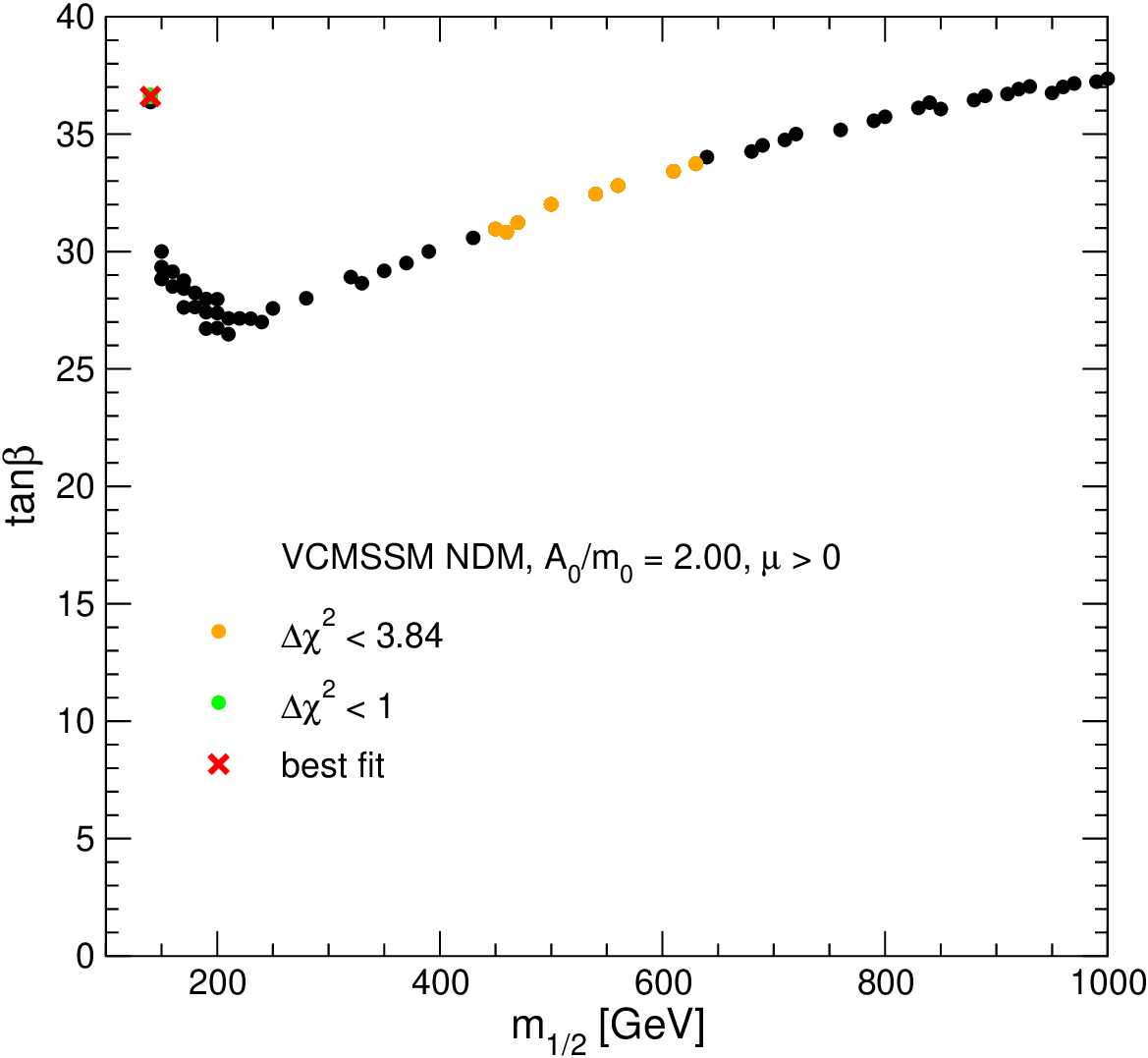}
\vspace{1em}
\begin{picture}(0,0)
\CBox(-190,410)(-130,350){White}{White}
\CBox(-420,165)(-350,120){White}{White}
\CBox(-190,090)(-130,050){White}{White}
\end{picture}
\caption{\it
Illustration of the preferred regions in the space of VCMSSM
models for (a) $A_0 = 0$, (b) $A_0/m_0 = 0.75$,
(c) $A_0/m_0 = 3 - \sqrt{3}$
and (d) $A_0/m_0 = 2$. In each case, the red points show the $\chi^2$
minimum, the green points have $\De \chi^2 < 1$, the orange points
have $\De\chi^2 < 3.84$, and the black points have larger $\chi^2$.
}
\label{fig:NDM}
\end{center}
\vspace{2em}
\end{figure}

The preferred ranges of $m_{1/2}$ seen in \reffi{fig:VCMSSMNDM}
correspond, through the VCMSSM vacuum conditions, to preferred ranges in
$\tb$. As seen in \reffi{fig:NDM}, these increase with the chosen value
of $A_0/m_0$, as does the correlation with $m_{1/2}$. For $A_0/m_0 = 0$
(top left panel), $\tb \sim 7$, increasing to $\tb \sim 10, 15, 32$ for
$A_0 / m_0 = 0.75, 3 - \sqrt{3}, 3$, respectively. In the last case,
descending the VCMSSM WMAP strip to lower $m_{1/2}$, whereas we see that
$\chi^2$ exceeds 10 for $m_{1/2} < 350 \gev$, we see again the isolated
dark (red) rapid-annihilation points with $m_{1/2} \sim
140 \gev$~\cite{Manuel}, which have relatively large $\tb \sim 37$.


We conclude that the extra constraint imposed in the VCMSSM modifies but
does not remove the preference found within the CMSSM for small $m_{1/2}$.
Within the VCMSSM with neutralino dark matter, the minimum of $\chi^2$
usually occurs along the generic WMAP coannihilation strip at $m_{1/2}
\sim 500 \gev$. However, when $A_0/m_0 = 2$, we find lower values of
$\chi^2$ in the rapid light-Higgs annihilation region with $m_{1/2} \sim
140 \gev$. The preferred value of $\tb$ varies between $\sim 7$ and $\sim
32$ on the generic WMAP strip, depending on the value of $A_0 / m_0$, but
$\tb \sim 37$ in the light Higgs-pole annihilation region for $A_0 / m_0 =
2$. These points offer prospects for a gluino discovery at the Tevatron:
all the other preferred parameter sets offer good prospects for observing
sparticles at the LHC and ILC(500).


\section{GDM Analysis}

The relation $A_0 = B_0 + m_0$ is just one of the further conditions on
supersymmetry-breaking parameters that would be imposed in minimal
supergravity (mSUGRA) models. The other is the equality between $m_0$ and
the gravitino mass. So far, we have implicitly assumed that the gravitino
is sufficiently heavy that the LSP is always the lightest neutralino
$\neu{1}$ and the 
cosmological constraints on gravitino decays are unimportant. However,
this is not always the case in mSUGRA models. Indeed, in generic mSUGRA
scenarios, as seen in the bottom right panel of \reffi{fig:VCMSSMobs},
in addition to a WMAP strip where the $\neu{1}$ is the LSP as we have
assumed so far, there is a wedge of parameter space at lower values of
$m_0$ (for given choices of $m_{1/2}$ and the other parameters), where the
gravitino is the LSP. In this case, there are important astrophysical and
cosmological constraints on the decays of the long-lived NLSP 
\cite{CEFO,eov,others}, which is generally
the lighter stau $\Staue$ in such mSUGRA
scenarios~%
\footnote{
There are also non-mSUGRA scenarios in which the NLSP
is the $\neu{1}$. Such models are subject to similar astrophysical and
cosmological constraints, but we do not consider them here.
}%
.

\begin{figure}[htb!]
\begin{center}
\includegraphics[width=.45\textwidth,height=5.4cm]{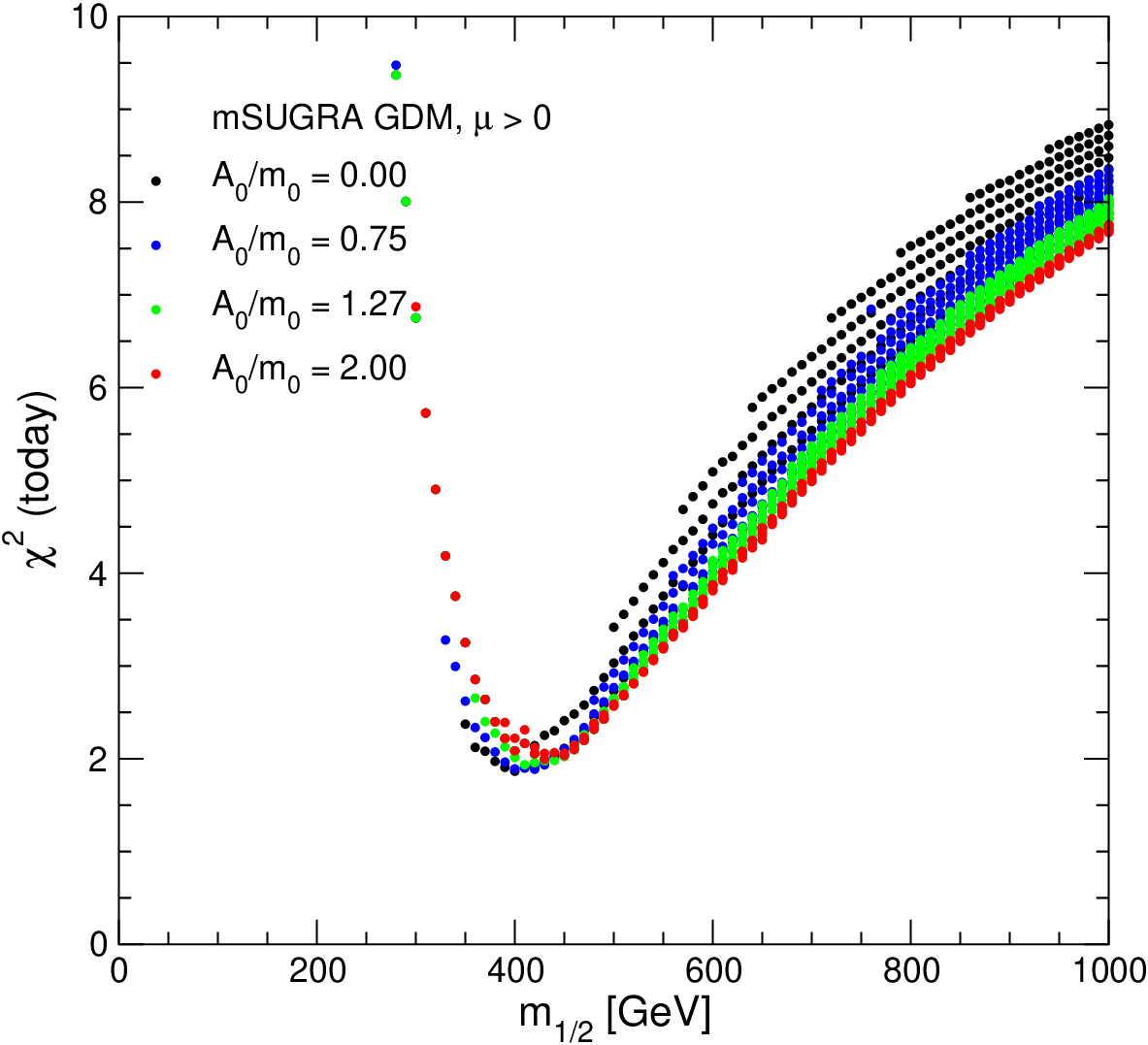}
\includegraphics[width=.45\textwidth,height=5.4cm]{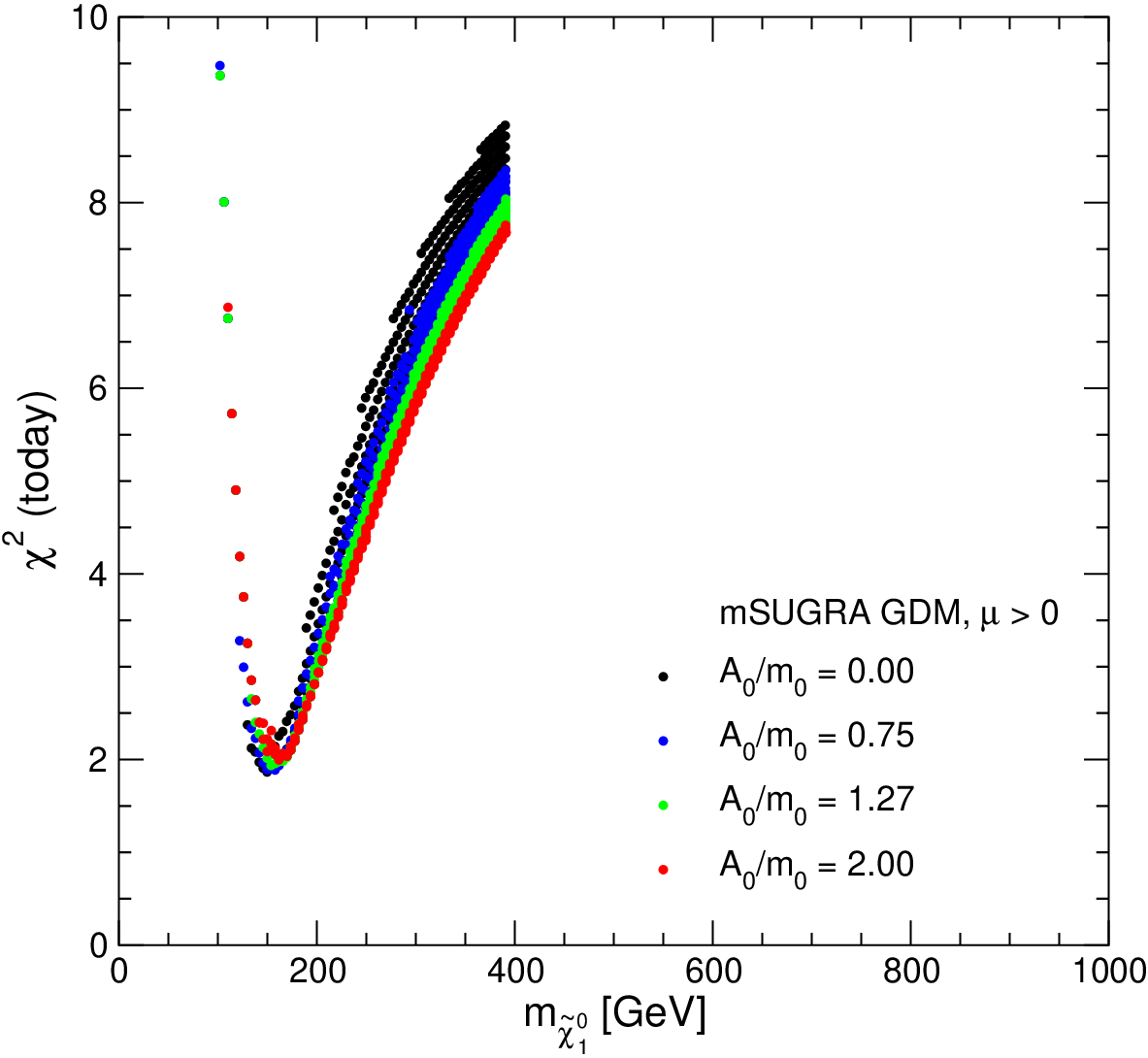}\\[3em]
\includegraphics[width=.45\textwidth,height=5.4cm]{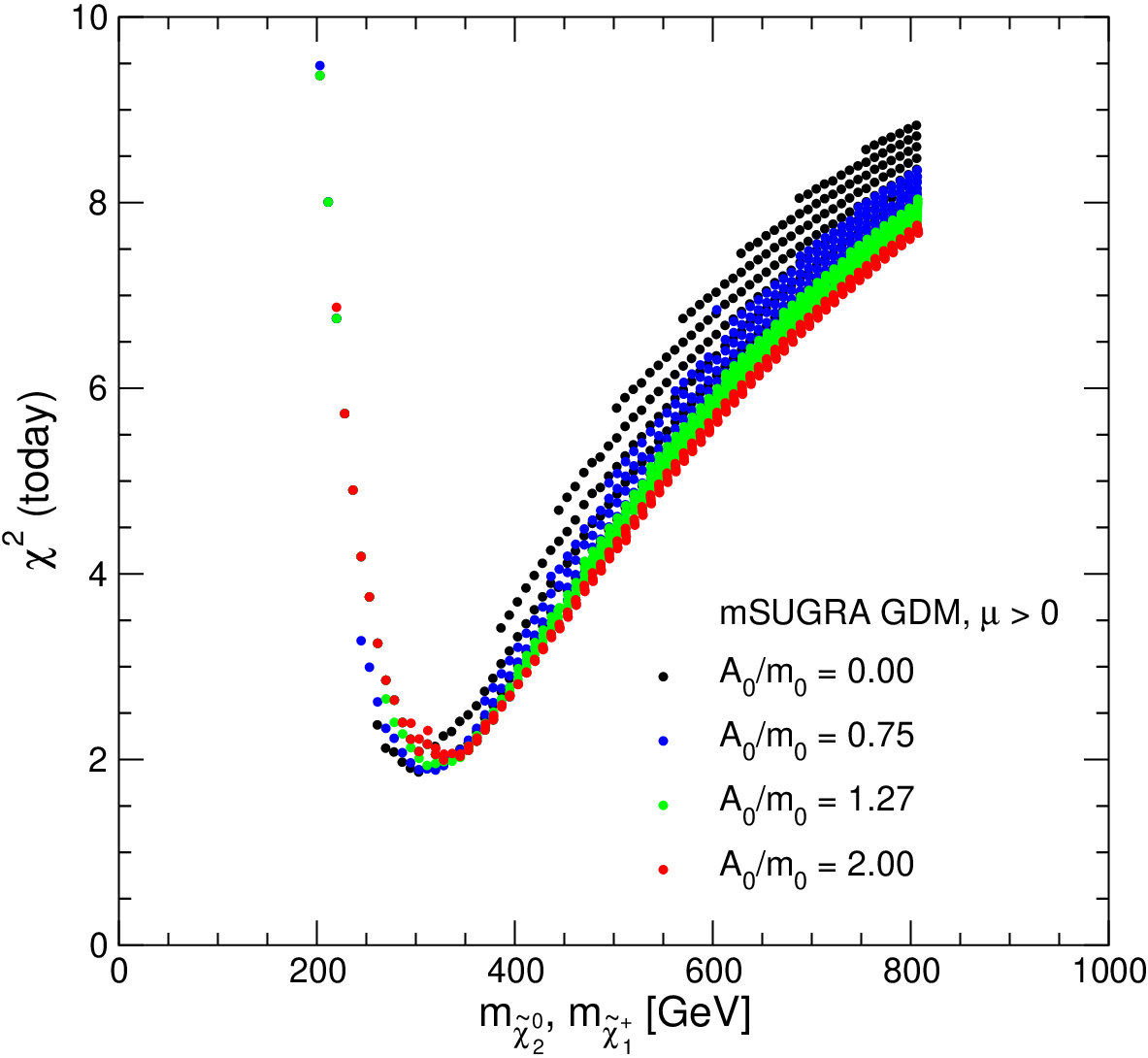}
\includegraphics[width=.45\textwidth,height=5.4cm]{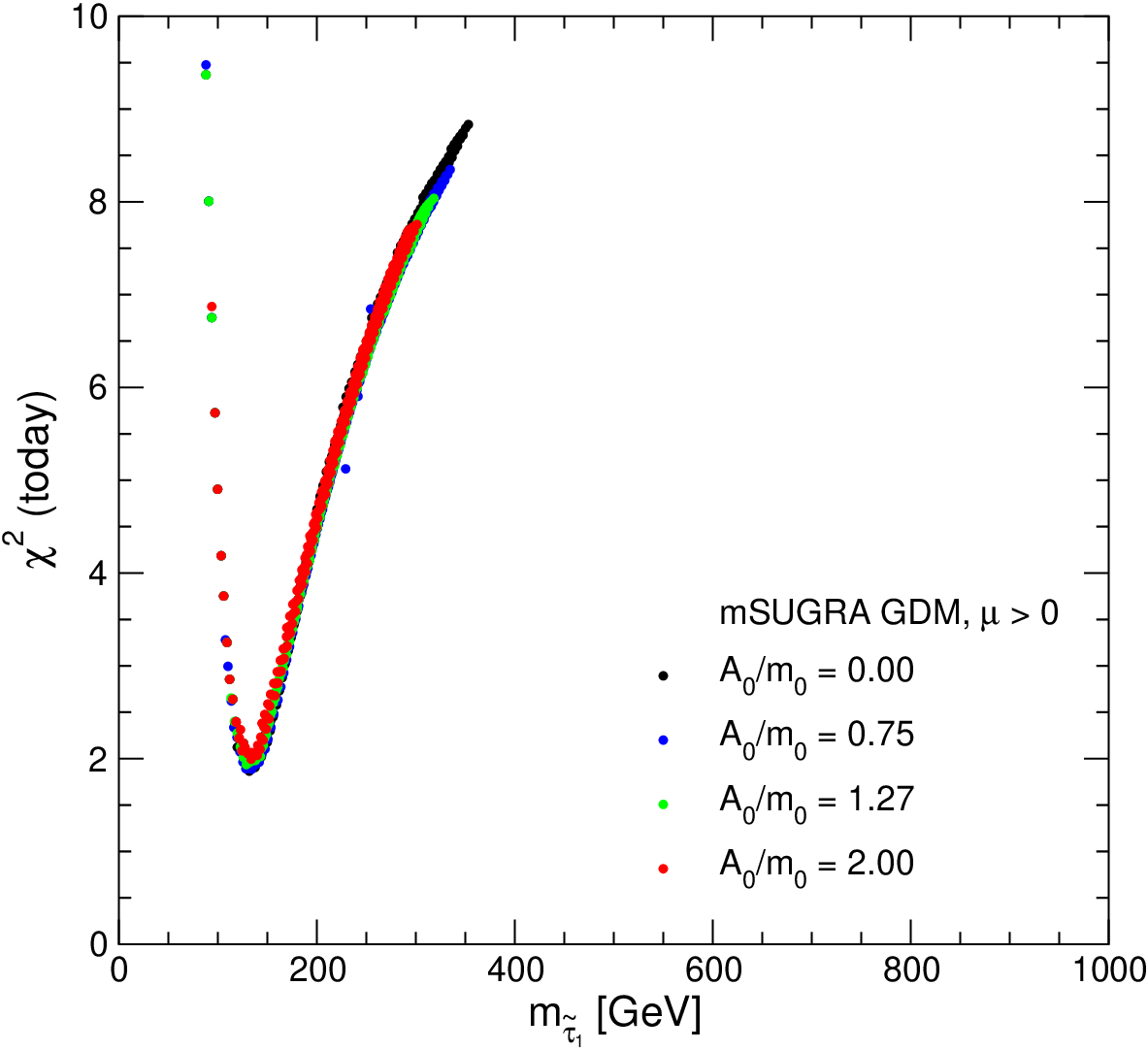}\\[3em]
\includegraphics[width=.45\textwidth,height=5.4cm]{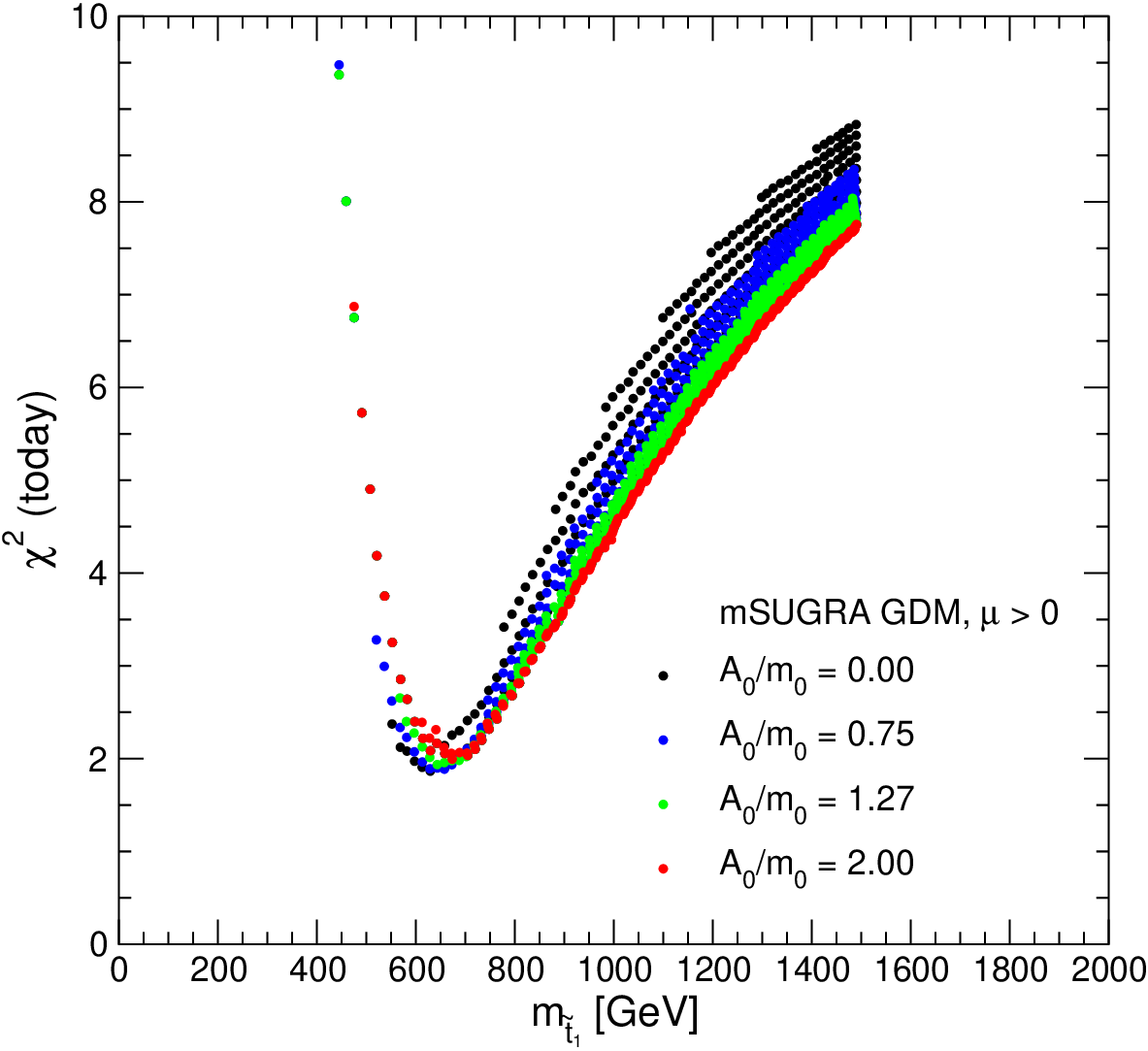}
\includegraphics[width=.45\textwidth,height=5.4cm]{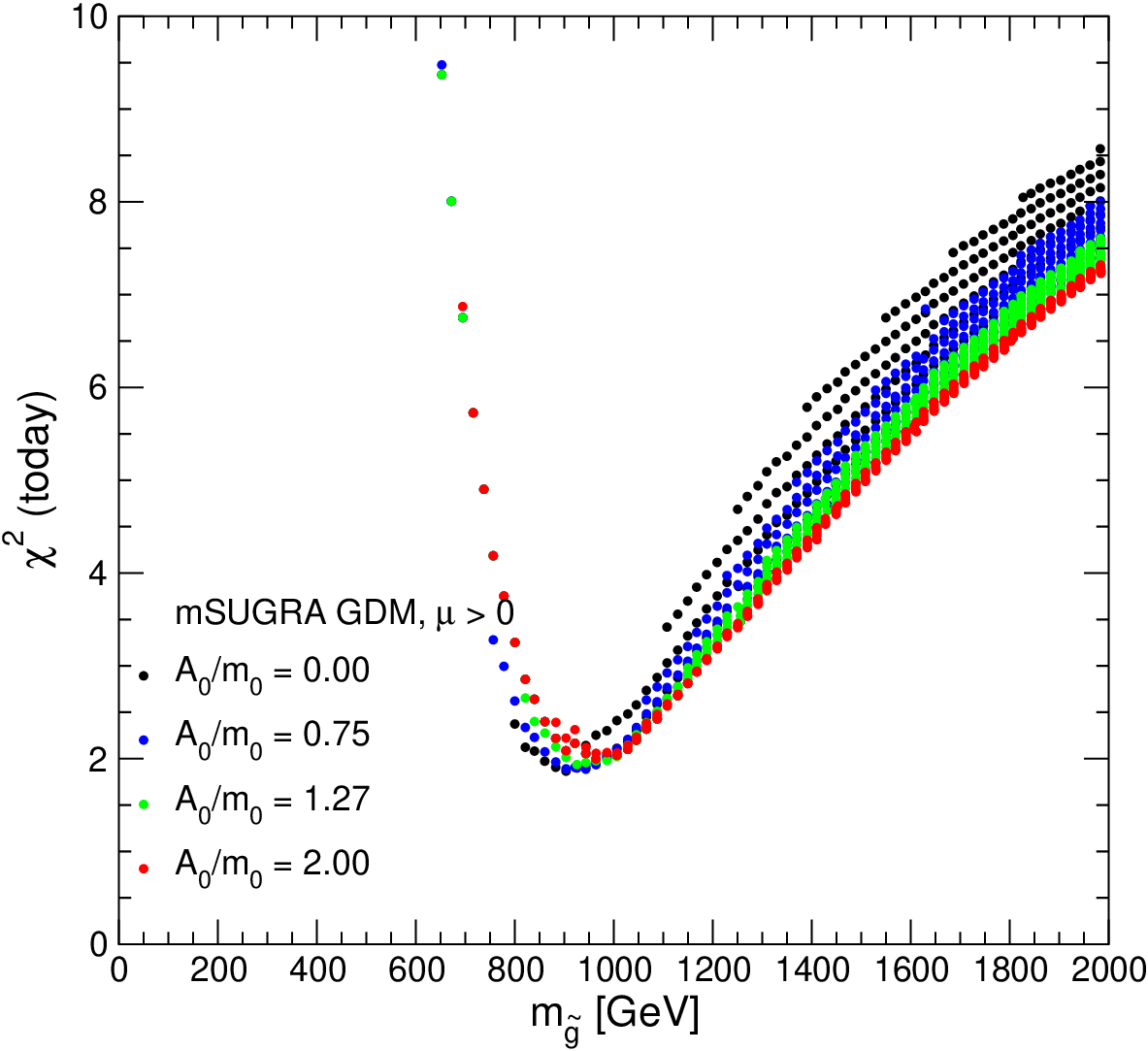}
\begin{picture}(0,0)
\CBox(-095,470)(-015,400){White}{White}
\CBox(-310,260)(-230,210){White}{White}
\CBox(-095,270)(-015,210){White}{White}
\CBox(-310,070)(-230,022){White}{White}
\CBox(-190,060)(-140,022){White}{White}
\end{picture}
\caption{
\it 
The dependence of the $\chi^2$ function on $m_{1/2}$ for
GDM scenarios with $A_0/m_0 = 0, 0.75, 3 - \sqrt{3}$ and $2$,
scanning the regions where the lighter stau $\Staue$ is the
NLSP, shown as a function of (a) $m_{1/2}$, (b) $\mneu{1}$,
(c) $\mneu{2}$ and $\mcha{1}$, (d) $\mstaue$, (e) 
$\mste$, and (f) $\mgl$.
}
\label{fig:VCMSSMGDM}
\end{center}
\vspace{-2em}
\end{figure}

\reffi{fig:VCMSSMGDM} displays the $\chi^2$ function for a sampling of
GDM scenarios obtained by applying the supplementary gravitino mass
condition to VCMSSM models for $A_0/m_0 = 0, 0.75, 3 - \sqrt{3}$ and 2,
and scanning the GDM wedges at low $m_0$. These wedges are scanned via a
series of points at fixed (small) $m_0$ and increasing $m_{1/2}$.  We note
that there is a marginal tendency for $\chi^2$ to increase with increasing
$m_0$, though this is not as marked as the tendency to increase with
$m_{1/2}$, and that the scan lines are more widely separated for the
smaller values of $A_0$. Comparing Figs.~\ref{fig:VCMSSMNDM}
and \ref{fig:VCMSSMGDM}, we see that lower $\chi^2$ values may be attained
in the GDM cases. The third panel of \reffi{fig:VCMSSMobs} and last
panel of \reffi{fig:NDM} 
illustrate how this comes about in the case $A_0/m_0 = 2$: there is a
large contribution to $\chi^2$ from $b \to s \ga$ in the NDM for small
$m_{1/2}$ that is absent in the GDM, which strongly prefers the
combination of smaller $m_0$ and smaller $\tb$ found in the GDM
models~%
\footnote{
The values of $\tb$ in these GDM models are also too small for $B_s
\to \mu^+ \mu^-$ currently to make any significant contribution to the
$\chi^2$ function~\cite{ournewBmumu}.
}%
.


\begin{figure}[htb!]
\vspace{1em}
\begin{center}
\includegraphics[width=.48\textwidth]{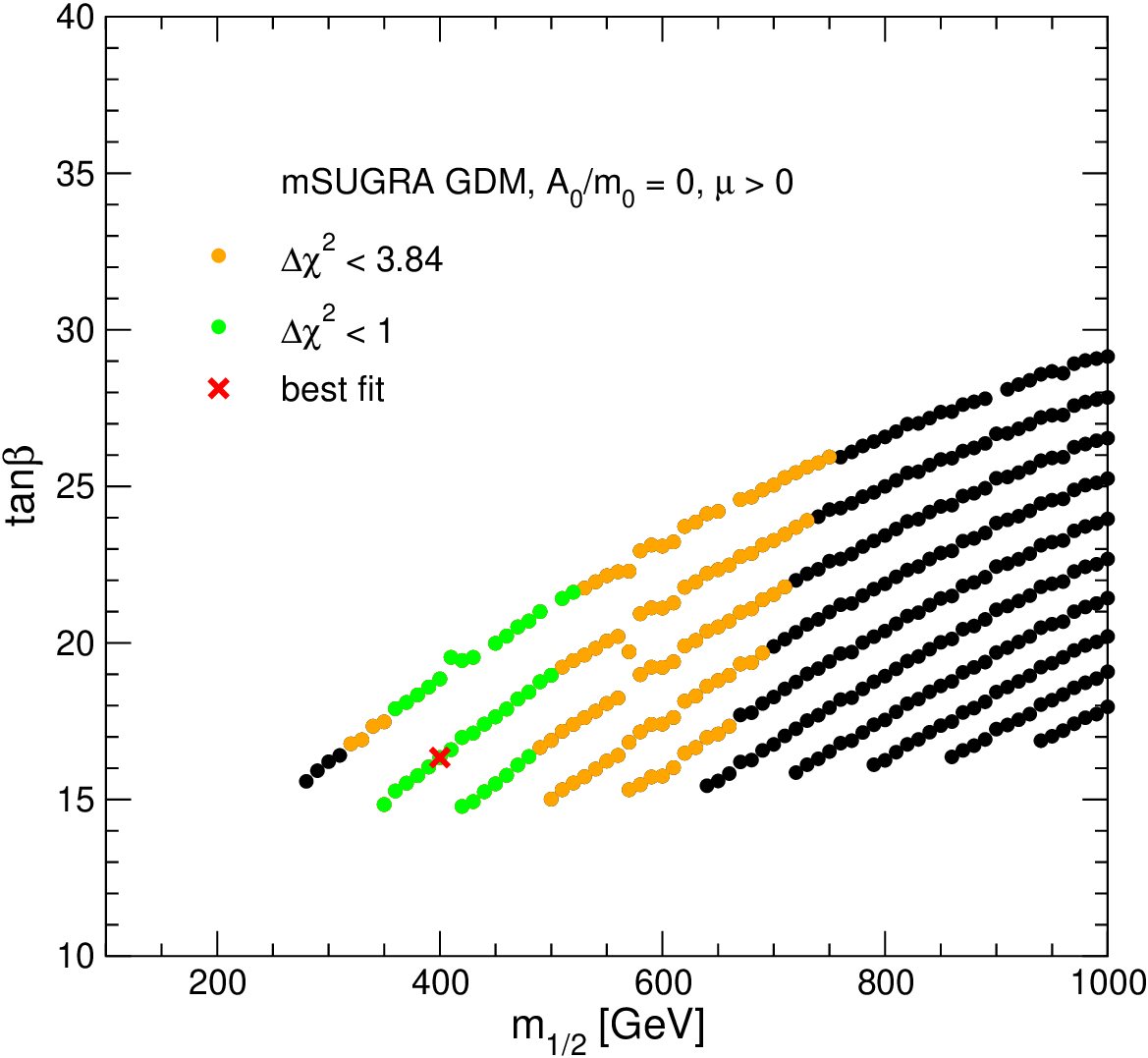}
\includegraphics[width=.48\textwidth]{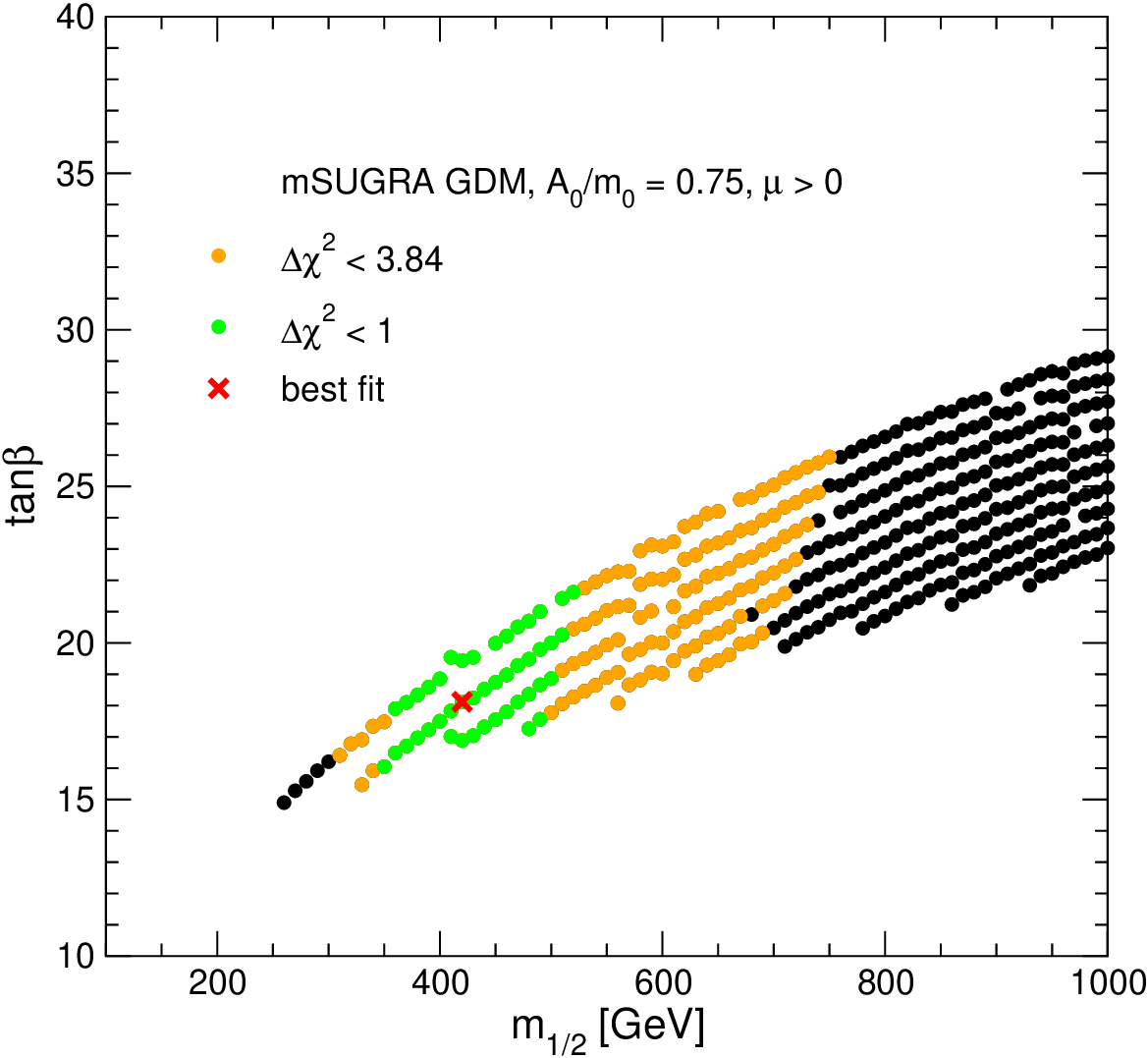}\\[3em]
\includegraphics[width=.48\textwidth]{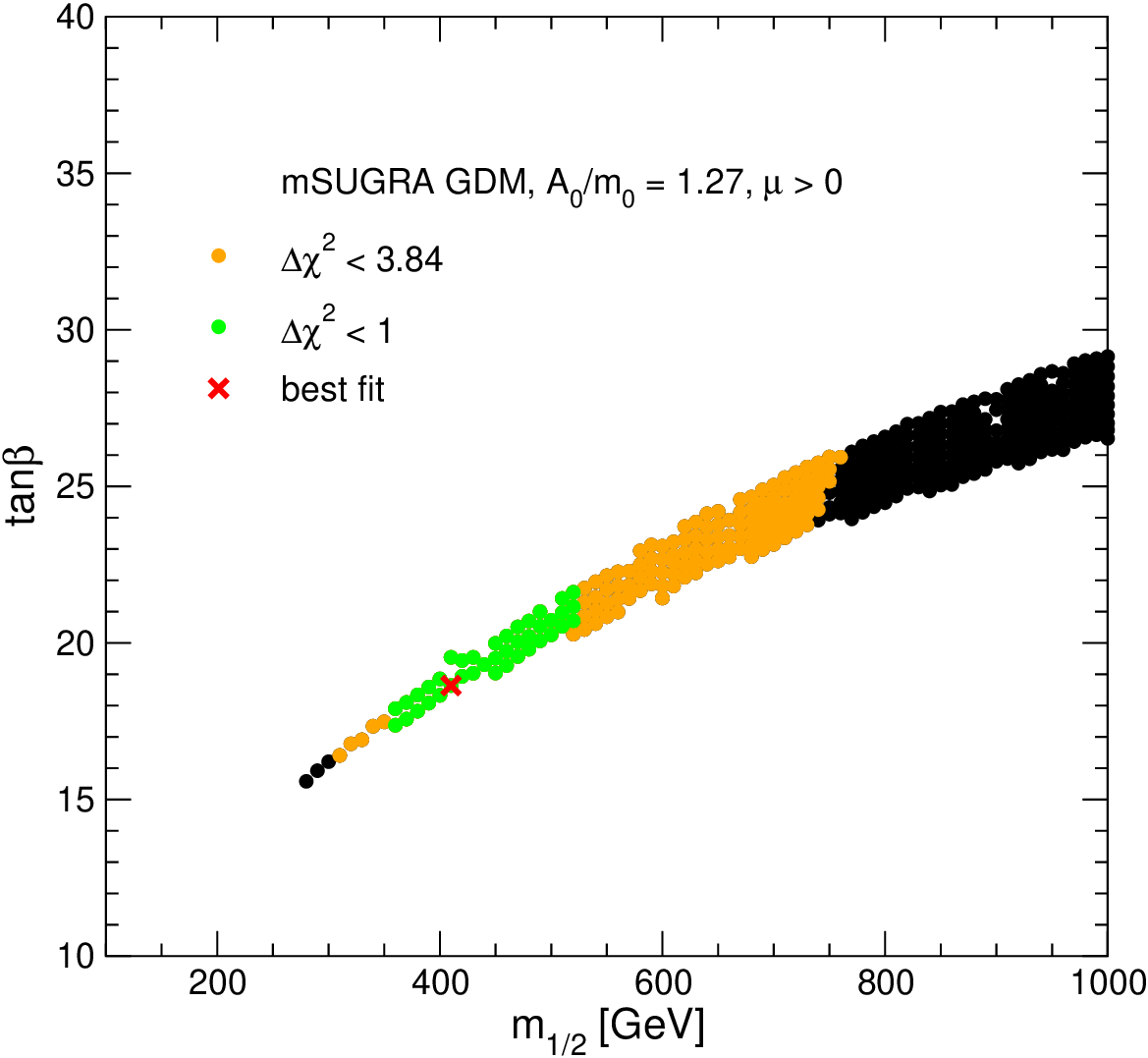}
\includegraphics[width=.48\textwidth]{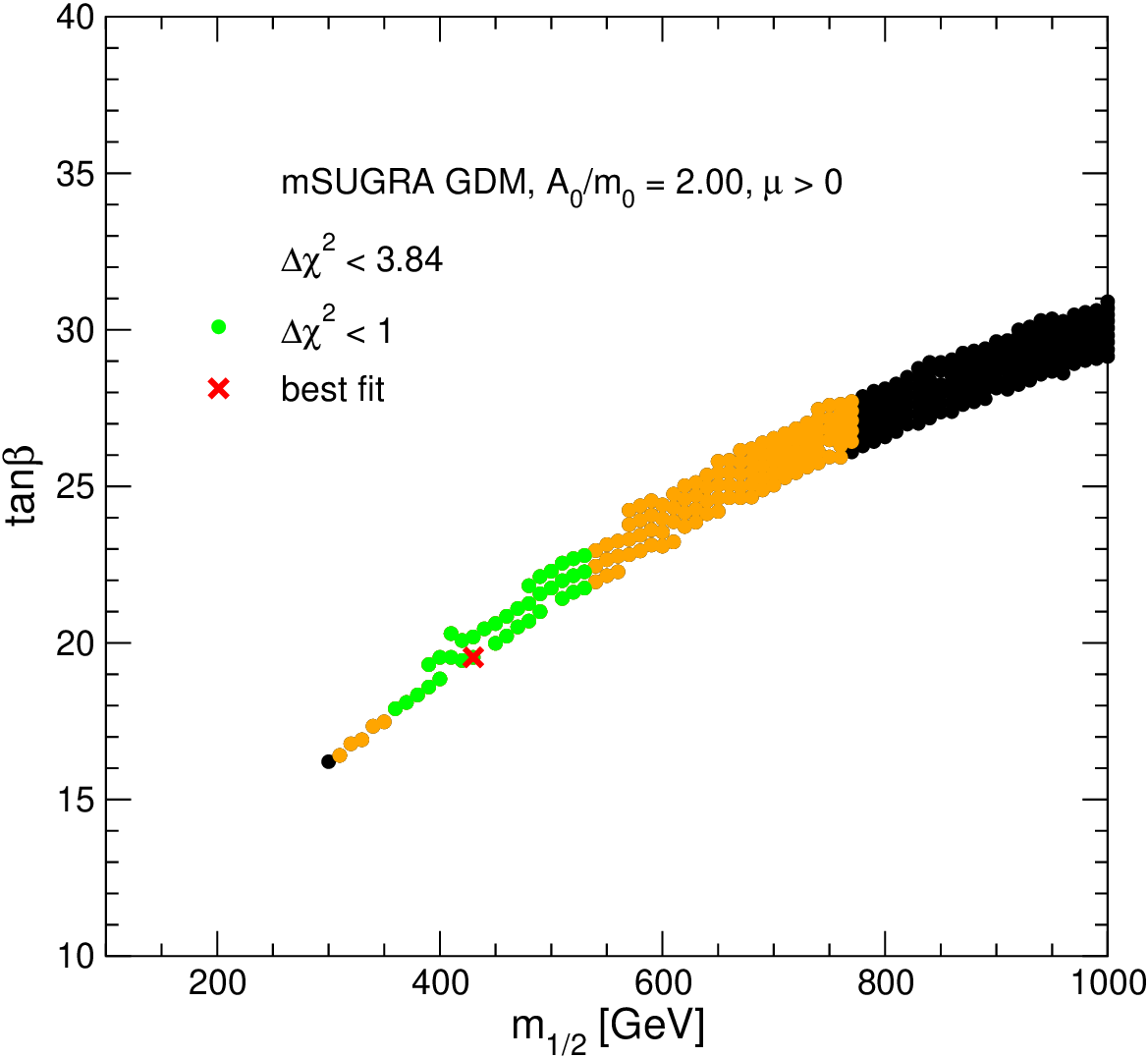}
\vspace{1em}
\begin{picture}(0,0)
\CBox(-190,410)(-130,350){White}{White}
\CBox(-420,165)(-350,120){White}{White}
\CBox(-190,165)(-130,120){White}{White}
\end{picture}
\caption{\it
Illustration of the preferred regions in the space of mSUGRA-motivated
GDM models for (a) $A_0 = 0$, (b) $A_0/m_0 = 0.75$, 
(c) $A_0/m_0 = 3 - \sqrt{3}$ 
and (d) $A_0/m_0 = 2$. In each case, the red points show the $\chi^2$
minimum, the green points have $\De \chi^2 < 1$, the orange points
have $\De\chi^2 < 3.84$, and the black points have larger $\chi^2$. 
} 
\label{fig:GDM}
\end{center}
\vspace{2em}
\end{figure}

As seen in \reffi{fig:VCMSSMGDM}, the global minimum of $\chi^2$ for
all the VCMSSM GDM models with $A_0/m_0 = 0, 0.75, 3 - \sqrt{3}$ and 2 is
at $m_{1/2} \sim 450 \gev$. However, this minimum is not attained for GDM
models with larger $m_0$, as they do not reach the low-$m_{1/2}$ tip of
the GDM wedge seen, for example, in the last panel of \reffi{fig:GDM}.
In general, we see in the different panels of \reffi{fig:VCMSSMGDM}
that, as in the CMSSM, there are good prospects for observing the $\gl$
and perhaps the $\Stope$ at the LHC, and that the ILC(500) has
good prospects for the $\neu{1}$ and $\Staue$, though these
diminish for larger $m_0$. The ILC(1000), again, offers much better
chances also for large $m_0$. We recall that, in these GDM scenarios, the
$\Staue$ is the NLSP, and that the $\neu{1}$ is heavier. The
$\Staue$ decays into the gravitino and a $\tau$, and is
metastable with a lifetime that may be measured in hours, days or weeks.
Specialized detection strategies for the LHC were discussed
in~\cite{Moortgat}: this scenario would offer exciting possibilities near
the $\Staue$ pair-production threshold at the ILC.

As discussed above, a feature of the class of GDM scenarios discussed here
is that the required value of $\tb$ increases with $m_{1/2}$.  Therefore,
the preference for relatively small $m_{1/2}$ discussed above maps into an
analogous preference for moderate $\tb$, as shown in \reffi{fig:GDM}.
The different panels are for the four choices $A_0/m_0 = 0, 0.75, 3 -
\sqrt{3}$ and 2.  In each case, the red point indicates the minimum of the
$\chi^2$ function, the green points have $\De \chi^2 < 1$ corresponding
to the 68 \% confidence level, the orange points have $\De \chi^2 <
3.84$ corresponding to the 95 \% confidence level, and the black points
have larger $\De \chi^2$. We see that, at the 95 \% confidence level
\begin{equation}
300 \gev \lsim m_{1/2} \lsim 800 \gev, \quad 15 \lsim \tb \lsim 27
\label{GDMlimits}
\end{equation}
in this mSUGRA class of GDM models. 

\section{Conclusions}

Precision electroweak data and rare processes have some sensitivity to the loop
corrections that might be induced by supersymmetric particles. As we discussed
previously in the context of the CMSSM~\cite{ehow3,LCWS05ehow3}, present data 
exhibit some preference for a relatively low scale of soft supersymmetry 
breaking: $m_{1/2} \sim 300 \ldots 600 \gev$. This preference is
largely driven by  
$(g - 2)_\mu$, with some support from measurements of $\MW$ and $\sweff$. In
this paper we have re-evaluated this preference, in the light of new
measurements of $m_t$ and $\MW$, and treating more completely the
information provided by the bound from the LEP direct searches for the Higgs
boson. The preference for $m_{1/2} \sim 300 \ldots 600 \gev$ is
maintained in the 
CMSSM, and also in other scenarios that implement different
assumptions for soft supersymmetry breaking. These include the less
constrained NUHM models in which the soft supersymmetry-breaking
scalar masses for the two Higgs multiplets are treated as 
free parameters as well as more constrained VCMSSM models in which the
soft trilinear and bilinear supersymmetry-breaking parameters are
related. The same preference is also maintained in GDM models
motivated by mSUGRA, where the LSP is the gravitino 
instead of being a neutralino as assumed in the other scenarios.

Whilst $m_{1/2}$ is quite constrained in our analysis, there are NUHM
scenarios in which $m_0$ could be considerably larger than the
corresponding values in the 
CMSSM, and significant variations in $\mu$ and $\MA$ are also possible. Within
the CMSSM and NUHM, we find no preference for any particular range of $\tb$,
but the preferred values of $m_{1/2}$ in the VCMSSM and GDM scenarios studied
here correspond to intermediate values of $\tb \sim 15$ to 30.

The ranges of $m_{1/2}$ that are preferred would correspond to gluinos and
other sparticles being light enough to be produced readily at the LHC. Many
sparticles would also be observable at the ILC in the preferred CMSSM, VCMSSM
and GDM scenarios considered, but the larger values of $m_0$ allowed in some of
the NUHM scenarios would reduce the number of sparticle species detectable at
the ILC, at least when operated at 500~GeV, whereas the ILC at 
$\sqrt{s} = 1000 \gev$ covers the full range for some sparticle
species. There are also prospects for
detecting supersymmetry at the Tevatron in some special VCMSSM models
with neutralino dark matter. 

We re-emphasize that our analysis depends in considerable part on the estimate
of the Standard Model contribution to $(g - 2)_\mu$ based on $e^+ e^-$
annihilation data, that we assume in this paper. Our conclusions would be
weakened if the Standard Model calculation were to be based on $\tau$ decay
data. Additional $e^+ e^-$ data are now coming available, and it will be
important to take into account whatever update of the Standard Model
contribution to $(g - 2)_\mu$ they may provide. However, the measurement of
$\MW$ is increasing in importance, particularly in the light of the recent
evolution of the preferred value of $m_t$. Future measurements of
$\MW$  and $\mt$ at the Tevatron will be particularly important in
this regard.


\subsection*{Acknowledgements}
S.H.\ and G.W.\ thank P.~Bechtle and K.~Desch for detailed
explanations on how to obtain $\chi^2$ values from the SM Higgs boson
searches at LEP. We thank A.~Read for providing the corresponding
$CL_s$ numbers. The work of S.H.\ was partially supported by CICYT
(grant FPA2004-02948) and DGIID-DGA (grant 2005-E24/2). The work of
K.A.O.\ was partially supported by DOE grant DE-FG02-94ER-40823.




\end{document}